\documentclass[11pt]{report}
\usepackage{slashed}
\usepackage{amsmath}
\usepackage{epsfig}
\usepackage{amssymb}
\usepackage{axodraw}
\newcommand{\la}[1]{\label{#1}}
\newcommand{\be}{\begin{equation}}
\newcommand{\ee}{\end{equation}}
\newcommand{\ba}{\begin{eqnarray}}
\newcommand{\ea}{\end{eqnarray}}
\newcommand{\rmi}[1]{{\mbox{\scriptsize #1}}}
\newcommand{\fig}{Fig.~}

\newcommand{\mc}{\mathcal}
\newcommand{\tr}{{\rm Tr\,}}
\newcommand{\nn}{\nonumber \\}
\newcommand{\fr}[2]{{\frac{#1}{#2}\,}}
\newcommand{\msbar}{{\overline{\mbox{\rm MS}}}}

\renewcommand{\(}{\left(}
\renewcommand{\)}{\right)}

\newcommand{\lk}{\left[}
\newcommand{\rk}{\right]}

\newcommand{\e}{\epsilon}

\newcommand{\mLARGE}[1]{\hbox{\LARGE $#1$}}
\def\sumint{\hbox{$\sum$}\!\!\!\!\!\!\!\int}

\renewcommand{\ln}{{\rm ln}}
\newcommand{\mubar}{\bar{\mu}}

\newcommand{\bmu}{\bar\Lambda}

\newcommand{\aG}{\alpha_\rmi{G}}

\newcommand{\aE}[1]{\alpha_\rmi{E#1}}
\newcommand{\bE}[1]{\beta_\rmi{E#1}}

\newcommand{\imathb}{i}

\newcommand{\pic}[1]{\;\parbox[c]{30pt}{\begin{picture}(30,30)(0,0)
\SetWidth{1.0}\SetScale{1.0} #1 \end{picture}}\;}

\newcommand{\picb}[1]{\;\parbox[c]{48pt}{\begin{picture}(45,30)(-9,0)
\SetWidth{1.0}\SetScale{1.0} #1 \end{picture}}\;}
\newcommand{\picc}[1]{\;\parbox[c]{45pt}{\begin{picture}(45,30)(0,0)
\SetWidth{1.0}\SetScale{1.0} #1 \end{picture}}\;}

\newcommand{\lpic}[1]{\;\parbox[c]{30pt}{\begin{picture}(15,0)(0,0)
\SetWidth{1.0}\SetScale{1.0} #1 \end{picture}}\;}
\newcommand{\pics}[1]{\;\parbox[c]{30pt}{\begin{picture}(60,60)(0,0)
\SetWidth{1.0}\SetScale{1.0} #1 \end{picture}}\;}
\newcommand{\piccc}[1]{\;\parbox[c]{45pt}{\begin{picture}(75,30)(0,0)
\SetWidth{1.0}\SetScale{1.0} #1 \end{picture}}\;}

\def\Lwidth{1}

\def\Agl(#1,#2)(#3,#4,#5){\PhotonArc(#1,#2)(#3,#4,#5){\Lwidth}
{6.283 #3 mul 360 div #4 #5 sub #4 #5 sub mul sqrt mul Ldensity mul}}
\def\Lgl(#1,#2)(#3,#4){\Photon(#1,#2)(#3,#4){\Lwidth}
{#1 #3 sub #1 #3 sub mul #2 #4 sub #2 #4 sub mul add sqrt Ldensity mul}}
\def\Agh(#1,#2)(#3,#4,#5){\DashArrowArc(#1,#2)(#3,#4,#5){1}}
\def\Aagh(#1,#2)(#3,#4,#5){\DashArrowArcn(#1,#2)(#3,#5,#4){1}}
\def\Lgh(#1,#2)(#3,#4){\DashArrowLine(#1,#2)(#3,#4){1}}
\def\Lagh(#1,#2)(#3,#4){\DashArrowLine(#3,#4)(#1,#2){1}}
\def\Ahh(#1,#2)(#3,#4,#5){\DashCArc(#1,#2)(#3,#4,#5){1}}
\def\Lhh(#1,#2)(#3,#4){\DashLine(#1,#2)(#3,#4){1}}
\def\Aqu(#1,#2)(#3,#4,#5){\ArrowArc(#1,#2)(#3,#4,#5)}
\def\Aaqu(#1,#2)(#3,#4,#5){\ArrowArcn(#1,#2)(#3,#5,#4)}
\def\Lqu(#1,#2)(#3,#4){\ArrowLine(#1,#2)(#3,#4)}
\def\Laqu(#1,#2)(#3,#4){\ArrowLine(#3,#4)(#1,#2)}
\def\Aqq(#1,#2)(#3,#4,#5){\CArc(#1,#2)(#3,#4,#5)}
\def\Lqq(#1,#2)(#3,#4){\ArrowLine(#1,#2)(#3,#4)}
\def\Asc(#1,#2)(#3,#4,#5){\ArrowArc(#1,#2)(#3,#4,#5)}
\def\Lsc(#1,#2)(#3,#4){\ArrowLine(#1,#2)(#3,#4)}
\def\DAsc(#1,#2)(#3,#4,#5){\DashCArc(#1,#2)(#3,#4,#5){3}}
\def\DLsc(#1,#2)(#3,#4){\DashLine(#1,#2)(#3,#4){3}}
\def\TAsc(#1,#2)(#3,#4,#5){\SetWidth{2.0}\CArc(#1,#2)(#3,#4,#5)\SetWidth{1.0}}
\def\TLsc(#1,#2)(#3,#4){\SetWidth{2.0}\ArrowLine(#1,#2)(#3,#4)\SetWidth{1.0}}

\def\TopoVRooT(#1,#2,#3){\;\pic{#1(15,15)(15,0,180) #1(15,15)(15,180,360)%
 \GCirc(0,15){3}{0} \GCirc(30,15){3}{0} \Text(5,15)[l]{{$\scriptstyle #2$}}%
 \Text(25,15)[r]{{$\scriptstyle #3$}}}\;}
\def\TopoVRorT(#1,#2,#3){\;\pic{#1(15,15)(15,0,180) #1(15,15)(15,180,360)%
 \GCirc(0,15){3}{0} \BBoxc(30,15)(6,6) \Text(5,15)[l]{{$\scriptstyle #2$}}%
 \Text(25,15)[r]{{$\scriptstyle #3$}}}\;}
\def\TopoVRoiT(#1,#2,#3){\;\pic{#1(15,15)(15,0,180) #1(15,15)(15,180,360)%
 \GCirc(0,15){3}{0} \GBoxc(30,15)(6,6){0} \Text(5,15)[l]{{$\scriptstyle #2$}}%
 \Text(25,15)[r]{{$\scriptstyle #3$}}}\;}
\def\TopoVRoooT(#1,#2,#3,#4){\pic{#1(15,15)(15,-30,90) #1(15,15)(15,90,210)%
 #1(15,15)(15,210,330) \GCirc(15,30){3}{0} \GCirc(2,7.5){3}{0}%
 \GCirc(28,7.5){3}{0} \Text(15,25)[t]{{$\scriptstyle #2$}}%
 \Text(6,9)[bl]{{$\scriptstyle #3$}} \Text(24,9)[br]{{$\scriptstyle #4$}}}}
\def\ToptVS(#1,#2,#3){\pic{#1(15,15)(15,0,180) #2(15,15)(15,180,360)%
 #3(30,15)(0,15)}}
\def\SToptVS(#1,#2,#3){\spic{#1(15,15)(15,0,180) #2(15,15)(15,180,360)%
 #3(30,15)(0,15)}}
\def\ToptVSTxt(#1,#2,#3,#4,#5){\;\pic{#1(15,15)(15,0,180)%
 #2(15,15)(15,180,360) #3(30,15)(0,15) \GCirc(0,15){3}{0} \GCirc(30,15){3}{0}%
 \Text(4,17)[bl]{{$\scriptstyle #4$}} \Text(26,13)[tr]{{$\scriptstyle #5$}}}\;}
\def\ToptVE(#1,#2){\picc{#1(15,15)(15,0,360) #2(45,15)(15,-180,180)}}
\def\SToptVE(#1,#2){\spicc{#1(15,15)(15,0,360) #2(45,15)(15,-180,180)}}
\def\ToprVM(#1,#2,#3,#4,#5,#6){\pic{#3(15,15)(15,-30,90) #1(15,15)(15,90,210)%
 #2(15,15)(15,210,330) #5(2,7.5)(15,15) #6(15,15)(15,30) #4(28,7.5)(15,15)}}
\def\SToprVM(#1,#2,#3,#4,#5,#6){\spic{#3(15,15)(15,-30,90)%
 #1(15,15)(15,90,210)%
 #2(15,15)(15,210,330) #5(2,7.5)(15,15) #6(15,15)(15,30) #4(28,7.5)(15,15)}}
\def\ToprVV(#1,#2,#3,#4,#5){\!\!\picb{#2(26.25,15)(15,256,76)%
 #3(30,30)(15,30) #1(18.75,15)(15,104,284) #4(15,30)(22.5,0)%
 #5(30,30)(22.5,0)}\!\!}
\def\SToprVV(#1,#2,#3,#4,#5){\!\!\spicb{#2(26.25,15)(15,256,76)%
 #3(30,30)(15,30) #1(18.75,15)(15,104,284) #4(15,30)(22.5,0)%
 #5(30,30)(22.5,0)}\!\!}
\def\ToprVB(#1,#2,#3,#4){\picb{#1(30,15)(15,-120,120) #2(30,15)(15,120,240)%
 #3(15,15)(15,60,300) #4(15,15)(15,-60,60)}}
\def\SToprVB(#1,#2,#3,#4){\spicb{#1(30,15)(15,-120,120) #2(30,15)(15,120,240)%
 #3(15,15)(15,60,300) #4(15,15)(15,-60,60)}}
\def\TopfVX(#1,#2,#3,#4,#5,#6,#7,#8,#9){\picb{#1(15,15)(15,90,270)%
 #2(30,15)(15,-90,90) #4(30,30)(15,30) #3(15,0)(30,0) #6(15,0)(15,15)%
 #5(15,15)(30,30) #8(15,30)(20,25) #8(25,20)(30,15) #7(30,15)(30,0)%
 #9(15,15)(30,15)}}
\def\STopfVX(#1,#2,#3,#4,#5,#6,#7,#8,#9){\spicb{#1(15,15)(15,90,270)%
 #2(30,15)(15,-90,90) #4(30,30)(15,30) #3(15,0)(30,0) #6(15,0)(15,15)%
 #5(15,15)(30,30) #8(15,30)(20,25) #8(25,20)(30,15) #7(30,15)(30,0)%
 #9(15,15)(30,15)}}
\def\TopfVH(#1,#2,#3,#4,#5,#6,#7,#8,#9){\picb{#1(15,15)(15,90,270)%
 #2(30,15)(15,-90,90) #4(30,30)(15,30) #3(15,0)(30,0) #6(15,0)(15,15)%
 #5(15,15)(15,30) #8(30,30)(30,15) #7(30,15)(30,0) #9(15,15)(30,15)}}
\def\STopfVH(#1,#2,#3,#4,#5,#6,#7,#8,#9){\spicb{#1(15,15)(15,90,270)%
 #2(30,15)(15,-90,90) #4(30,30)(15,30) #3(15,0)(30,0) #6(15,0)(15,15)%
 #5(15,15)(15,30) #8(30,30)(30,15) #7(30,15)(30,0) #9(15,15)(30,15)}}
\def\TopfVW(#1,#2,#3,#4,#5,#6,#7,#8){\pic{#1(15,15)(15,90,180)%
 #3(15,15)(15,180,270) #2(15,15)(15,270,360) #4(15,15)(15,0,90)%
 #5(15,15)(15,30) #7(15,15)(15,0) #6(0,15)(15,15) #8(30,15)(15,15)}}
\def\STopfVW(#1,#2,#3,#4,#5,#6,#7,#8){\spic{#1(15,15)(15,90,180)%
 #3(15,15)(15,180,270) #2(15,15)(15,270,360) #4(15,15)(15,0,90)%
 #5(15,15)(15,30) #7(15,15)(15,0) #6(0,15)(15,15) #8(30,15)(15,15)}}
\def\TopfVV(#1,#2,#3,#4,#5,#6,#7,#8){\!\!\picb{#2(26.25,15)(15,256,346)%
 #3(26.25,15)(15,-14,76) #4(30,30)(15,30) #1(18.75,15)(15,104,284)%
 #7(22.5,0)(15,30) #6(30,30)(26.25,15) #8(26.25,15)(22.5,0)%
 #5(25.25,15)(39.8,11.4)}\!\!}
\def\STopfVV(#1,#2,#3,#4,#5,#6,#7,#8){\!\!\spicb{#2(26.25,15)(15,256,346)%
 #3(26.25,15)(15,-14,76) #4(30,30)(15,30) #1(18.75,15)(15,104,284)%
 #7(22.5,0)(15,30) #6(30,30)(26.25,15) #8(26.25,15)(22.5,0)%
 #5(25.25,15)(39.8,11.4)}\!\!}
\def\TopfVB(#1,#2,#3,#4,#5,#6,#7){\picb{#2(30,15)(15,-120,120)%
 #6(30,15)(15,120,180) #5(30,15)(15,180,240) #1(15,15)(15,60,300)%
 #4(15,15)(15,-60,0) #3(15,15)(15,0,60) #7(30,15)(15,15)}}
\def\STopfVB(#1,#2,#3,#4,#5,#6,#7){\spicb{#2(30,15)(15,-120,120)%
 #6(30,15)(15,120,180) #5(30,15)(15,180,240) #1(15,15)(15,60,300)%
 #4(15,15)(15,-60,0) #3(15,15)(15,0,60) #7(30,15)(15,15)}}
\def\TopfVN(#1,#2,#3,#4,#5,#6,#7){\picb{#1(15,15)(15,90,270)%
 #2(30,15)(15,-90,90) #4(30,30)(15,30) #3(15,0)(30,0)%
 #5(15,0)(15,30) #6(30,30)(30,0) #7(15,30)(30,0)}}
\def\STopfVN(#1,#2,#3,#4,#5,#6,#7){\spicb{#1(15,15)(15,90,270)%
 #2(30,15)(15,-90,90) #4(30,30)(15,30) #3(15,0)(30,0)%
 #5(15,0)(15,30) #6(30,30)(30,0) #7(15,30)(30,0)}}
\def\TopfVU(#1,#2,#3,#4,#5,#6,#7){\pic{#3(15,15)(15,0,90)%
 #2(15,15)(15,90,180) #4(15,15)(15,180,270) #1(15,15)(15,270,360)%
 #6(0,15)(15,30) #7(15,0)(0,15) #5(30,15)(15,0)}}
\def\STopfVU(#1,#2,#3,#4,#5,#6,#7){\spic{#3(15,15)(15,0,90)%
 #2(15,15)(15,90,180) #4(15,15)(15,180,270) #1(15,15)(15,270,360)%
 #6(0,15)(15,30) #7(15,0)(0,15) #5(30,15)(15,0)}}
\def\TopfVT(#1,#2,#3,#4,#5,#6){\pic{#1(15,15)(15,90,210)%
 #2(15,15)(15,210,330) #3(15,15)(15,-30,90) #4(2,7.5)(15,30)%
 #6(28,7.5)(2,7.5) #5(15,30)(28,7.5)}}
\def\STopfVT(#1,#2,#3,#4,#5,#6){\spic{#1(15,15)(15,90,210)%
 #2(15,15)(15,210,330) #3(15,15)(15,-30,90) #4(2,7.5)(15,30)%
 #6(28,7.5)(2,7.5) #5(15,30)(28,7.5)}}
\def\STopEVa(#1,#2){\spicb{#1(15,15)(15,90,270)%
 #1(30,15)(15,-90,90) #2(30,30)(15,30) #2(15,0)(30,0) #2(15,0)(15,30)%
 #2(30,30)(30,0) #2(15,15)(30,15) #2(22.5,15)(22.5,0)}}
\def\STopEVb(#1,#2){\spicb{#1(15,15)(15,90,270)%
 #1(30,15)(15,-90,90) #2(30,30)(15,30) #2(15,0)(30,0) #2(15,0)(15,30)%
 #2(30,30)(30,0) #2(15,10)(30,10) #2(15,20)(30,20)}}
\def\STopEVc(#1,#2){\spicb{#1(15,15)(15,90,270)%
 #1(30,15)(15,-90,90) #2(30,30)(15,30) #2(15,0)(30,0) #2(15,0)(15,30)%
 #2(30,30)(30,0) #2(15,10)(30,20) #2(15,20)(19.5,17) #2(25.5,13)(30,10)}}
\def\STopEVd(#1,#2){\spicb{#1(15,15)(15,90,270)%
 #1(30,15)(15,-90,90) #2(30,30)(15,30) #2(15,0)(30,0) #2(15,0)(15,30)%
 #2(30,30)(30,0) #2(15,11.25)(30,11.25) #2(22.5,9)(22.5,0)%
 #1(15,11.25)(7.5,15,90)}}
\def\STopEVe(#1,#2){\spic{#1(15,15)(15,0,90)%
 #1(15,15)(15,90,180) #1(15,15)(15,180,270) #1(15,15)(15,270,360)%
 #2(0,15)(15,30) #2(10,15)(10,1) #2(30,15)(0,15) #2(20,15)(20,1)}}
\def\STopEVf(#1,#2){\spic{#1(15,15)(15,0,90)%
 #1(15,15)(15,90,180) #1(15,15)(15,180,270) #1(15,15)(15,270,360)%
 #2(0,15)(15,30) #2(15,6)(24,15) #2(30,15)(0,15) #2(15,15)(15,0)}}
\def\STopEVg(#1,#2){\spicb{#1(15,15)(15,90,270)%
 #1(30,15)(15,-90,90) #2(30,30)(15,30) #2(15,0)(30,0) #2(15,0)(15,30)%
 #2(30,30)(30,0) #2(15,15)(45,15)}}
\def\STopEVh(#1,#2){\!\!\spicb{#1(26.25,15)(15,256,346)%
 #1(26.25,15)(15,-14,76) #2(30,30)(15,30) #1(18.75,15)(15,104,284)%
 #2(22.5,0)(15,30) #2(30,30)(26.25,15) #2(26.25,15)(22.5,0)%
 #2(25.25,15)(39.8,11.4) #2(19.75,15)(5.2,11.4)}\!\!}
\def\STopEVi(#1,#2){\spic{#1(15,15)(15,0,90)%
 #1(15,15)(15,90,180) #1(15,15)(15,180,270) #1(15,15)(15,270,360)%
 #2(15,15)(15,30) #2(0,15)(30,15) #2(15,0)(10,15) #2(15,0)(20,15)}}
\def\STopEVj(#1,#2){\spicb{#1(15,15)(15,90,270)%
 #1(30,15)(15,-90,90) #2(30,30)(15,30) #2(15,0)(30,0) #2(15,0)(15,20)%
 #2(15,20)(22.5,30) #2(22.5,30)(30,20) #2(30,20)(30,0)%
 #2(15,10)(30,20) #2(15,20)(19.5,17) #2(25.5,13)(30,10)}}
\def\STopEVk(#1,#2){\spicb{#1(15,15)(15,90,270)%
 #1(30,15)(15,-90,90) #2(30,30)(15,30) #2(15,0)(30,0) #2(15,0)(15,30)%
 #2(30,0)(30,30) #2(15,0)(30,24) #2(20,30)(24.5,19.2) #2(26.5,14.4)(30,6)}}
\def\STopEVl(#1,#2){\!\!\spicb{#1(26.25,15)(15,256,346)%
 #1(26.25,15)(15,-14,76) #2(30,30)(15,30) #1(18.75,15)(15,104,284)%
 #2(22.5,0)(5.2,11.4) #2(30,30)(26.25,15) #2(26.25,15)(22.5,0)%
 #2(26.25,15)(39.8,11.4) #2(15,30)(5.2,11.4) }\!\!}
\def\STopEVm(#1,#2){\spicb{#1(30,15)(15,-120,120) #1(30,15)(15,120,240)%
 #1(15,15)(15,60,300) #1(15,15)(15,-60,60) #2(22.5,15)(30,15)%
 #2(22.5,15)(17,22.5) #2(22.5,15)(17,7.5) }}
\def\STopEVn(#1,#2){\spicb{#1(15,15)(15,90,270)%
 #1(30,15)(15,-90,90) #2(30,30)(15,30) #2(15,0)(30,0)%
 #2(15,0)(15,30) #2(30,30)(30,0) #2(15,30)(30,0) #2(30,15)(45,15)}}
\def\STopEVo(#1,#2){\spicb{#1(30,15)(15,-120,120) #1(30,15)(15,120,240)%
 #1(15,15)(15,60,300) #1(15,15)(15,-60,60)%
 #2(16,20.5)(29,20.5) #2(16,9.5)(29,9.5) }}
\def\STopEVp(#1,#2){\spic{#1(15,15)(15,0,90)%
 #1(15,15)(15,90,180) #1(15,15)(15,180,270) #1(15,15)(15,270,360)%
 #2(0,15)(15,30) #2(0,15)(30,15) #2(15,0)(10,15) #2(15,0)(20,15)}}
\def\STopEVq(#1,#2){\spic{#1(15,15)(15,0,90)%
 #1(15,15)(15,90,180) #1(15,15)(15,180,270) #1(15,15)(15,270,360)%
 #2(0,15)(15,30) #1(19,15)(4,0,360) #2(30,15)(23,15) #2(15,15)(0,15)%
 #2(15,15)(15,0)}}
\def\STopEVr(#1,#2){\spic{#1(15,15)(15,0,90)%
 #1(15,15)(15,90,180) #1(15,15)(15,180,270) #1(15,15)(15,270,360)%
 #2(0,15)(15,30) #1(26,15)(4,0,360) #2(22,15)(0,15)%
 #2(15,15)(15,0)}}
\def\STopEVs(#1,#2){\spic{#1(15,15)(15,0,90)%
 #1(15,15)(15,90,180) #1(15,15)(15,180,270) #1(15,15)(15,270,360)%
 #2(0.75,10.35)(29.25,10.35) #2(15,0)(15,10.35)%
 #2(0.75,10.35)(6.2,27.1) #2(29.25,10.35)(23.8,27.1)}}
\def\STopEVt(#1,#2){\spic{#1(15,15)(15,0,90)%
 #1(15,15)(15,90,180) #1(15,15)(15,180,270) #1(15,15)(15,270,360)%
 #2(0,15)(15,30) #1(15,4)(4,0,360) #2(30,15)(0,15) #2(15,15)(15,8)}}
\def\STopEVu(#1,#2){\spic{#1(15,15)(15,0,90)%
 #1(15,15)(15,90,180) #1(15,15)(15,180,270) #1(15,15)(15,270,360)%
 #2(0,15)(15,30) #2(15,7.5)(30,15) #2(30,15)(0,15) #2(15,15)(15,0)}}
\def\STopEVv(#1,#2){\spic{#1(15,15)(15,0,90)%
 #1(15,15)(15,90,180) #1(15,15)(15,180,270) #1(15,15)(15,270,360)%
 #2(0,15)(30,15) #2(0,15)(15,7.5) #2(15,0)(15,15) #2(15,0)(22.5,15)}}
\def\STopEVw(#1,#2){\spicb{#1(30,15)(15,-120,120) #1(30,15)(15,120,240)%
 #1(15,15)(15,60,300) #1(15,15)(15,-60,60) #2(0,15)(15,15) #2(30,15)(45,15)}}
\def\STopEVx(#1,#2){\spicb{#1(15,15)(15,90,270)%
 #1(30,15)(15,-90,90) #2(30,30)(15,30) #2(15,0)(30,0) #2(15,0)(15,15)%
 #2(15,0)(30,15) #2(30,0)(30,15) #2(15,15)(20.5,9.5) #2(24.5,5.5)(30,0)%
 #2(15,15)(22.5,22.5) #2(30,15)(22.5,22.5) #2(22.5,30)(22.5,22.5)}}
\def\STopEVy(#1,#2){\spicb{#1(15,15)(15,90,270)%
 #1(30,15)(15,-90,90) #2(30,30)(15,30) #2(15,0)(30,0) #2(15,0)(15,30)%
 #2(30,0)(30,30) #2(15,0)(30,24) #2(15,30)(22.375,18.2) #2(26.375,11.8)(30,6)}}
\def\STopEVz(#1,#2){\spic{#1(15,15)(15,0,90)%
 #1(15,15)(15,90,180) #1(15,15)(15,180,270) #1(15,15)(15,270,360)%
 #2(15,0)(0.75,10.35) #2(15,0)(29.25,10.35)%
 #2(0.75,10.35)(6.2,27.1) #2(29.25,10.35)(23.8,27.1)}}
\def\STopEVaa(#1,#2){\spic{#1(15,15)(15,0,90)%
 #1(15,15)(15,90,180) #1(15,15)(15,180,270) #1(15,15)(15,270,360)%
 #2(0,15)(15,30) #2(15,30)(30,15) #2(30,15)(0,15) #2(15,15)(15,0)}}
\def\STopEVab(#1,#2){\spic{#1(15,15)(15,0,90)%
 #1(15,15)(15,90,180) #1(15,15)(15,180,270) #1(15,15)(15,270,360)%
 #2(15,0)(0.75,10.35) #2(15,0)(23.8,27.1)%
 #2(0.75,10.35)(6.2,27.1) #2(29.25,10.35)(23.8,27.1)}}
\def\STopEVac(#1,#2){\spic{#1(15,15)(15,0,90)%
 #1(15,15)(15,90,180) #1(15,15)(15,180,270) #1(15,15)(15,270,360)%
 #2(0,15)(15,30) #2(15,0)(30,15) #2(30,15)(0,15) #2(15,15)(15,0)}}
\def\STopEVad(#1,#2){\!\!\spicb{#1(26.25,15)(15,256,346)%
 #1(26.25,15)(15,-14,76) #2(30,30)(15,30) #1(18.75,15)(15,104,284)%
 #2(22.5,0)(15,30) #2(30,30)(26.25,15) #2(26.25,15)(22.5,0)%
 #1(33,13.2)(7,0,360)}\!\!}
\def\STopEVae(#1,#2){\spic{#1(15,15)(15,90,180)%
 #1(15,15)(15,180,270) #1(15,15)(15,270,360) #1(15,15)(15,0,90)%
 #2(15,15)(15,30) #2(15,15)(15,0) #2(0,15)(15,15) #2(30,15)(15,15)%
 #2(15,0)(30,15)}}
\def\STopEVaf(#1,#2){\spicb{#1(15,15)(15,90,270)%
 #1(30,15)(15,-90,90) #2(30,30)(15,30) #2(15,0)(30,0)%
 #2(15,0)(15,30) #2(30,30)(30,0) #2(15,30)(22.5,0) #2(22.5,0)(30,30)}}
\def\STopEVag(#1,#2){\spicb{#1(15,15)(15,90,270)%
 #1(30,15)(15,-90,90) #2(30,30)(15,30) #2(15,0)(30,0) #2(15,0)(15,15)%
 #2(15,15)(22.5,30) #2(22.5,30)(30,15) #2(30,15)(30,0) #2(15,15)(30,15)%
 #2(15,0)(30,15)}}
\def\STopEVah(#1,#2){\spic{#1(15,15)(15,0,90)%
 #1(15,15)(15,90,180) #1(15,15)(15,180,270) #1(15,15)(15,270,360)%
 #2(0,15)(15,30) #2(15,0)(0,15) #2(30,15)(15,0) #2(30,15)(15,30)}}
\def\STopEVai(#1,#2){\spicb{#1(30,15)(15,-120,120) #1(30,15)(15,120,240)%
 #1(15,15)(15,60,300) #1(15,15)(15,-60,60) #1(22.5,15)(7.5,0,360) }}

\begin{document}

\begin{titlepage}
\begin{flushright}
hep-ph/0402242\\
HU-P-D109\\
\end{flushright}
\begin{centering}
\vfill

{\large {\bf The pressure of QCD at finite temperature and quark number density}}

\vspace{0.8cm}

{A. Vuorinen\footnote{aleksi.vuorinen@helsinki.fi}}

\vspace{0.8cm}

{\em
Department of Physical Sciences and Helsinki Institute of Physics,\\
P.O. Box 64,
FI-00014 University of Helsinki,
Finland\\}

\vspace*{1.4cm}

\end{centering}

\noindent

This paper is a slightly modified version of the introductory part of a doctoral dissertation that contained also three original articles, hep-ph/0212283, hep-ph/0305183 and hep-ph/0311323. Our purpose is to review the history and present status of finite-temperature perturbation theory as applied to the context of determining the equilibrium properties of quark-gluon plasma, most notably the pressure of QCD at finite temperatures and quark chemical potentials. We first introduce the general formalism of finite-temperature field theory and perturbation theory, then follow through the evaluation of the pressure order by order, and finally proceed to analyze the most recent, order $g^6\ln\,g$ results by comparing the perturbative predictions with lattice data. In the appendix we provide a somewhat pedagogical introduction to the most important computational techniques used in the perturbative framework, namely the analytic evaluation of multi-loop vacuum diagrams both in full QCD and in its three-dimensional high-$T$ effective theories.

\vfill
\noindent

\vspace*{1cm}

\noindent

\vfill

\end{titlepage}

\tableofcontents

\thispagestyle{empty}

\newpage

%

\setcounter{page}{1}
\pagenumbering{arabic}

\chapter{Introduction}

Quantum chromodynamics, or QCD, is the theory of the strong interaction. One of its most remarkable features can be seen in the way a hadronic system behaves, when it is either strongly heated or compressed. The system initially composed of individual baryons and mesons --- particles whose substructure can at lower energies be studied only indirectly --- transforms at an energy density of roughly 1 GeV/fm$^3$ into a plasma of deconfined, though strongly interacting, quarks and gluons. This transition is today at the focus of considerable interest in the high energy physics community due to the fact that quark-gluon plasma, the phase of hot deconfined matter, is currently being produced in ultra-relativistic heavy-ion collisions at RHIC in Brookhaven. A similar large-scale experiment is also being prepared at LHC in CERN.

The rapid progress in experimental heavy-ion physics has set an important challenge for theorists. One needs to have a solid understanding of the processes that take place in the nuclear collisions and in particular obtain accurate numerical predictions for the different quantities measured. Whether or not the plasma produced in the present-day experiments has had time to thermalize and reach an equilibrium state, it is clear that one of the most important and fundamental quantities describing the deconfined phase of QCD matter is the grand potential $\Omega=-pV$ of quark-gluon plasma. Its value is of relevance both to the study of the evolution of the heavy-ion collision products in terms of ideal hydrodynamics, and to cosmology, as e.g.~the cooling rate of the very early universe depends on the energy and entropy densities of its content. 

The purpose of the present paper is to study the perturbative evaluation of the QCD pressure at finite quark chemical potentials, i.e.~keeping the net number densities of the different flavors arbitrary. It was originally written as the introductory part of a doctoral thesis regarding the subject and consisting of three articles \cite{avsusc,avpres,avyork}, and we have therefore set as one of our main objectives to present a review of the theoretical and mathematical background behind this work. This includes following through the construction of the functional integral representation of the QCD partition function, deriving from it the finite-temperature Feynman rules of the theory, and finally in some detail outlining the calculation of the most recent $\mc{O}(g^6\ln\,g)$ result for the pressure at finite $T$ and $\mu$ \cite{avpres}. We will also analyze the renormalization scale dependence of the perturbative results and study their agreement with current lattice data, and in doing so will extend the treatment of the original articles somewhat.

The paper is organized as follows. In chapter 2 the theoretical structure of QCD is reviewed, and some of the most
fundamental results of finite-temperature field theory are presented. A special emphasis is given to the construction of
the partition function of non-Abelian gauge field theories and to the derivation of generic rules for finite-temperature perturbation theory. Chapter 3 is then devoted to describing the history and present status of the perturbative determination of the QCD pressure, and we analyze in particular how the description of the full four-dimensional theory can at high temperatures be reduced into studying three-dimensional effective ones. The special features encountered when trying to approach the limit of small temperatures and large chemical potentials in the perturbative expansion are highlighted as well.

Chapter 4 contains a review and analysis of the results of Refs.~\cite{avsusc,avpres}. We extend the discussion
of Ref.~\cite{avpres} by studying the renormalization scale dependence of the perturbative expansion of the pressure and
by making comparisons with the results of a very recent lattice computation of Allton \textit{et al.} \cite{karsch2}. In
chapter 5 we then finally draw conclusions, while the appendices contain an introduction to the analytic methods used in
Ref.~\cite{avpres} to compute three-loop diagrams in finite-temperature QCD and in Ref.~\cite{avyork} to evaluate massive
four-loop diagrams in three-dimensional field theories. Due to the highly technical and rather lengthy nature of this
part we have wanted to keep it separate from the main body of the text.

In the following we will always work in natural units, i.e.~we set $c\;=\;\hbar\;=\;k_B=\;1$. Other notational
conventions will be fixed at the end of chapter 2.

\chapter{Basics of QCD thermodynamics}
In this chapter we will present a brief introduction to the general properties of QCD and to the basic principles of
statistical mechanics. We will especially concentrate on deriving a proper functional integral representation
for the QCD partition function, and end the chapter by listing the finite-temperature Feynman rules for the theory. More
detailed presentations on the subject can be found e.g.~in Refs.~\cite{kap,leb,gpy}.

\section{General properties of the theory}
Quantum chromodynamics is a renormalizable non-Abelian gauge field theory developed in the early 1970's to describe the
observed features of the strong interaction \cite{QCD1}. Its symmetry group is the color group SU(3), but for the sake of
completeness we will in this work consider a more general case of SU($N$) keeping the number of colors $N$ a free
parameter. The basic constituents of the theory are then the $N^2-1$ gauge bosons, gluons, and the $n_f$ flavors of
spin-1/2 fermions, quarks, whose bound states the hadrons are. The gluons belong to the adjoint representation of the
gauge group and are denoted by $A_{\mu}^a$, where $\mu = 0,...,3$ is a Lorentz and $a=1,...,N^2-1$ a color index. The
quarks on the other hand transform according to the fundamental representation of SU($N$) and are here assembled into a
single $4Nn_f$-component spinor $\psi$ with Dirac, color and flavor indices suppressed. In
nature the number of flavors is of course six, corresponding to the up, down, strange, charm, top and bottom quarks. The
last three have, however, masses above the GeV-scale, and as the particles with masses larger than the energy scale of
interest (here a few hundred MeV) very efficiently decouple, we have in all practical applications $n_f \leq 3$. In
this work we will furthermore regard all relevant quark flavors as massless, making $n_f=2$ often the most realistic
value of the parameter.

One of the most fundamental properties of QCD is its asymptotic freedom \cite{growil,pol}, which states that at large
energies the gauge coupling constant $g$ of the theory approaches zero. This can be seen most easily from the running of
the coupling as obtained from the leading order solution to the renormalization group equation,
\ba
g^2(\Lambda)&=&\fr{24\pi^2}{(11N-2n_f)\ln(\Lambda / \Lambda_\rmi{QCD})},\label{gres1l}
\ea
where $\Lambda$ is the renormalization scale and $\Lambda_\rmi{QCD}\sim 150$ MeV a free parameter corresponding to the
characteristic energy scale of the theory. Asymptotic freedom implies that at very small distances the behavior of QCD
tends to that of a free field theory making the use of perturbation theory feasible in the description of hard processes
such as deep inelastic scattering. From the thermodynamic point of view this means that at least in the limit of
asymptotically high temperatures or chemical potentials one might expect a perturbative approach to be fruitful in the
computation of the partition function. This is an important observation, and its validity will be examined in the
following chapters.

\begin{figure}[t]

\centerline{\epsfxsize=12cm\epsfysize=9cm \epsfbox{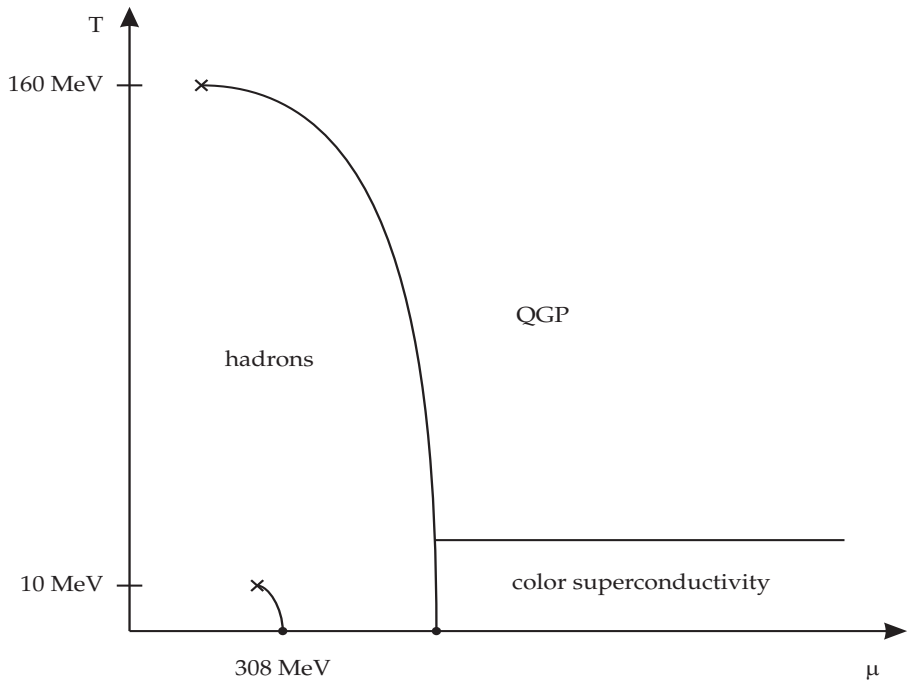}}

\caption[a]{A schematic phase diagram of QCD \cite{dris} in terms of the independent variables temperature and quark
chemical potential $\mu=\mu_\rmi{B}/3$.}
\end{figure}

Before moving on to a more quantitative description of QCD thermodynamics let us briefly review the present understanding
of the phase structure of the theory, which is summarized in Fig.~2.1. The ($\mu$-$T$)-plane is divided into three parts corresponding to the three fundamental phases of strongly
interacting matter. At temperatures below $T_c\sim 160$ MeV and chemical potentials below $\mu_c\sim 350$ MeV, QCD matter
appears in the form of hadrons in either a liquid or a gaseous form depending on the values of $\mu$ and $T$. There is a
first order liquid-gas transition line originating from the ground state of nuclear matter at $(\mu,T)=(308,0)$ MeV and
ending at a critical endpoint at $T\sim 10$ MeV. Above this temperature there is merely a smooth cross-over from the
dilute gaseous phase into the more dense liquid phase.

At temperatures and chemical potentials higher than $T_c$ and $\mu_c$, strongly interacting matter appears in a deconfined
phase, where the quarks are liberated from their confinement. There
is again a first order transition line dividing the hardonic and deconfined phases, which starts from
approximately $(350,0)$ MeV and ends at a critical point at roughly $(240,160)$ MeV. If the matter is even more dilute
than this, the transition is believed to become a cross-over. The details of this pattern are, however, strongly dependent on the
number of flavors as well as the values of the quark masses (see e.g.~Ref.~\cite{laeph}).

The deconfined phase of QCD is further composed of two distinct parts. At high temperatures strongly interacting matter
forms a color-neutral plasma of almost freely moving quarks and gluons --- the quark-gluon plasma --- the equilibrium
thermodynamics of which we will study in the present work. At low temperatures and large chemical potentials
the properties of the system are on the other hand completely different, as its behavior is dominated by an
attractive interaction between two quarks lying on the Fermi surface. Analogously to the electromagnetic case these
quarks form Cooper pairs, which now constitute the ground state of a third fundamental phase of strongly interacting
matter, the color superconductor \cite{rajwil}. The superconducting phase has a rich internal structure on the
($\mu$-$T$)-plane composed of e.g.~a color-flavor locked and a crystalline phase, and it terminates in a melting transition at a
temperature of the order of several MeV (for a review, see Ref.~\cite{rajwil2}). The description of these phenomena is,
however, outside the scope of our present treatment.

\section{Quantum statistical mechanics}
To prepare for the main topic of our presentation, the perturbative evaluation of the QCD pressure, let us first consider
the thermodynamics of a general grand canonical ensemble, i.e.~a system which can freely exchange energy and particles
with its surroundings. Its most important equilibrium properties can be straightforwardly obtained as soon as one has
computed the corresponding partition function, which is defined as the trace of the density operator taken over all the quantum states of the system,
\ba
Z(V,T,\{\mu_i\})&\equiv&\tr \rho\;\;\,=\;\;\,\tr {\rm exp}\big[-\beta\big(H-\sum_i \mu_i N_i\big)\big]. \label{zdef}
\ea
Here $H$ is the Hamiltonian, $N_i$ are the conserved number operators, $\mu_i$ the corresponding chemical potentials and
$\beta\equiv 1/T$. Quantities such as the pressure, entropy and different particle numbers are available as first
derivatives of $Z$ through the fundamental relations
\ba
P &=& T\fr{\partial\,\ln Z}{\partial V}, \\
S &=& \fr{\partial (T\ln Z)}{\partial T}, \\
N_i &=& T\fr{\partial\,\ln Z}{\partial \mu_i},
\ea
and the internal energy can furthermore be expressed as a linear combination of these. Different response functions, such as the susceptibilities,
\ba
\chi_{ijk...}&\equiv&\fr{\partial^n P}{\partial\mu_i \partial\mu_j \partial\mu_k ...},
\ea
are as well directly obtainable as higher derivatives of the partition function, and thermal averages of different operators can be determined through the formula
\ba
\langle A\rangle &=&\fr{1}{Z}\tr (\rho A).
\ea

Apart from a few special cases, such as non-interacting systems, the partition function is unfortunately only
rarely obtainable by
analytically performing the trace of Eq.~(\ref{zdef}) as an explicit sum over the different energy eigenstates. In field
theories a useful general representation for $Z$ can, however, be obtained by using an obvious analogy to the functional
integral form of the transition amplitude in zero-temperature quantum field theory. The trace appearing in
the definition of the partition function can namely be written as an integral\footnote{To be specific, a functional
integral over the space of normalizable and suitably smooth coordinate space functions.} over the states of the system as
\ba
Z&=& \int {\rm d}\phi \;\langle \phi\!\mid {\rm exp}\big[-\beta\big(H-\sum_i \mu_i N_i\big)\big]\mid \!\phi \rangle ,
\ea
and regarding the integrand as a transition amplitude over an imaginary time $-i\beta$ we get for a neutral
scalar field ($\tau = it$)
\ba
Z&=&\int_{per.}\mathcal{D}\phi \,{\rm exp}\big[-\int_0^{\beta}{\rm d} \tau \int {\rm d}^3 x \,\mathcal{L}\big], \label{zbos}
\ea
and for a charged Dirac field
\ba
Z&=&\int_{antiper.}\mathcal{D}\bar{\psi}\mathcal{D}\psi\,{\rm exp}
\big[-\int_0^{\beta}{\rm d} \tau \int {\rm d}^3 x \,\big({\mathcal L}- \mu\psi^{\dagger}\psi\big)\big]. \label{zferm}
\ea
As in indicated here, the boundary conditions require the bosonic field to be periodic on the interval
$[0,\beta]$, whereas the fermionic field is antiperiodic,
\ba
\phi(\tau=0)&=&\phi(\tau=\beta), \\
\psi(\tau=0)&=&-\psi(\tau=\beta), \\
\bar{\psi}(\tau=0)&=&-\bar{\psi}(\tau=\beta).
\ea
While the first of these relations is trivial, the origin of the latter two can be found from the Grassmannian nature
of the integration variables in Eq.~(\ref{zferm}).

Above we have adopted a notation, in which $\psi^{\dagger}$ is regarded as a function of $\bar{\psi}$,
\ba
\psi^{\dagger}&\equiv&\bar{\psi}\gamma_0,
\ea
and all trivial normalization factors of the functional integrals are suppressed. The choice of the imaginary time
formalism, on the other hand, implies that we from now on work exclusively in Euclidean metric with the
identification $x_0\rightarrow -ix_0$. This means that in all standard Minkowskian equations we must make the substitutions
\ba
x^{\mu}x_{\mu}\;\;\,\equiv\;\;\, x_0^2-x_i^2&\rightarrow &-x_0^2-x_i^2\;\;\,\equiv\;\;\,-x_{\mu}x_{\mu} \\
\gamma_i&\rightarrow&-i\gamma_i,
\ea
which renders all $\gamma$-matrices Hermitean according to the convention chosen. The time variable $\tau$ will henceforth be always denoted by $x_0$.

It is well known that the analytic evaluation of functional integrals appearing in interacting field theories is in
general impossible. Apart from a numerical approach, the only method available for us to compute the integrals of
Eqs.~(\ref{zbos}) and (\ref{zferm}) for a given Lagrangian is usually to resort to a perturbative approach, i.e.~to
expand the interaction part of the exponentiated action in a series in the coupling constant. This reduces the
determination of the partition function to the evaluation of the connected vacuum diagrams of the theory with the main
difference to the zero-temperature case being just the replacement of the momentum space loop integrals by
sum-integrals --- a direct consequence of the finiteness of the temporal direction in the coordinate space. In the next
sections we will see an explicit example of the finite-temperature machinery in action, as we now move on to apply the
above general discussion to the special case of QCD starting from the very basics, i.e.~Eq. (\ref{zdef}).

\section{Construction of the QCD partition function}
We define massless quantum chromodynamics by the bare Euclidean Lagrangian density
\ba
{\cal L}_\rmi{QCD} & = &  \fr14 F_{\mu\nu}^a F_{\mu\nu}^a + \bar\psi\slashed{D}\psi,
\ea
where the field strength tensor and the covariant (fermionic) derivative read
\ba
F_{\mu\nu}^a &\equiv& \partial_\mu A_\nu^a - \partial_\nu A_\mu^a + g f^{abc} A_\mu^b A_\nu^c,\\
D_\mu &\equiv& \partial_\mu - i g A_\mu \;\;\,\equiv\;\;\, \partial_\mu - i g A_\mu^a T^a \label{covd}
\ea
and where we have, as usual, defined
\ba
\slashed{D}&\equiv&\gamma_{\mu}D_{\mu}.
\ea
The symbols $T^a$, $a=1,...,N^2-1$, denote the generators of the fundamental representation of SU($N$) and are $N\times N$ matrices in color space. They are used to define the usual antisymmetric and symmetric structure coefficients $f^{abc}$ and $d^{abc}$ through the relations
\ba
\big[T^a,T^b\big]&=&if^{abc}T^c, \\
\{T^a,T^b\}&=&id^{abc}T^c,
\ea
and the antisymmetric coefficients furthermore act as the generators of the adjoint representation of the gauge group as
\ba
(\tau^a)^{bc}&\equiv&if^{abc}, \\
\big[\tau^a,\tau^b\big]&=&if^{abc}\tau^c.
\ea
In this work we, however, stick to the convention of writing them out explicitly; e.g.~the covariant derivative $D$ will
in the following always refer to that of the fundamental representation, Eq.~(\ref{covd}). Another convention we have
chosen is to write all indices of the adjoint representation as superscripts.

As can be straightforwardly verified, the QCD action is invariant under the local gauge transformations
\ba
A_{\mu}\;\;\,\equiv\;\;\,A_{\mu}^aT^a&\rightarrow &\Omega^{-1}A_{\mu}\Omega+\fr{i}{g}\(\partial_{\mu}\Omega^{-1}\)\Omega, \label{atrans}\\
\psi&\rightarrow & \Omega^{-1}\psi,
\ea
where the transformation matrix $\Omega$ reads
\ba
\Omega&=&{\rm exp}\big[igT^a\alpha^a\big]
\ea
and $\alpha^a$ are arbitrary smooth functions. This causes a problem when trying to evaluate the partition function
of the theory, as a proper gauge condition must be imposed on the fields in order to avoid an overcounting of degrees of
freedom: we must be able to choose a unique representative from all gauge orbits to contribute to the quantity. In the
following we will briefly explain how this can be implemented in the
functional integral representation of $Z_\rmi{QCD}$ starting from its definition
\ba
Z_\rmi{QCD}&=&\tr {\rm exp}\big[-\beta\big(H-\sum_f \mu_f N_f\big)\big]. \label{zqdef}
\ea
Here the sum runs over all relevant quark flavors, and the corresponding chemical potentials $\mu_f$ may be assumed independent as long as we are
dealing with energies sufficiently below the scale of the flavor-mixing weak interactions, $m_W$.

A convenient choice for the present purposes is to work in the $A_0=0$ (temporal) gauge, which leads to a relatively simple
algebra but has the disadvantage of not fixing the gauge completely (see e.g.~Refs.~\cite{gpy,tt2}). The Lagrangian now reads
\ba
L_\rmi{temp}&=&\int {\rm d}^3 x \bigg\{\fr{1}{2}\dot{A}_i^a\dot{A}_i^a+\fr14 F_{ij}^a F_{ij}^a + \bar\psi\gamma_iD_i\psi +\psi^{\dagger}\partial_0\psi \bigg\},
\ea
and defining the canonical momenta corresponding to the coordinates $A_i$ in the usual way,
\ba
\Pi_i^a&\equiv&\fr{\delta L}{\delta \dot{A}_i^a}\;\;\,=\;\;\,\dot{A}_i^a,
\ea
we obtain as the Hamiltonian
\ba
H_\rmi{temp}\;\;\,\equiv\;\;\,\int {\rm d}^3x\,\mc{H}&=&\int {\rm d}^3x\bigg\{\fr{1}{2}\Pi_i^a\Pi_i^a-\fr14 F_{ij}^a F_{ij}^a - \bar\psi\gamma_iD_i\psi -\psi^{\dagger}\partial_0\psi\bigg\}.
\ea

Despite the gauge fixing, we still have the freedom to make gauge transformations independent of $x_0$. This is a
consequence of the fact that the Gauss law,
\ba
G^a&\equiv&\partial_iF^a_{i0}+gf^{abc}A^b_iF^c_{i0}+T^a\psi^{\dagger}\psi\;\;\,=\;\;\,0, \label{gauss}
\ea
where $G^a$ is the generator of time-independent gauge transformations, doesn't appear in the Hamiltonian equations of
motion
\ba
\dot{\Pi}_i^a&=&-\fr{\delta H}{\delta A_i^a},\;\;\;
\dot{A}_i^a\;\;\,=\;\;\,\fr{\delta H}{\delta \Pi_i^a}.
\ea
When evaluating the partition function we thus must impose Eq.~(\ref{gauss}) explicitly on the states
contributing to the result, which is most easily accomplished by including the operator
\ba
P&=& \int_{\Lambda(\infty)=0}\mathcal{D}\Lambda\,{\rm exp}\big[i\beta\int{\rm d}^3x\Lambda^aG^a\big]
\ea
into the definition of the partition function, where it acts as a projector onto the space of physical
states. As is indicated above, only fields $\Lambda$, which vanish at spatial infinity, contribute to $P$.

Following the treatment of Refs.~\cite{gpy,tt2}, we now substitute $P$ into Eq.~(\ref{zqdef}) in the form
\ba
Z_\rmi{QCD}&=&\lim_{n\rightarrow\infty}\tr\bigg( P{\rm exp}\Big[-\beta\big(H_\rmi{temp}-\sum_f \mu_f N_f\big)/n\Big]\bigg)^n,
\ea
which after the repeated use of the standard completeness relation and the renaming of the integration variable $\Lambda$
as $A_0$ leads to the functional integral representation
\ba
\label{zqcdweyl2}
Z_\rmi{QCD}&=&\int_{A_{\mu}\, per.\atop \psi\, antip.}\!\!\!\!\mathcal{D}A_{\mu}\mathcal{D}\Pi_{i}\mathcal{D}\bar{\psi}\mathcal{D}\psi\\
&\times&{\rm exp}
\Big[-\int_0^{\beta}{\rm d} x_0 \int {\rm d}^3 x \,\Big\{\mathcal{H}-i\Pi_i^a\partial_0A_i^a-iA_0^a(\partial_i\delta^{ab}-gf^{abc}A_i^c)\Pi_i^b
-\psi^{\dagger}\(iA_0^aT^a+{\mbox{\boldmath$\mu$}}\)\psi\Big\}\Big]. \nonumber
\ea
Here we have chosen periodic boundary conditions\footnote{This is based on mere computational convenience. It has been
shown by Montonen (see Ref.~\cite{tt2}) that in the temporal gauge the partition function is actually independent of the boundary
conditions of $\Lambda$.} for the $A_0$ field and have furthermore denoted
\ba
{\mbox{\boldmath$\mu$}}&\equiv& {\rm diag}(\mu_1,\mu_2,...,\mu_{n_f}).
\ea

The integral over the conjugate momenta in Eq.~(\ref{zqcdweyl2}) is of a trivial Gaussian form and its subsequent
evaluation leads us to the familiar expression
\ba
Z_\rmi{QCD}&=&\int_{A_{\mu}\, per.\atop \psi\, antip.}\!\!\!\!\mathcal{D}A_{\mu}
\mathcal{D}\bar{\psi}\mathcal{D}\psi\,{\rm exp}
\big[-\int_0^{\beta}{\rm d} x_0 \int {\rm d}^3 x \,\big(\mathcal{L}_\rmi{QCD}- \psi^{\dagger}{\mbox{\boldmath$\mu$}}
\psi\big)\big], \label{zqcdweyl}
\ea
analogous to the general results of Eqs.~(\ref{zbos}) and (\ref{zferm}). This form of the partition function however
still contains excess gauge freedom, as it --- due to the periodicity choice made above --- is invariant under
SU($N$) gauge transformations periodic in the temporal direction,
\ba
\Omega(x_0=\beta)&=&\Omega(x_0=0). \label{per1}
\ea
This renders the functional integral singular due to the infinite volume of the corresponding local gauge
group,\footnote{Despite the compactness of SU($N$), the locality of the gauge invariance makes the volume of the group
infinite. The integration measure is defined in such a way that in an infinitesimal region we may write
\ba
\mathcal{D}\Omega&=&\prod_{x,a}{\rm d}\alpha_a(x) \label{infiint}
\ea
and that the relation
\ba
\int \mathcal{D}\Omega\,f[\Omega]&=&\int\mathcal{D}\Omega\,f[\Omega\Omega '] \label{groupint}
\ea
holds, where $\Omega'$ denotes an independent gauge transformation (see Refs.~\cite{rothe,bl,taylor}).}
\ba
V_{\Omega}&\equiv&\int_{\Omega\,per.\atop\in\,{\rm SU}(N)}\!\!\!\!\mathcal{D}\Omega\;\;\,=\;\;\,\infty,
\ea
and implies that we must attempt to modify Eq.~(\ref{zqcdweyl}) in order to be able to factor out a piece proportional to
this integral.

To fix the gauge freedom remaining in the above form of $Z_\rmi{QCD}$, let us now introduce the covariant gauge
condition
\ba
F^a[A]&\equiv&\partial_{\mu}A_{\mu}^a-f^a\;\;\,=\;\;\,0,
\ea
where $f^a$ is an undetermined function of $x$. Following the standard Faddeev-Popov procedure we insert the
functional
\ba
\Delta[A]&\equiv&\int_{\Omega\,per.\atop\in\,{\rm SU}(N)}\!\!\!\!\mathcal{D}\Omega \,\delta[F^a[A^{\Omega}]],
\ea
where $A^{\Omega}$ denotes the gauge transformed field of Eq.~(\ref{atrans}), into Eq.~(\ref{zqcdweyl}) in the form $\Delta\Delta^{-1}=1$. Using the gauge invariance properties of the action
\ba
S&\equiv&\int_0^{\beta}{\rm d} x_0 \int {\rm d}^3 x \,\big(\mathcal{L}_\rmi{QCD}- \psi^{\dagger}{\mbox{\boldmath$\mu$}}
\psi\big),
\ea
of the functional $\Delta$ and of the integration measure $\mc{D}A_{\mu}$, we then obtain
\ba
Z_\rmi{QCD}&=&\int_{A_{\mu}\, per.\atop \psi\, antip.}\!\!\!\!\mathcal{D}A_{\mu}\mathcal{D}\bar{\psi}\mathcal{D}\psi\,{\rm exp}
\big[-S[A]\big] \nn
&=&\int_{A_{\mu}\, per.\atop \psi\, antip.}\!\!\!\!\mathcal{D}A_{\mu}\mathcal{D}\bar{\psi}\mathcal{D}\psi\,\Delta^{-1}[A]\int_{\Omega\,per.\atop\in\,{\rm SU}(N)}\!\!\!\!\mathcal{D}\Omega \,\delta[F^a[A^{\Omega}]]\,{\rm exp} \big[-S[A]\big]\nn
&=&\int_{\Omega\,per.\atop\in\,{\rm SU}(N)}\!\!\!\!\mathcal{D}\Omega \,\int_{A^{\Omega}_{\mu}\, per.\atop \psi\, antip.}\!\!\!\!\mathcal{D}A^{\Omega}_{\mu}\mathcal{D}\bar{\psi}\mathcal{D}\psi\,\Delta^{-1}[A^{\Omega}]\delta[F^a[A^{\Omega}]]\,{\rm exp} \big[-S[A^{\Omega}]\big],
\ea
where we in the final stage may clearly drop the $\Omega$-integral as a trivial constant factor. This finally gives us the result
\ba
Z_\rmi{QCD}&=&\int_{A_{\mu}\, per.\atop \psi\, antip.}\!\!\!\!\mathcal{D}A_{\mu}\mathcal{D}\bar{\psi}\mathcal{D}\psi\,\Delta^{-1}[A]\delta[F^a[A]]\,{\rm exp} \big[-S[A]\big], \label{covg}
\ea
from which all excess gauge freedom has been removed.

What remains to be done is to express the functional $\Delta^{-1}$ in a more practical form, which can be achieved by
writing it as a Faddeev-Popov determinant
\ba
\Delta^{-1}[A]&=&\det\left.\(\fr{\delta F^a(x)}{\delta\alpha^b(x')}\)\right|_{F^a=0}\;\;\,\equiv\;\;\,\det M^{ab},
\ea
a relation straightforwardly verifiable using Eq.~(\ref{infiint}). The determinant can, on the other hand, be expressed in
the form of a standard Gaussian integral over Grassmannian `ghost' fields $\eta,\,\bar{\eta}$,
\ba
\det M^{ab}&=&\int_{\eta\, per.}\!\!\!\!\mathcal{D}\bar{\eta}\mathcal{D}\eta\,{\rm exp}\bigg[-\int_0^{\beta}{\rm d} x_0 \int {\rm d}^3 x \int_0^{\beta}{\rm d} y_0 \int {\rm d}^3 y \,\bar{\eta}^a(x)M^{ab}(x,y)\eta^b(y) \bigg],
\ea
where the integration variables obey periodic boundary conditions on the interval $[0,\beta]$ as a result of the
invariance of the partition function under periodic gauge transformations. This appearance of anticommuting
but periodic fields in the theory is thus merely a simple consequence of the choice we made for the boundary conditions
of $A_0$. Other choices would have been possible, too, but would result in computational problems later.

Inserting the above ghost field representation of $\Delta^{-1}$ into Eq.~(\ref{covg}) together with the explicit
form of $M^{ab}$,
\ba
M^{ab}(x,y)&=&\partial_{\mu}\Big\{\(\partial_{\mu}\delta^{ab} + gf^{abc}A_{\mu}^c\)\delta(x-y)\Big\},
\ea
we have now successfully reduced the partition function into a well-defined form. It can be applied to practical
calculations as soon as the remaining $\delta$-functional has been removed, which is easily achieved by
multiplying the functional integral by the factor
\ba
{\rm exp}\bigg[-\fr{1}{2\xi}\int_0^{\beta}{\rm d} x_0 \int {\rm d}^3 x (f^a(x))^2  \bigg],
\ea
and subsequently integrating over the functions $f^a$, on which the partition function clearly cannot depend. This
leads us to the final result
\ba
Z_\rmi{QCD}&=&\int_{A_{\mu}\, per.\atop \psi\, antip.}\!\!\!\!\mathcal{D}A_{\mu}\mathcal{D}\bar{\psi}\mathcal{D}
\psi\mathcal{D}\bar{\eta}\mathcal{D} \eta\,{\rm exp} \bigg[-\int_0^{\beta}{\rm d} x_0 \int {\rm d}^3 x
\,\mathcal{L}_\rmi{eff}  \bigg], \label{zres}
\ea
where the effective Lagrangian density $\mathcal{L}_\rmi{eff}$ is given by
\ba
\mathcal{L}_\rmi{eff}&=&\mathcal{L}_\rmi{QCD}+\fr{1}{2\xi}(\partial_{\mu}A_{\mu}^a)^2 -\psi^{\dagger}{\mbox{\boldmath$\mu$}}\psi + \bar{\eta}^a\(\partial^2\delta^{ab}  + gf^{abc}A_{\mu}^c\partial_{\mu}\)\eta^b.
\ea
From this point on this functional integral serves as the definition of the partition function for us.

\section{Feynman rules for finite-temperature QCD}
Not surprisingly, it is impossible to analytically evaluate the integrals in Eq.~(\ref{zres}). The formula can, however,
be used as the starting point of a perturbative treatment, which relies on the small value of the coupling constant at
large energies and amounts to expanding the integrand in a power series in $g$. We will now demonstrate the procedure by
first considering the case of a generic scalar field theory defined by its action $S[\phi]$.

Let us start by dividing the action into a free quadratic part $S_0$ and an interaction part $S_I$ by
\ba
S&=&S_0+S_I.
\ea
An expansion of the partition function in powers of $S_I$,
\ba
Z&=&\int\mathcal{D}\phi\,{\rm exp} \big[-(S_0+S_I)\big] \;\;\,=\;\;\,\sum_{n=0}^{\infty}\fr{(-1)^n}{n!}
\int\mathcal{D}\phi\,{\rm exp} \big[-S_0\big]S_I^n,
\ea
then reduces the determination of the functional integral into the evaluation of expectation values of different
products of field operators. Using Wick's theorem, these functions can be represented in terms of ordinary coordinate
space integrals, which contain in the integrands the two-point functions, or propagators,
\ba
\Delta(x-y)&\equiv&\fr{1}{Z}\int\mathcal{D}\phi\,{\rm exp} \big[-S_0\big]\,\phi(x)\phi(y).
\ea
The integrals are graphically represented by the famous Feynman diagrams, in which each line represents the propagator
$\Delta(x-y)$ and each vertex, which correspond to the different terms of the interaction part of the Lagrangian, a
four-dimensional
integration variable $x$. As can be straightforwardly verified (see e.g.~Ref.~\cite{kap}), the computation of the partition
function is equivalent to evaluating all connected vacuum diagrams that can be drawn using the propagators and vertices
of the theory. In practice almost all diagrammatic calculations are however performed in momentum space, where the
propagators represent momenta flowing through the graph.

Returning to the special case of QCD, the separation of the action gives
\ba
S_0&=&
\int_0^{\beta}{\rm d} x_0 \int {\rm d}^3
x\bigg\{-\fr{1}{2}A^a_{\mu}\Big(\delta_{\mu\nu}\partial^2 -
\Big(1-\fr{1}{\xi}\Big)\partial_{\mu}\partial_{\nu}\Big)A^a_{\nu}+\bar{\psi}\big(\slashed{\partial}
- \gamma_0{\mbox{\boldmath$\mu$}}\big)\psi+\bar{\eta}^a\partial^2\eta^a\bigg\}, \label{kinet} \\
S_I&=&S_\rmi{eff}-S_0\;\;\,\equiv\;\;\, \int_0^{\beta}{\rm d} x_0 \int {\rm d}^3x\,\mathcal{L}_\rmi{eff} -S_0  \label{inter}\\
&=&g\int_0^{\beta}{\rm d} x_0 \int {\rm d}^3
x\bigg\{\fr{g}{4}f^{abc}f^{ade}A_{\mu}^bA_{\nu}^cA_{\mu}^dA_{\nu}^e
+ f^{abc}(\partial_{\mu}A^a_{\nu})A^b_{\mu}A^c_{\nu}
-i\bar{\psi}\slashed{A}\psi + f^{abc}\bar{\eta}^a
A_{\mu}^c\partial_{\mu}\eta^b\bigg\},\nonumber
\ea
from where we can immediately read off the inverse propagators in coordinate space recalling the result of the Gaussian functional integral (as well as its fermionic counterpart)
\ba
\fr{1}{Z}\int\mathcal{D}\phi\,{\rm exp}
\bigg[-\fr{1}{2}\int_0^{\beta}{\rm d} x_0 \int {\rm d}^3 x
\phi(x)\Delta^{-1}(x-y)\phi(x)
\bigg]\,\phi(x)\phi(y)&=&\Delta(x-y).
\ea
The momentum space representations of the actual propagators can then be obtained by solving the corresponding Green's function equations, which is
easily accomplished by means of Fourier-transforming. In the finite-temperature case a transform consists of a
three-dimensional integral over the spatial part of the momentum and a sum over the discrete values of its zero-component.
Suppressing the diagonal color parts of the propagators we easily\footnote{Getting all the signs right in the Feynman
rules can, however, be surprisingly tricky. See Ref. \cite{ippdis} for a thorough analysis of this point.} get
\ba
(\Delta_A)_{\mu\nu}(P)&\equiv&\lpic{\Lgl(0,0)(30,0)}\;\;\,=\;\;\,\fr{1}{P^2}\bigg(\delta_{\mu\nu}-(1-\xi)\fr{P_{\mu}P_{\nu}}{P^2}\bigg), \\
\Delta_{\psi}(P)&\equiv&\lpic{\Lsc(0,0)(30,0)}\;\;\,=\;\;\,-\fr{1}{\slashed{P}}\;\;\,=\;\;\,-\fr{\slashed{P}}{P^2}, \\
\Delta_{\eta}(P)&\equiv&\lpic{\Lgh(0,0)(30,0)}\;\;\,=\;\;\,-\fr{1}{P^2},
\ea
where the zero-components of the bosonic (including ghosts) and fermionic momenta read
\ba
(p_0)_\rmi{bos.}&=&2\pi nT,\,n\in\mathbb{Z}, \\
(p_0)_\rmi{ferm.}&=&(2n+1)\pi T-i\mu,\,n\in\mathbb{Z},
\ea
respectively. The exact forms of these Matsubara frequencies are dictated by the periodicity conditions of the different
fields in the temporal direction.

The vertex functions of the theory are available from Eq.~(\ref{inter}) and read in momentum space
\ba
\nn
\pics{\Lgl(30,30)(60,60)\Lgl(30,30)(0,0)\Text(-6,66)[c]{$\alpha$}\Text(6,66)[c]{$a$}%
 \Text(-6,-7)[c]{$\delta$}\Text(6,-7)[c]{$d$}%
 \Text(66,68)[c]{$b$}\Text(54,66)[c]{$\beta$} \Text(66,-7)[c]{$c$}\Text(54,-7)[c]{$\gamma$}
 \Lgl(0,60)(30,30)\Lgl(60,0)(30,30)}\;\;\;\;\;\;\;\;\;\;\;\;\;\; &=& -g^2\bigg\{
 f^{eab}f^{ecd}(\delta_{\alpha\gamma}\delta_{\beta\delta}-\delta_{\alpha\delta}\delta_{\beta\gamma}) + \({b\atop\beta}\leftrightarrow{c\atop\gamma} \leftrightarrow {d\atop\delta}\)\bigg\},\label{4vert}\\\nn\nn
\pics{\Lgl(30,30)(60,0)\Line(25,48)(30,43)\Line(35,48)(30,43)\Line(15,15)(15,8)\Line(15,15)(8,15)%
\Line(45,15)(45,8)\Line(45,15)(52,15)\Lgl(30,30)(0,0)\Text(24,66)[c]{$\alpha$}\Text(36,66)[c]{$a$}  %
\Text(-6,-8)[c]{$\gamma$}\Text(6,-7)[c]{$c$}\Text(54,-7)[c]{$\beta$}\Text(66,-7)[c]{$b$}\Lgl(30,30)(30,60)
\Text(19,47)[c]{$P$}\Text(2,20)[c]{$R$}\Text(58,20)[c]{$Q$}}\;\;\;\;\;\;\;\;\;\;\;\;\;\;&=&
igf^{abc}\Big\{\delta_{\alpha\beta}(P-Q)_{\gamma}+\delta_{\beta\gamma}(Q-R)_{\alpha} +
\delta_{\alpha\gamma}(R-P)_{\beta}\Big\},\\\nn\nn
\pics{\Lsc(30,30)(60,0)%
\Lsc(0,0)(30,30)\Text(24,66)[c]{$\alpha$}\Text(36,66)[c]{$a$}  %
\Text(-4,-8)[c]{$j$}\Text(64,-7)[c]{$i$}\Lgl(30,30)(30,60)}\;\;\;\;\;\;\;\;\;\;\;\;\;\;&=&
-g\gamma_{\alpha}(T^a)_{ij}, \\\nn\nn
\pics{\Lgh(30,30)(60,0)%
\Lgh(0,0)(30,30)\Text(24,66)[c]{$\alpha$}\Text(36,66)[c]{$a$}  %
\Text(-4,-8)[c]{$c$}\Text(64,-7)[c]{$b$}\Lgl(30,30)(30,60)
\Text(58,20)[c]{$P$}}\;\;\;\;\;\;\;\;\;\;\;\;\;\;&=&
igf^{abc}P_{\alpha},\\\nn\nonumber
\ea
where the notation in Eq.~(\ref{4vert}) implies a cyclic permutation of the index pairs $(b,\beta)$, $(c,\gamma)$ and
$(d,\delta)$. Apart from trivial factors of $i$, there are only two differences to the familiar
zero-temperature Feynman rules, which both are due to the discrete nature of the zero-components of the momenta: the
$p_0$ $\delta$-functions appearing in the vertices are replaced by Kronecker $\delta$-symbols,
\ba
\delta^{(4)}(P)&\rightarrow&\fr{\beta}{2\pi}\delta^{(3)}({\mbox{\boldmath$p$}})\delta_{p_0,0},
\ea
which ensures the conservation of momenta. In addition, conventional four-dimensional loop integrals are replaced by
sum-integrals,
\ba
\int\fr{{\rm d}^4p}{(2\pi)^4}&\rightarrow&T\sum_{p_0/\{p_0\}}\int\fr{{\rm d}^3{\mbox{\boldmath$p$}}}{(2\pi)^3},
\ea
where $p_0$ stands for bosonic and $\{p_0\}$ for fermionic momenta. Symmetry factors on the other hand remain
unchanged at finite temperature, as does the rule of assigning a factor $-1$ to each fermion and ghost loop. This is
again merely a trivial consequence of the anti-commuting nature of the Grassmannian variables.

Having now defined a functional integral representation for the partition function and obtained the necessary Feynman
rules, it is natural to start considering the computation of specific diagrams. The analytic
evaluation of sum-integrals is an
interesting field of study in itself, and the techniques used in the calculations are constantly evolving. The
traditional method relies on transforming the $p_0$ sums into contour integrals by the Residue theorem, but this
becomes increasingly difficult when the number of loops in a diagram increases. In the appendix B of the present paper
we will introduce a more evolved scheme developed for the calculation of multi-fold sum-integrals at vanishing
chemical potentials in Ref.~\cite{az}
and generalized to $\mu\neq 0$ in Refs.~\cite{avpres,antti}, and give specific examples of its application. The method consists
of Fourier-transforming the spatial integrals into coordinate space, which makes it possible to evaluate the sums
analytically
and leads to the computation of one-dimensional hyperbolic integrals in the end. In the next three chapters we,
however, restrict ourselves to only quoting the results for the different graphs and leave the explicit computations to
the appendices. For now we instead concentrate on the physics involved in determining the partition function of QCD.

\section{Notation}
To finish the chapter we introduce some notation that will be in frequent use in the following. Using the
generators of the fundamental representation of SU($N$) as well as the antisymmetric structure coefficients, we define
four group theory factors (see e.g.~Ref.~\cite{cutsiv}) through
\ba
C_A \delta^{cd} & \equiv & f^{abc}f^{abd} \;\,\;=\;\,\; N \delta^{cd}, \\
C_F \delta_{ij} & \equiv & (T^a T^a)_{ij} \;\,\;=\;\,\; \fr{N^2-1}{2N}\delta_{ij},\\
T_F \delta^{ab} & \equiv & \tr T^a T^b \;\,\;=\;\,\; \fr{n_f}{2} \delta^{ab}, \\
D \delta^{cd} &\equiv & d^{abc}d^{abd} \;\;\,=\,\;\; \fr{N^2-4}{N}\delta^{cd},
\ea
where the trace is taken over both color and flavor indices. The dimensions of the adjoint and fermionic representations of the gauge group will henceforth be denoted by
\ba
d_A &\equiv& \delta^{aa} \;\;\,=\;\;\,  N^2 - 1, \\
d_F &\equiv& \delta_{ii} \;\;\,=\;\;\, d_A T_F/C_F \;\;\,=\;\;\, N n_f.
\ea

For certain frequently occurring combinations of special functions we will apply the abbreviations
\ba
\zeta'(x,y) &\equiv& \partial_x \zeta(x,y), \label{specf1}\\
\aleph(n,z) &\equiv& \zeta'(-n,z)+\(-1\)^{n+1}\zeta'(-n,z^{*}), \\
\aleph(z) &\equiv& \Psi(z)+\Psi(z^*),\label{specf3}
\ea
where $\zeta$ denotes the Riemann zeta function and $\Psi$ is the digamma function
\ba
\Psi(w)&\equiv&\fr{\Gamma '(w)}{\Gamma(w)}
\ea
The properties of these functions will be examined in the appendix B of the present paper.

The chemical potentials will henceforth usually appear in the dimensionless combinations
\ba
\mubar &\equiv& \mu/(2\pi T), \\
z &\equiv& 1/2-\imathb\mubar,
\ea
and in the context of considering the zero temperature partition function the following abbreviation will be used:
\ba
\sum_f\mu_f^2\;\;\,\equiv\;\;\,{\mbox{\boldmath$\mu$}}^2.
\ea
In sums over a single flavor index the subscript $f$ in $\mu_f$ is usually suppressed.

Even though we have so far been working strictly in four dimensions, we will throughout the following chapters apply
dimensional regularization and use $d=4-2\e$. The momentum integration measure and the notation used for sum-integrals
from here onwards read
\ba
\int_p \;\;\,\equiv\;\;\, \int\! \fr{{\rm d}^{d-1} p}{(2\pi)^{d-1}} &=& \Lambda^{-2\e}\(\!\fr{e^{\gamma}\bar{\Lambda}^2}{4\pi}\!\!\)^{\!\!\e}\!\!\int\!
\fr{{\rm d}^{d-1} p}{(2\pi)^{d-1}}, \label{asdfg}\\
\sumint_{P/\{P\}} &\equiv& T \sum_{p_0/\{p_0\}} \int_p,
\ea
where $\bar{\Lambda}\equiv (4\pi e^{-\gamma})^{1/2}\Lambda$ is the scale parameter of the $\msbar$ renormalization scheme
we have adopted and $\gamma$ is the Euler-Mascheroni constant. The excess factor of $\Lambda^{-2\e}$ in Eq.~(\ref{asdfg}) will be suppressed in the appendices.

\def\ToptVE(#1,#2){\picc{#1(15,15)(15,0,360) #2(45,15)(15,-180,180)}}

\def\Elmeri(#1,#2,#3){{\pic{#1(15,15)(15,0,180)%
 #2(15,15)(15,180,360)%
 #3(0,15)(30,15)}}}

\def\Petteri(#1,#2,#3,#4,#5,#6){\pic{#3(15,15)(15,-30,90)%
 #1(15,15)(15,90,210)%
 #2(15,15)(15,210,330) #5(2,7.5)(15,15) #6(15,15)(15,30) #4(15,15)(28,7.5)}}

\def\Jalmari(#1,#2,#3,#4,#5,#6){\picc{#1(15,15)(15,90,270)%
 #2(30,15)(15,-90,90) #4(30,30)(15,30) #3(15,0)(30,0) #5(15,0)(15,30)%
 #6(30,30)(30,0) }}

\def\Oskari(#1,#2,#3,#4,#5,#6,#7,#8){\picc{#1(15,15)(15,90,270)%
 #2(30,15)(15,-90,90) #4(30,30)(15,30) #3(15,0)(30,0) #6(15,0)(15,15)%
 #5(15,15)(15,30) #8(30,30)(30,15) #7(30,15)(30,0) }}

\def\Sakari(#1,#2,#3){\picb{#1(15,15)(15,30,150)%
#1(15,15)(15,210,330) #2(0,15)(7.5,-90,90) #2(0,15)(7.5,90,270) %
#3(30,15)(7.5,-90,90) #3(30,15)(7.5,90,270) }}

\def\Maisteri(#1,#2){\picb{#1(15,15)(15,0,150)%
#1(15,15)(15,210,360) #2(0,15)(7.5,-90,90) #2(0,15)(7.5,90,270) #1(37.5,15)(7.5,0,360) }}

\def\Pietari(#1){\picb{#1(7.5,15)(7.5,0,360)%
#1(30,15)(15,0,360) #1(52.5,15)(7.5,0,360) }}

\def\Jari(#1){\pic{#1(15,15)(15,0,360)}}

\chapter{Perturbative evaluation of the QCD pressure}
We are now ready to start applying the perturbative machinery to the evaluation of the QCD partition function. This is
done assuming the energy density of the system to be large enough to guarantee that we work in the deconfined phase and
that the coupling constant is sufficiently small for the expansion of the functional integral to be well-defined. In this
chapter we review some of the most important work that has been performed during the last three decades to drive
perturbation theory to new orders, but at the same time keep the main focus on the present status of the computations.
The motivation of the multi-loop calculations will be addressed in a later chapter by quantitatively examining the
convergence properties of the perturbative series of the pressure.

\section{A historical overview}
As soon as the status of QCD as the correct\footnote{According to our present understanding.} theory of the strong interactions had been established and the formalism necessary to deal with non-Abelian gauge field theories at finite temperatures developed, interest towards the
perturbative determination of the QCD pressure began to rise. The leading order result for the quantity could be obtained
immediately, as at vanishing coupling the theory reduces to a trivial one containing just non-interacting quarks and
gluons. The pressure of such a system is given by the Stefan-Boltzmann law and reads
\ba
p_\rmi{QCD}(g=0)&\equiv&p_{\mbox{\tiny SB}}\;\;\,=\;\;\,\fr{\pi^2 T^4}{45}\Big(N^2-1+\fr{7N n_f}{4}\Big)
+\fr{N}{6}T^2\sum_f\mu_f^2 + \fr{N}{12\pi^2}\sum_f\mu_f^4.
\ea
The first non-trivial finite-temperature computation was on the other hand the determination of the order $g^2$
correction to this result, which was carried out independently by Shuryak and Chin in 1978 \cite{es}. This calculation
consisted of evaluating the one- and two-loop vacuum graphs of the theory (see Fig.~3.1) and produced a sizable
contribution to the pressure,
\ba
\Delta p_\rmi{QCD}&=&-\fr{(N^2-1)g^2}{16}\bigg\{\fr{T^4}{9}\Big(N+\fr{5n_f}{4}\Big)+\fr{T^2}{2\pi^2}\sum_f\mu_f^2 +
\fr{1}{4\pi^4}\sum_f\mu_f^4\bigg\}.
\ea
The $T=0$ part of this expression was, however, known already beforehand from a computation of Freedman and McLerran
\cite{fmcl}, to which we will return soon.

\begin{figure}[t]
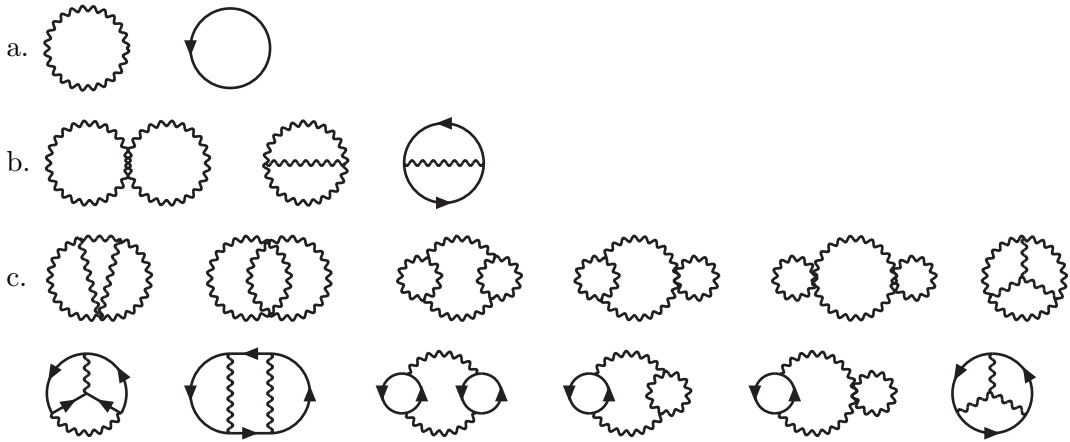

\centering
\ba \nonumber
\begin{array}{llllll}
\rm{a}.~\Jari(\Agl) &
\!\!\!\!\Jari(\Asc)&&&& \nn\nn
\rm{b}.~\ToptVE(\Agl,\Agl)&
\;\;\;\;\;\;\;\ToptVS(\Agl,\Agl,\Lgl)&
\;\;\ToptVS(\Asc,\Asc,\Lgl) &&&
\nn\nn
\rm{c}.~\!\!\!\!\ToprVV(\Agl,\Agl,\Lgl,\Lgl,\Lgl)&
\!\!\!\!\!\ToprVB(\Agl,\Agl,\Agl,\Agl)&
\;\Sakari(\Agl,\Agl,\Agl)&
\;\Maisteri(\Agl,\Agl)&
\;\Pietari(\Agl)&
\;\;\;\;\;\;\;\;\ToprVM(\Agl,\Agl,\Agl,\Lgl,\Lgl,\Lgl)\nn\nn
\;\;\;\;\ToprVM(\Asc,\Agl,\Asc,\Lsc,\Lsc,\Lgl)&
\!\!\!\!\Jalmari(\Asc,\Asc,\Lsc,\Lsc,\Lgl,\Lgl)&
\!\!\Sakari(\Agl,\Asc,\Asc) &
\Sakari(\Agl,\Asc,\Agl)&
\;\Maisteri(\Agl,\Asc)&
\;\;\;\;\ToprVM(\Asc,\Asc,\Asc,\Lgl,\Lgl,\Lgl)
\end{array}
\ea
\caption[a]{The one-, two- and three-loop Feynman diagrams contributing to the QCD partition function. Diagrams
containing ghost lines have been suppressed.}
\end{figure}

Motivated by the magnitude of the $\mathcal{O}(g^2)$ term, it was natural to attempt evaluating the pressure further to
three loops. It was however well understood, based on an analogy with scalar field theories, that
conventional perturbation theory ceases to work at this order due to uncancelled infrared (IR) divergences
originating from the ring diagrams of Fig.~3.1~c. The problem can be cured only through an all-orders summation of the IR
sensitive graphs, which gives a finite contribution to the pressure, but at the same time
produces the first signs of non-analytic behavior in $g^2$ to the perturbative expansion. The resulting order $g^3$ and
$g^4\ln\,g$ terms for the high-temperature pressure were first obtained by Kapusta \cite{jk} and Toimela \cite{tt} in 1979 and 1983, respectively.

The next $\mathcal{O}(g^4)$ and $\mathcal{O}(g^5)$ corrections to the pressure at vanishing chemical potentials were
computed over a decade later by Arnold, Zhai and Kastening \cite{az,zk}, who made a significant contribution to the
field of finite-temperature field theory by radically developing the computational methods used to evaluate
sum-integrals. This was soon followed by another important work of Braaten and Nieto \cite{bn1}, who were the first to
fully exploit the machinery of dimensional reduction \cite{dr,klrs2} in this particular problem by showing how all terms
in the perturbative series non-analytic in $g^2$ could be obtained using effective three-dimensional theories. Recently
Kajantie, Laine, Rummukainen and Schr\"oder \cite{klry,klry2} have used the same approach to drive the expansion of the
pressure even one step further, order $g^6\ln\,g$, in an impressive calculation involving four-loop diagrammatics in the
effective theories and an extensive use of the Form program \cite{form} designed for symbolic manipulation. This result
has since then been generalized also to the case of finite chemical potentials by the present author in Ref.~\cite{avpres},
where the perturbative orders $g^4$ through $g^6\ln\,g$ were simultaneously covered by generalizing the calculational
methods of Ref.~\cite{az} to finite $\mu$ and taking full advantage of the existing $\mc{O}(g^6\ln\,g)$ result at $\mu=0$.

As has been known for a long time \cite{linde}, high-temperature perturbation theory runs into serious infrared problems
again at order $g^6$. This time they can be overcome only by applying genuinely non-perturbative methods, as can be most
straightforwardly verified by investigating the IR structure of diagrams of the general type
\ba
\piccc{\Agl(15,15)(15,90,270)%
 \Agl(60,15)(15,-90,90) \Lgl(60,30)(15,30) \Lgl(15,0)(60,0) \Lgl(15,0)(15,30)%
 \Lgl(30,30)(30,0) \Text(45,15)[c]{....}  \Lgl(60,30)(60,0) \Text(80,15)[c]{.} }
\ea
A simple power-counting (see e.g.~Ref.~\cite{kap}) reveals that all similar graphs with at least 5 loops contribute to the partition function starting at order $g^6$, which necessitates the computation of an infinite number of diagrams in order to obtain the
next term in the expansion of the pressure. Unlike in the case of the ring sum, no computational method has been found to
enable the simultaneous treatment of the whole class of IR sensitive graphs, making it impossible to proceed further
without resorting to lattice QCD. The effective theory approach has, however, helped one to isolate the
non-perturbativity to a three-dimensional pure Yang-Mills theory, where lattice studies are considerably easier and less
expensive than in the full four-dimensional QCD. We will return to the implications of this observation in the next
section, where we present a brief introduction to the method of dimensional reduction.

At zero temperature and large chemical potentials the history of perturbation theory is completely different from the
high-$T$ case. The expansion of the pressure has been known to $\mathcal{O}(g^4)$ already since 1977, when Freedman and
McLerran \cite{fmcl} performed the summation of the ring diagrams in this limit. Since then there has been very little progress\footnote{Recently, the result of Ref.~\cite{fmcl} has been verified in an analytic computation \cite{avpres}, which removed all numerical uncertainties
from the result. No new perturbative orders have, however, been obtained.} in the field, as the most interesting
phenomena occurring in this region of the ($\mu$-$T$)-plane have turned out to be of fundamentally
non-perturbative origin \cite{rajwil}. Due to this reason and certain computational difficulties, the region of large
chemical potentials and small but non-zero temperatures has actually been left completely untouched since the early
work of Shuryak and Chin \cite{es}. Progress towards an order $g^4$ result for the pressure applicable throughout the
deconfined phase of quantum chromodynamics is nevertheless currently under way \cite{ikrv}.

Apart from the traditional perturbative approach, numerous other methods have been applied to the computation of
the QCD pressure. These include most importantly four-dimensional lattice simulations (see e.g.~Refs.~\cite{karsch2,karsch1,fodor,rummu1,gup4}), which have produced plenty of reliable results at small chemical potentials
and temperatures close to $T_c$, and different resummation schemes of perturbation theory, such as the hard thermal and
dense loop approximations \cite{birn1,htl2loop,htl2loopq,andstr}. While a detailed description of these methods is out
of the scope of our treatment, we want to end the section by mentioning two computational schemes of
special interest to us.

Moore, Ipp and Rebhan have recently evaluated the QCD pressure non-perturbatively in the limit of a large number of
flavors --- where the theory simplifies significantly --- both at vanishing chemical potentials \cite{moore} and at
finite $\mu$ \cite{ippreb}. In doing so they have been able to extract the coefficients of the perturbative expansion of
the quantity up to order $g^6$ \cite{ippreb}, which in the limit of large $n_f$ provides an accurate numerical check for
the results of Refs.~\cite{avpres,klry} and in addition gives an estimate for the magnitude of the yet undetermined
$\mc{O}(g^6)$ contribution to the $\mu=0$ pressure. In Ref. \cite{avippreb} the analysis of the perturbative results has
been extended to cover the entire ($\mu$-$T$)-plane, and one has in particular been able to quantitatively examine the
breakdown of dimensional reduction in the $\mc{O}(g^6\ln\,g)$ pressure \cite{avpres} in the limit of large $\mu/T$.
Remarkably, methods similar to those used in the large-$N_f$ context have also been used to study the non-Fermi-liquid
behavior of the specific heats of the full theory at low temperatures \cite{igr}.

Another very interesting approach to perturbative computations is the `numerical stochastic perturbation theory' developed by Di
Renzo \textit{et al.} \cite{parma}, which is based on finding numerical solutions to the Langevin equation that the quark
and gluon fields are required to obey. The form of the equation determines the evolution of the fields in stochastic time
and provides a natural way of expanding the solutions in power series in $g$. This method has recently provided important results
that can be used to extract the non-perturbative $\mc{O}(g^6)$ contribution to the QCD pressure originating from an effective three-dimensional Yang-Mills theory.

\section{The pressure at high temperatures}
Let us now move on to take a closer look into the most recent perturbative computations of Refs.~\cite{avpres,klry}.
This part of the paper is divided into two sections corresponding to different regions on the ($\mu$-$T$)-plane. In this
first section we cover the region of high temperatures and relatively small chemical potentials giving special emphasis
to explaining how three-dimensional effective theories can be applied in the computations. In the latter one we then
review the current status of perturbation theory in the limit of large chemical potentials and small temperatures, and
outline some possible future developments.

\subsection{Dimensional reduction and effective theories}
If the temperature of a strongly interacting system is well above the critical temperature of the deconfinement
transition and the chemical potentials are negligible in comparison\footnote{This implies $\mu_f\ll 2\pi T$ for all
flavors. For a more detailed analysis, see chapter 4 and Refs.~\cite{avpres,hlp}.}, the description of thermodynamics simplifies somewhat.
This is due to a phenomenon called dimensional reduction, which refers to the observation that the large distance static
correlation functions of many field theories can at sufficiently high temperatures be obtained via effective
lower-dimensional theories \cite{dr}. It is based on the reduction of the relevant degrees of freedom to the zeroth
Matsubara modes of the bosonic fields, as we will demonstrate in the following. The method has played a central role in
making the recent $\mc{O}(g^6\ln\,g)$ perturbative computations of the QCD pressure possible.

To introduce the concept, let us first consider a generic field theory, in which the masses of the
different fields are of negligible magnitude in comparison with the temperature, and all chemical potentials are
taken to be zero. Just as we have seen in the previous chapter, the functional integral representation of the partition
function leads to the requirement of periodic and antiperiodic boundary conditions for the bosonic and fermionic fields
on the interval $x_0\in[0,\beta]$. This in turn causes their Fourier representations to be of the form
\ba
\phi\(x\) &=& T\sum_{n = - \infty}^\infty e^{i\omega_n x_0}\phi_n\(\textbf{x}\),
\ea
where the bosonic and fermionic Matsubara frequencies read
\ba
\omega^\rmi{b}_n &=& 2n\pi T, \\
\omega^\rmi{f}_n &=& \(2n+1\)\pi T,
\ea
respectively. Assuming the kinetic part of the Lagrangian to be of the usual quadratic form, the momentum space propagators of the
field modes are then always proportional to $1/(\omega_n^2+p^2)$, and it is obvious that the non-static modes gain
effective
masses that grow linearly with the temperature. It would therefore seem natural to expect some sort of decoupling to take
place for these fields at high temperatures, which would leave the zero-modes $\phi_0$ of the bosonic fields as the true
degrees of freedom.

To verify that the decoupling indeed takes place, we refer to a well-known theorem of Appelquist and Carazzone \cite{ac}, stating that for a renormalizable (zero-temperature) field theory containing two types of fields with masses $m_1 \ll m_2$, the Green's functions with typical momentum scales $p \ll m_2$ may up to corrections of the orders $(m_1/m_2)^n$, $(p/m_2)^n$, $n\geq 1$, be calculated using a Lagrangian with the heavy fields removed. In the new Lagrangian the coupling constants have modified values and new, possibly non-renormalizable, interaction terms often need to be introduced in order to reproduce the results of the old theory.

The application of the decoupling theorem to the computation of the finite-temperature partition function seems at first
sight extremely straightforward, as we are dealing with diagrams with vanishing external momenta, and the light
fields --- which now clearly are the bosonic zero modes --- are furthermore massless. Here we, however, must take into
account also the possible generation of thermal masses for these fields, which leads to corrections to the partition
function proportional to positive powers of the coupling constant. Assuming our theory to be asymptotically free in the
ultraviolet (UV), it is anyhow clear that with increasing temperature the static fields eventually become the dominant degrees
of freedom, meaning that the non-static modes can be integrated out from the description of equilibrium thermodynamics at
least in the framework of perturbation theory. This implies that we are left with an effective theory with a Lagrangian
independent of $x_0$, which makes it a genuinely three-dimensional one. In addition, the new theory generally has a
simpler structure than the original one due to the lack of fermions, and practical calculations are often easier to
perform there, as the tedious frequency sums disappear from the Feynman rules. On the downside, it has to be remembered
that the effective theory parameters can usually be obtained only through the original theory.

Returning to massless QCD, it is now obvious that the dominant degrees of freedom contributing to the high-$T$ pressure
are the different Lorentz components of the zero mode of the gluon field's Fourier expansion,
\ba
\label{klubi}
A_\mu^a(x) &=& T\sum_{n = - \infty}^\infty {\rm exp}\big[i\omega^\rmi{b}_n x_0\big]A_{\mu, n} ^a\(\textbf{x}\).
\ea
They can up to a multiplicative factor be identified with the electrostatic scalar field $A_0^a\(\textbf{x}\)$ and the
magnetostatic gauge field $A_i^a\(\textbf{x}\)$ of a three-dimensional effective theory, electrostatic QCD (EQCD). The
Lagrangian of the new theory is most easily obtained by first writing down the most general one respecting all the correct
symmetries and then determining the different parameters through matching computations in full QCD. The eventual
super-renormalizable Lagrangian density reads \cite{hlp}
\ba
{\cal L}_\rmi{E}& = & \fr12 \tr F_{ij}^2 + \tr [D_i,A_0]^2 + m_\rmi{E}^2\tr A_0^2 + \fr{\imathb g^3}{3\pi^2}
\sum_f\mu_f \,\tr A_0^3  \nn
&+&\lambda_\rmi{E}^{(1)} (\tr A_0^2)^2
 +\lambda_\rmi{E}^{(2)} \tr A_0^4
\label{leqcd}
\ea
with
\ba
F_{ij}^a&=&\partial_i A_j^a - \partial_j A_i^a + g_\rmi{E} f^{abc} A_i^b A_j^c,\\
D_i&=&\partial_i-ig_\rmi{E}A_i,
\ea
and while there could in principle appear also higher-order, possibly non-renormalizable operators in Eq. (\ref{leqcd}), it
can be easily seen that they would only contribute to the pressure at order $g^7$ or
higher \cite{klry}. We have therefore neglected them in our analysis.

The parameters appearing in the above Lagrangian are determined by demanding that the effective theory produces the correlation functions of the original one at distances $1/(gT)$ and higher, which in the perturbative framework this leads to power series expansions in $g^2$ for them. The determination of the coefficients in the expansions, however, often becomes a highly laborious task in practise; see e.g. Ref. \cite{bn1} for details.

Using the effective theory approach, we can now write the pressure of QCD in the form
\ba
p_\rmi{QCD}\!\(T,\mu\) &=& p_\rmi{E}\!\(T,\mu\)+\fr{T}{V} \ln \! \int {\cal D}A_i^a \,
{\cal D}A_0^a \exp\Big\{\!-\!S_\rmi{E}\Big\}, \label{asdf}
\ea
where the function $p_\rmi{E}$ is regarded as a parameter of the effective theory and is obtainable by computing the
strict perturbation expansion\footnote{Meaning a pure diagrammatic expansion, where no form of resummation
has been applied. It leads to a power series in $g^2$.} of the pressure in the full theory. The pressure has here been
divided into two parts corresponding to the contributions of different
momentum scales: $p_\rmi{E}$ represents the hard scale $2\pi T$, which originates from the lowest non-zero Matsubara modes, while the partition function of the effective theory describes the effects of the soft scales, $gT$ and lower, corresponding to the screening of the static gluon fields. All non-perturbativity naturally resides in the latter part.

As soon as gauge fixing has been taken care of, the partition function of the effective theory can be determined by
non-perturbative methods. One may, however, further simplify its computation by noticing that the theory still
contains two dynamical momentum scales, $gT$ and $g^2T$. The non-perturbativity manifesting itself at order $g^6$ in the
expansion of the pressure originates completely from the magnetostatic sector, which corresponds to the fields $A_i$
and to the scale $g^2T$ proportional to their non-perturbative screening masses. This means that we can further divide the partition function of the effective theory into two parts by writing
\ba
\fr{T}{V} \ln \! \int {\cal D}A_i^a \,
{\cal D}A_0^a \exp\Big\{\!-\!S_\rmi{E}\Big\} &=& p_\rmi{M}+\fr{T}{V} \ln  \!\int {\cal D}A_i^a
\exp\Big\{\!-\!S_\rmi{M}\Big\} \nn
&\equiv&p_\rmi{M}+p_\rmi{G},
\ea
where the last non-perturbative term denotes the pressure of a new effective theory describing the magnetostatic sector
(MQCD following the terminology of Ref.~\cite{bn1}). This theory has been obtained by integrating out the $A_0$ field and is defined by the Lagrangian density
\ba
{\cal L}_\rmi{M} & = & \fr12 \tr F_{ij}^2,\\
F_{ij}&=&\partial_i A_j^a - \partial_j A_i^a + g_\rmi{M} f^{abc} A_i^b A_j^c,
\ea
which is parametrized by the coupling constant $g_\rmi{M}$. This parameter is determined in the electrostatic
effective theory (EQCD), and analogously to the case of $p_\rmi{E}$, the function $p_\rmi{M}$ is obtainable through
the determination of the strict perturbation expansion of the EQCD partition function. This is the last step in the
series of effective theories, as all scales softer than $g^2T$ are absent from MQCD due to its confining nature.

\subsection{The result to order $g^6$}
In our present treatment we do not wish to go into great details in explaining how the parameters of the
effective theories can be computed. Instead, we restrict ourselves to presenting the general
structure of the results relevant for the computation of the QCD pressure, and start this by first inspecting the part originating from the hard momenta of order $2\pi T$.
This scale enters the perturbative expansion exclusively through the parameters of EQCD, which can be shown to have the forms \cite{klry}
\ba
 \fr{p_\rmi{E}(T,\mu)}{T \Lambda^{-2 \epsilon} } & = & T^3 \Bigl[\aE{1}
 + g^2
 \Bigl(\aE{2} + {\cal O}(\epsilon)\Bigr) \nn
 &+ &
  \frac{g^4}{(4\pi)^2}
 \Bigl(\aE{3} + {\cal O}(\epsilon)\Bigr)
 + \frac{g^6}{(4\pi)^4}
 \Bigl(\bE{1} + {\cal O}(\epsilon)\Bigr) +\mc{O}(g^8)
 \Bigr], \la{pe}
 \ea
 \ba
 m_\rmi{E}^2 & = & T^2 \Bigl[ g^2
 \Bigl( \aE{4} +
 \aE{5} \epsilon + {\cal O}(\epsilon^2) \Bigr)
 + \frac{g^4}{(4\pi)^2}
 \Bigl( \aE{6} + \bE{2} \epsilon +
 {\cal O}(\epsilon^2) \Bigr) + {\cal O}(g^6) \Bigr], \hspace*{0.5cm} \\
 g_\rmi{E}^2 & = & T \Bigl[ g^2 + \frac{g^4}{(4\pi)^2}
 \Bigl( \aE{7} +
 \bE{3} \epsilon + {\cal O}(\epsilon^2) \Bigr)
 + {\cal O}(g^6) \Bigr] , \\
 \lambda_\rmi{E}^{(1)} & = & T \Bigl[
 \frac{g^4}{(4\pi)^2} \Bigl( \bE{4} +
 {\cal O}(\epsilon) \Bigr) + {\cal O}(g^6) \Bigr] , \\
 \lambda_\rmi{E}^{(2)} & = & T \Bigl[
 \frac{g^4}{(4\pi)^2}
 \Bigl( \bE{5} +
 {\cal O}(\epsilon) \Bigr) + {\cal O}(g^6) \Bigr], \la{lE2}
 \ea
where the $\alpha$'s and $\beta$'s are `matching' coefficients with \textit{a priori} unknown values. The $\alpha$'s are
needed in determining the pressure up to order $g^6\ln\,g$ and have been computed in Refs.~\cite{az,bn1,klrs2,huli} at $\mu=0$ and
in Refs.~\cite{avpres,hlp} at $\mu\neq 0$. The most non-trivial one of them is the coefficient $\alpha_\rmi{E3}$,
which represents the contribution of three-loop diagrams to the strict perturbation expansion of the pressure
(see Fig.~3.1 c). Its computation at arbitrary chemical potentials is
explained in great detail in Ref.~\cite{avpres} and some of the techniques used in this calculation are also demonstrated in
the appendix B of the present paper. The $\beta$ coefficients, on the other hand, are not needed until order $g^6$, and
one of them, $\beta_\rmi{E1}$, still remains unknown. Not surprisingly, it will also be by far the hardest one to evaluate.

\begin{figure}[t]

\centerline{\epsfxsize=16cm \epsfbox{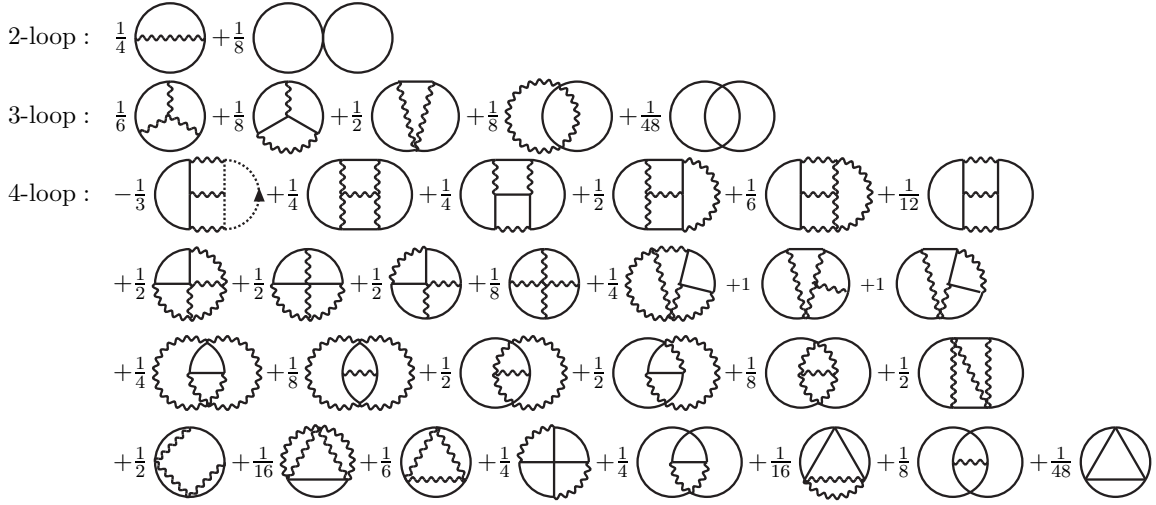}}

\caption[a]{The two-, three- and four-loop skeleton diagrams contributing to $p_\rmi{M}$ \cite{klry2}. The solid, wiggly and dotted lines stand respectively for the adjoint scalar $A_0$, the gauge boson $A_i$ and the ghost fields.}
\end{figure}

The soft scale $g T$ contributes to the pressure through the parameters of MQCD, which we write in the forms
\cite{avpres,klry}
 \ba
 \frac{p_\rmi{M}(T,\mu)}{T \Lambda^{-2 \epsilon} } & = &
 \frac{1}{(4\pi)}
 d_A  m_\rmi{E}^3
 \bigg[\fr13 + {\cal O}(\epsilon) \bigg] \nn
 & + &
 \frac{1 
 }{(4\pi)^2}
 d_A C_A
 g_\rmi{E}^2 m_\rmi{E}^2
 \bigg[-\frac{1}{4\epsilon} - \fr34 -\ln\frac{\bar{\Lambda}}{2 m_\rmi{E}}
 + {\cal O}(\epsilon) \bigg] \nn
 & + &
 \frac{1 
 }{(4\pi)^3}
 d_A C_A^2
 g_\rmi{E}^4 m_\rmi{E}
 \bigg[-\frac{89}{24} - \fr16 \pi^2 + \frac{11}{6}\ln\,2
 + {\cal O}(\epsilon) \bigg]
 \ea
 \ba
 & + &
 \frac{1 
 }{(4\pi)^4}
 d_A C_A^3
 g_\rmi{E}^6\bigg[\alpha_\rmi{M1}\Big(\fr{1}{\e}+8\,\ln\frac{\bar{\Lambda}}{2 m_\rmi{E}}\Big)+\beta_\rmi{M1}
  +  {\cal O}(\epsilon) \bigg] \nn
   & + &
 \frac{1 
 }{(4\pi)^2}
 d_A (d_A + 2)
 \lambda_\rmi{E}^{(1)} m_\rmi{E}^2
 \bigg[ - \fr14 + {\cal O}(\epsilon)
 \bigg] \nn
 & + &
 \frac{1 
 }{(4\pi)^2}
 d_A (2 d_A - 1) N^{-1}
 \lambda_\rmi{E}^{(2)} m_\rmi{E}^2
 \bigg[ - \fr14 + {\cal O}(\epsilon)
 \bigg] \label{pm}\\
& + &
\frac{1 
}{(4\pi)^4}
d_A D T_F^2
g_\rmi{E}^6 \bigg(\fr{1}{n_f}\sum_f \mubar\bigg)^{\!2}\bigg[\alpha_\rmi{M2}\Big(\fr{1}{\e}+4\,\ln\frac{\bar{\Lambda}}{2 m_\rmi{E}}\Big)+\beta_\rmi{M2}
+ {\cal O}(\epsilon) \bigg] + {\mathcal O}(g^7), \nn
g_\rmi{M}^2 &=& g_\rmi{E}^2+ {\mathcal O}(g^3).
\ea
As explained in the previous subsection, the above expression for $p_\rmi{M}$ has been obtained by computing
the diagrammatic expansion of the EQCD
partition function \cite{bn1,klry2} up to four loops, which was a long-standing and highly non-trivial problem. It was
finished
only very recently with the determination of the last two coefficients $\alpha_\rmi{M1}$ and $\beta_\rmi{M1}$ by Kajantie
\textit{et al.} \cite{klry2}, a task made possible by an efficient use \cite{yorkll} of the Form program \cite{form} designed for
large-scale symbolic calculations. The general strategy of the computation was to first carry out all tensorial
contractions in the diagrams and then reduce the number of necessary integrals into a small class of scalar master ones
by applying an efficient algorithm developed by Laporta \cite{lapo}. The analytic and numerical evaluation of
Euclidean scalar integrals analogous to these is the topic of Ref.~\cite{avyork}, and it is also briefly dealt with in
the appendix B of the present paper. The skeleton (two-particle irreducible) vacuum diagrams of EQCD are depicted in Fig.~3.2. and the scalar
diagrams that remain to be computed after the scalarization process in Fig.~3.3. In the latter graphs the propagators
have the usual scalar form $1/(p^2+m^2)$, where the mass parameter has the values $m=m_\rmi{E}$ and $m=0$ corresponding
to the $A_0$ and $A_i$ fields, respectively. The vertex functions in the graphs are simply constants.

\begin{figure}[t]
\centerline{\epsfxsize=11cm \epsfbox{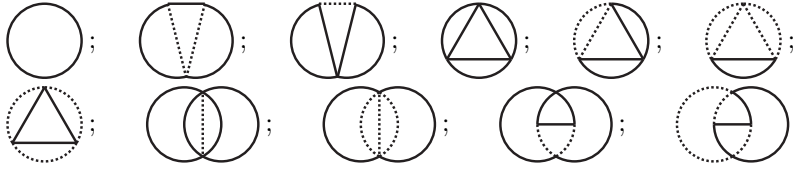}}
\caption[a]{The one-, two-, three- and four-loop scalar master diagrams remaining after scalarization \cite{klry2}. The solid lines correspond to massive and dotted to massless scalar propagators.}
\end{figure}

Finally, the ultrasoft scale $g^2T$ contributes to the QCD pressure through the partition function of the three-dimensional pure
gauge theory MQCD. In terms of the parameters of the theory the perturbative expansion of this quantity reads
\ba
 \frac{p_\rmi{G}(T)}{T \Lambda^{-2 \epsilon} } &=&
 d_A C_A^3 \frac{g_\rmi{M}^6}{(4\pi)^4}\bigg[\alpha_\rmi{G}\Big(\fr{1}{\e}+8\,\ln\frac{\bar{\Lambda}}{2 m_\rmi{G}}\Big)+\beta_\rmi{G}
  +  {\cal O}(\epsilon) \bigg]+ {\mathcal O}(g^7), \label{pg}
\ea
where the dependence on chemical potentials is of order $g^7$ or higher and where the magnetic mass $m_\rmi{G}$ is given by
\ba
m_\rmi{G}&=&C_Ag_\rmi{E}^2 +\mc{O}(g^3).
\ea
The coefficient $\alpha_\rmi{G}$ appearing in Eq.~(\ref{pg}) has been computed perturbatively in Ref.~\cite{klry}, but $\beta_\rmi{G}$ can only be determined using lattice methods due to the non-perturbative nature of $p_\rmi{G}$. Its computation is a highly non-trivial task, as alone the conversion of the lattice results into continuum regularization requires complicated multi-loop calculations in lattice perturbation theory (see Ref.~\cite{klry}). In this process the recent results of Ref.~\cite{parma} may turn out to be very useful.

Adding up the different parts of the full theory pressure and expanding in $g$, we are now ready to write down its
perturbative result as
\ba
 \frac{p_\rmi{QCD}(T,\mu)}{T^4 \Lambda^{-2 \epsilon} } & = &
 \frac{p_\rmi{E}(T,\mu) + p_\rmi{M}(T,\mu) + p_\rmi{G}(T)}{T^4 \Lambda^{-2 \epsilon} } \nn
 & = &
 g^0 \bigg\{ \aE{1} \bigg\}
  +
 g^2 \bigg\{ \aE{2} \bigg\}
  +
 \frac{g^3}{(4\pi)}
 \bigg\{ \frac{d_A}{3} \aE{4}^{3/2} \bigg\} \nn
 & + &
 \frac{g^4}{(4\pi)^2} \bigg\{
 \aE{3} - d_A C_A
 \bigg[
 \aE{4} \bigg(
 \frac{1}{4\epsilon} + \fr34 + \ln\frac{\bar{\Lambda}}{2 g T \aE{4}^{1/2}}
 \bigg)
 + \fr14 \aE{5} \bigg] \bigg\} \nn
 & + &  \frac{g^5}{(4\pi)^3} \bigg\{ d_A \aE{4}^{1/2} \bigg[
 \fr12 \aE{6} - C_A^2
 \bigg(
 \frac{89}{24} + \frac{\pi^2}{6} - \frac{11}{6} \ln\,2
 \bigg)
 \bigg] \bigg\}  \nn\label{pres}
  & + &  \frac{g^6}{(4\pi)^4} \bigg\{
 \bE{1} - \aE{4} \frac{d_A}{4}
 \bigg[
 \frac{2 d_A - 1}{N} \bE{5} +
 (d_A + 2) \bE{4}
 \bigg] \nn
 &-& d_A C_A \bigg[
 \fr14 \Big( \aE{6} + \aE{5} \aE{7} + 3 \aE{4} \aE{7} + \bE{2}
 + \aE{4} \bE{3} \Big) \nn
 &+& \Big( \aE{6} + \aE{4} \aE{7} \Big)
 \bigg(\frac{1}{4\epsilon} + \ln \frac{\bmu}{2 g T \aE{4}^{1/2}} \bigg)
 \bigg] \nn
 &+&
 d_A C_A^3 \bigg[
 \beta_\rmi{M1} + \beta_\rmi{G}
 + \alpha_\rmi{M1} \bigg( \frac{1}{\epsilon} + 8\, \ln \frac{\bmu}{2 g T \aE{4}^{1/2}}
 \bigg)
 +\aG \bigg( \frac{1}{\epsilon} + 8 \,\ln \frac{\bmu}{2 g^2 T C_A} \bigg)
 \bigg]
 \nn
 &+&d_A D T_F^2
\bigg(\fr{1}{n_f}\sum_f \mubar\bigg)^{\!2}\bigg[\alpha_\rmi{M2}\Big(\fr{1}{\e}+4\,\ln\frac{\bar{\Lambda}}{2 m_\rmi{E}}\Big)+\beta_\rmi{M2}
 \bigg] \bigg\}
 +   {\cal O}(g^7).
\ea
Here $g$ is the renormalized gauge coupling of the theory and the explicit $1/\e$ terms cancel with the ones hidden in
the matching coefficients. The result contains both an explicit and an implicit (through the running
of $g$) dependence on the scale of dimensional regularization, $\bar{\Lambda}$, which gets cancelled order-by-order but
completely vanishes only through an all-orders summation of perturbation theory. This induces an arbitrariness to the
result, which we may try to reduce by choosing the value of $\bar{\Lambda}$ in a way that would minimize the effects
of the higher order contributions. This is, however, a highly non-trivial problem, and we will return to it in the next chapter. Here we merely note that even before determining the exact values of the
coefficients $\beta_\rmi{E1}$ and $\beta_\rmi{G}$, we can easily find the explicit form of the scale dependence of the order
$g^6$ term. This is achieved by applying the two-loop running of $g$,
\ba
g^2\!\(\Bar{\Lambda}\) &=&
g^2\!\(\Lambda\)\bigg[1+\fr{1}{6}\Big(11\,C_A-4\,T_F\Big)\ln\fr{\Lambda}{\Bar{\Lambda}}\fr{g^2\!\(\Lambda\)}{4\pi^2} \nn
&+&\fr{1}{12}\!\(17\,C_A^2-10\,C_AT_F-6 \fr{d_FC_F^2}{d_A} +
\fr{1}{3}\(11\,C_A-4\,T_F\)^2\ln\fr{\Lambda}{\Bar{\Lambda}}\)
\ln\fr{\Lambda}{\Bar{\Lambda}}\bigg(\!\fr{g^2}{4\pi^2}\!\!\!\bigg)^{\!\!2}\,\bigg] \label{rengroupa} \nn
&\equiv& g^2\!\(\Lambda\)\bigg[1+\alpha_\rmi{R1}\fr{g^2\!\(\Lambda\)}{(4\pi)^2}
+\(\alpha_\rmi{R2} + \alpha_\rmi{R3}\,\ln\fr{\Lambda}{\Bar{\Lambda}}\)
\ln\fr{\Lambda}{\Bar{\Lambda}}\bigg(\!\fr{g^2}{(4\pi)^2}\!\!\!\bigg)^{\!\!2}\,\bigg],
\ea
to Eq.~(\ref{pres}) and demanding that the $\bar{\Lambda}$ dependence of the result is beyond order $g^6$. The result we
hereby get for the $\bar{\Lambda}$ dependent part of the $g^6$ term reads
\ba
\delta p_\rmi{QCD}&=&\fr{g^6}{(4\pi)^4}\Bigg[\aE{2}\bigg\{\alpha_\rmi{R2}+\alpha_\rmi{R3}\,\ln\fr{\bar{\Lambda}}{T}\bigg\}\nn
&+& \alpha_\rmi{R1}\bigg\{
 \aE{3} - d_A C_A
 \bigg[
 \aE{4} \bigg(
 \frac{1}{4\epsilon} + \fr34 + \ln\frac{\bar{\Lambda}}{2 g T \aE{4}^{1/2}}
 \bigg)
 + \fr14 \aE{5} \bigg] \bigg\} \Bigg]\ln\fr{\bar{\Lambda}}{T}. \nonumber
\ea

Combined with this result, the above expression for the pressure, Eq.~(\ref{pres}), represents the current status of
high-temperature perturbation theory and in a clear way shows what is missing from the
complete order $g^6$ result for the quantity. The matching coefficients $\alpha$, which are necessary to obtain the
pressure to order $g^6\ln\,g$, are listed in appendix A, while two of the $\beta$'s, $\beta_\rmi{E1}$ and
$\beta_\rmi{G}$, still remain unknown at present. For a more detailed derivation of the above results and the matching
coefficients, we refer the reader to the original work performed in Refs.~\cite{avpres,bn1,klry,klry2,hlp,bir}.

\section{The pressure at low temperatures}
A graphic analysis of the properties of the above high-temperature expansion of the QCD pressure will be carried out in
the next chapter, where we will also quantitatively address the question, at which
values of chemical potentials the dimensionally reduced result is applicable. It is, however, clear already
beforehand that at least in the limit, where the temperatures are well below $T_c$ but the chemical potentials large
enough to keep us in the deconfined phase, the assumption of the temperature being the dominant energy scale can no
longer be correct. This makes the reliability of the above result highly questionable, as it was derived assuming
the zeroth Matsubara modes of the bosonic fields to be the dominant degrees of freedom. At low $T$ this clearly is not
the case, and the mechanism responsible for the screening of the IR divergencies is furthermore
more involved than in the high-temperature case (see Refs.~\cite{fmcl,az}). In the region of large
$\mu/T$ we therefore have no effective lower-dimensional theories to work with, which radically alters the computational
methods available. We will now briefly describe the status of the perturbative calculations in this limit,
but at the same time keep in mind
that our treatment is restricted to the ordinary quark-gluon plasma phase. It cannot be applied to the study of color
superconductivity, which in nature is the phase of strongly interacting matter near the $T=0$ line.

At small but finite temperatures the only existing perturbative result for the pressure is of order $g^2$ \cite{es}. To
determine the next $\mc{O}(g^4\ln\,g)$ and $\mc{O}(g^4)$ terms in the expansion would require an explicit summation of
the full theory ring diagrams\footnote{Those containing the one-loop gluon polarization tensor (see Fig.~3.4 d) as loop
insertions.} to an arbitrarily high loop order, which leads to the
computation of a very complicated logarithmic sum-integral. This has so far been achieved only in the limits of large $T$
\cite{tt} and $T=0$ \cite{fmcl}, but a project aimed at determining the $\mc{O}(g^4)$ pressure at all values of $\mu/T$
is currently being completed \cite{ikrv}. As has been
argued in Ref.~\cite{avpres}, the quantitative difference between the outcome of this new computation and the present
dimensionally reduced result may, however, turn out to be small in most parts of the ($\mu$-$T$)-plane. The motivation for
this claim will be studied in the next chapter.

\begin{figure}[t]
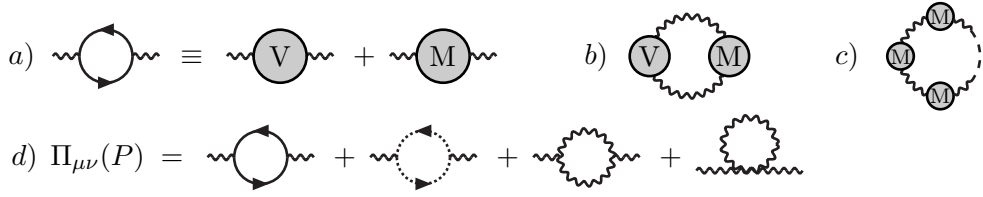

\centering
\ba \nonumber
\begin{array}{lll}
a)~ \pic{\Lgl(0,15)(10,15)%
 \Asc(20,15)(10,0,180) \Asc(20,15)(10,180,360) \Lgl(30,15)(40,15)}\;\;\;\;\equiv\; \pic{\Lgl(0,15)(10,15)%
\Lgl(30,15)(40,15)\GCirc(20,15){10}{0.8} \Text(20,15)[c]{V}} \;\;\;\;+
\pic{\Lgl(0,15)(10,15)\GCirc(20,15){10}{0.8}\Text(20,15)[c]{M}%
 \Lgl(30,15)(40,15)} \;\;\;\;\;\;\;\;\;\;
&b)~ \picb{\Agl(15,15)(15,30,150)%
\Agl(15,15)(15,210,330) \GCirc(0,15){7.5}{0.8}\Text(0,15)[c]{V} %
\GCirc(30,15){7.5}{0.8}\Text(30,15)[c]{M}}
\;\;\;\;\;\;
&c)~ \picb{\Agl(15,15)(15,110,160)%
\Agl(15,15)(15,200,250) %
\GCirc(0,15){5}{0.8} \GCirc(15,30){5}{0.8} \GCirc(15,0){5}{0.8} \Agl(15,15)(15,40,70)\Agl(15,15)(15,290,320)
\DAsc(15,15)(15,-40,40)\Text(15,0)[c]{{${\mbox{\scriptsize{M}}}$}}
\Text(15,30)[c]{{${\mbox{\scriptsize{M}}}$}}\Text(0,15)[c]{{${\mbox{\scriptsize{M}}}$}} }
\end{array}\nn\nn
d)\;\,\Pi_{\mu\nu}(P)\;=\;\pic{\Lgl(0,15)(10,15)%
 \Asc(20,15)(10,0,180) \Asc(20,15)(10,180,360) \Lgl(30,15)(40,15)}\;\;\;\;+\pic{\Lgl(0,15)(10,15)%
 \Agh(20,15)(10,0,180) \Agh(20,15)(10,180,360) \Lgl(30,15)(40,15)}\;\;\;\;+\pic{\Lgl(0,15)(10,15)%
 \Agl(20,15)(10,0,180) \Agl(20,15)(10,180,360) \Lgl(30,15)(40,15)}\;\;\;\;+\pic{\Lgl(0,10)(20,10)%
 \Agl(20,20)(10,0,360) \Lgl(20,10)(40,10)}\;\;\;\;\;\;\;\;\;\;\;\;\;\;\;\;\;\;\;\; \;\;\;\;\;\;\;\;\;\;\nonumber
\ea
\caption[a]{a) The fermionic part of the one-loop gluon polarization tensor divided into its vacuum ($T=\mu=0$) and
matter (vacuum-subtracted) parts. \\
b) An IR safe diagram contributing to the  pressure at $T=0$. \\
c) The generic form of the zero-temperature ring diagrams. \\
d) The whole one-loop gluon polarization tensor.}
\end{figure}

At exactly zero temperature several simplifications occur in the computation of the pressure \cite{avpres,fmcl}: sum-integrals reduce to ordinary four-dimensional integrals and,
more importantly, only diagrams containing fermion loops give non-zero contributions to the ring sum. This is due to the fact that
apart from the scale parameter, the only dimensionful numbers appearing in the diagrams at $T=0$ are the chemical
potentials, which enter exclusively through the fermionic propagators. When the temperature is zero, the scale-free
gluon (are ghost) diagrams simply vanish under dimensional regularization.

Following the treatment of Ref.~\cite{avpres}, the zero-temperature pressure can up to order $g^4$ be observed to receive
contributions from three classes of diagrams. These include the infrared safe fermionic one-, two- and
three-loop graphs of Fig.~3.1, the single IR safe diagram of Fig.~3.4~b.~as well as the ring sum of Fig.~3.4 c, where
the grey circle denotes the vacuum-subtracted part of the fermionic one-loop gluon polarization tensor. The values
obtained for the IR safe diagrams at arbitrary $T$ and $\mu$ in Ref.~\cite{avpres} can be straightforwardly continued to the
limit of zero temperature, while the ring sum must be separately computed at $T=0$ \cite{avpres,fmcl}. Adding the different parts
together, we then obtain as the final result of the $T=0$ pressure in the $\msbar$ scheme
\ba
\label{pt0res}
p_\rmi{QCD}(T=0)
&=&\fr{1}{4\pi^2}\Bigg(\sum_f \mu^4\bigg\{\fr{N}{3}-d_A\bigg(\!\fr{g}{4\pi}\!\!\!\bigg)^{\!\!2}
-d_A\bigg(\!\fr{g}{4\pi}\!\!\!\bigg)^{\!\!4}\bigg[\fr{2}{3}\(11N-2n_f\)\ln\fr{\bar{\Lambda}}{\mu} +
\fr{16}{3}\ln\,2 \nn
&+&\fr{17}{4}\fr{1}{N} + \fr{1}{36}\(415-264\,\ln\,2\)N-\fr{4}{3}\(\fr{11}{6}-\ln\,2\)n_f\bigg]\bigg\} \nn
&-& d_A\bigg(\!\fr{g}{4\pi}\!\!\!\bigg)^{\!\!4}\bigg\{\Big(4\,\ln\,\fr{g}{4\pi} -
\fr{22}{3} +\fr{16}{3}\,\ln\,2\(1-\ln\,2\) + \delta+ \fr{2\pi^2}{3}\Big)({\mbox{\boldmath$\mu$}}^2)^2+F({\mbox{\boldmath$\mu$}})\bigg\}\Bigg) \nn
&+&\mathcal{O}(g^6\ln\,g),
\ea
where the function $F$ is defined by
\ba
F({\mbox{\boldmath$\mu$}})&=& -2{\mbox{\boldmath$\mu$}}^2\sum_f \mu^2\,\ln\,\fr{\mu^2}{{\mbox{\boldmath$\mu$}}^2}
+\fr{2}{3}\sum_{f>g}\bigg\{(\mu_f-\mu_g)^2\ln\,\fr{|\mu_f^2-\mu_g^2|}{\mu_f\mu_g} \nn
&+& 4\mu_f\mu_g(\mu_f^2+\mu_g^2)\ln\,\fr{(\mu_f+\mu_g)^2}{\mu_f\mu_g}-(\mu_f^4-\mu_g^4)\ln\,\fr{\mu_f}{\mu_g}\bigg\}
\ea
and the constant $\delta$ has the approximative value $\delta\approx-0.85638320933...$ This constant possesses a simple
one-dimensional integral respresentation \cite{avpres}, the analytic value of which we however have not been able to
obtain even using the PSLQ algorithm \cite{pslq}.

\chapter{Analysis of the perturbative results}
The correctness of most of the terms in the $\mc{O}(g^6\ln\,g)$ perturbative result obtained for the QCD pressure in the previous section has been verified in several independent computations in the limits of $\mu=0$ \cite{az,bn1}, $T=0$ \cite{fmcl} and large $n_f$ \cite{ippreb}. The practical value of the result is, however, solely determined by its predictive power, i.e.~in which parts of the ($\mu$-$T$)-plane we can trust perturbation theory and, more specifically, dimensional reduction to be applicable. As there is no experimental data available for the pressure, this question is a very difficult one to address. There are nevertheless several methods we may try to apply.

Four-dimensional lattice simulations (see e.g.~Refs.~\cite{karsch2,gup4,boyd,gup3}) have during the past decade produced a number of accurate results for the pressure, but are even today restricted to small chemical potentials, $\mu \lesssim T$, and to temperatures relatively close to the critical one, $T\lesssim 5T_c$. This is due to the fact that at finite $\mu$ the exponentiated action appearing in Eq.~(\ref{zres}) becomes complex and thus spoils importance sampling, while at large values of $T$ the lattice treatment rapidly becomes more and more time-consuming. The lattice computations may nevertheless provide a good check of our results in the region of relatively small $T$ and $\mubar$, but it is \textit{a priori} not at all clear whether perturbation theory can be meaningfully applied so close to the phase transition line. The perturbative expansions have after all been derived assuming the coupling constant to have an in principle arbitrarily small value, which is only guaranteed to be the case in the limit of asymptotically large temperatures or densities.

Another useful way to investigate the reliability of perturbation theory --- this time on the whole ($\mu$-$T$)-plane --- is to analyze the convergence properties of the perturbative expansions themselves. Assuming that a consistent all-orders summation of perturbation theory would lead to a correct result for the pressure, the question of determining the accuracy of an expansion carried out to a specific order in $g$ can naturally be reduced to analyzing the rate at which the corresponding series converges. This can, on the other hand, be often easily and efficiently examined through only the first few terms of the expansion, as we will soon observe.

In our analysis of the perturbative results we will from now on restrict the treatment to the physical case of $N=3$. We
find that the most illustrative way to proceed is to plot the results as functions of temperature and chemical potentials
and then to analyze the graphs in a qualitative fashion. This is done separately for three different cases: first for the
pressure at zero chemical potential, then for the $\mu$-dependent part of the pressure at high temperatures, and lastly
for the zero-temperature pressure. In each region we investigate the convergence properties and scale parameter
dependence of the results and for the first two we also provide a comparison with recent lattice results. The analysis of
the $\mu$-dependence of the high-temperature pressure is of special interest to us, as there we have both the pressure
and the quark number susceptibilities (see Ref.~\cite{avsusc}) to compare with the lattice data.

When investigating the plots of the following sections, it should be noted that while in Secs.~4.1 and 4.3 we consistently
apply the one-loop result, Eq.~(\ref{gres1l}), for the gauge coupling constant $g$,
this will not be the case in Sec.~4.2. There we will in analogy with Refs.~\cite{avsusc,avpres} use a three-loop result
for the quantity, as we have wanted to make the comparison with the figures of these papers as straightforward as possible. The
effect of this
choice on the plots is numerically relatively small, but the question is nevertheless in principle important. Using a three-loop
result for $g$
is somewhat inconsistent in light of the fact that one only needs the leading order term of its running to verify that
the renormalization scale dependence of the perturbative result, Eq.~(\ref{pres}), cancels at order $g^6\ln\,g$.

\begin{figure}[t]
\centerline{\epsfxsize=7.4cm\epsfysize=6.8cm\epsfbox{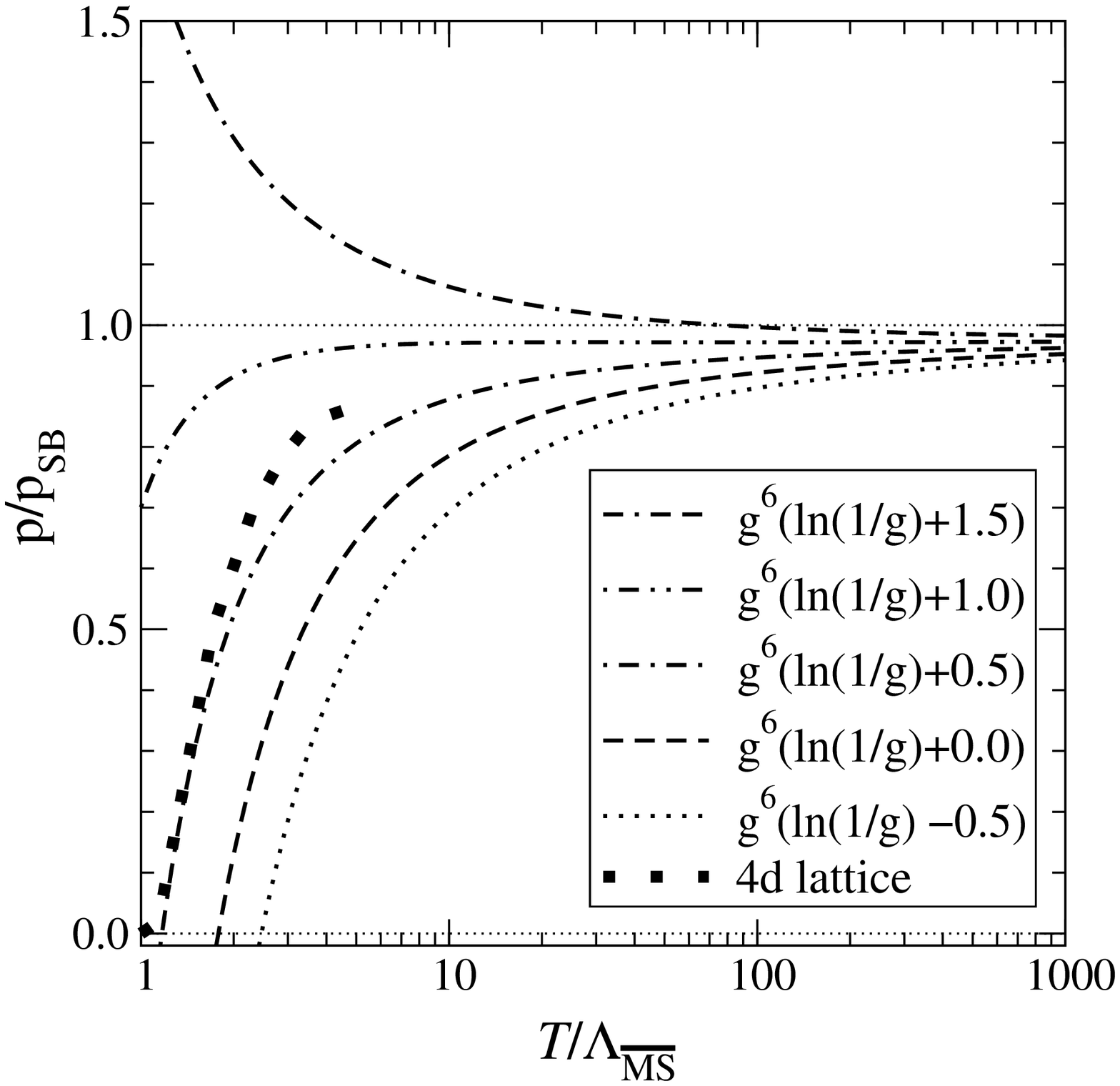}\;\;\;\;\;\;\epsfxsize=7.4cm\epsfysize=6.8cm\epsfbox{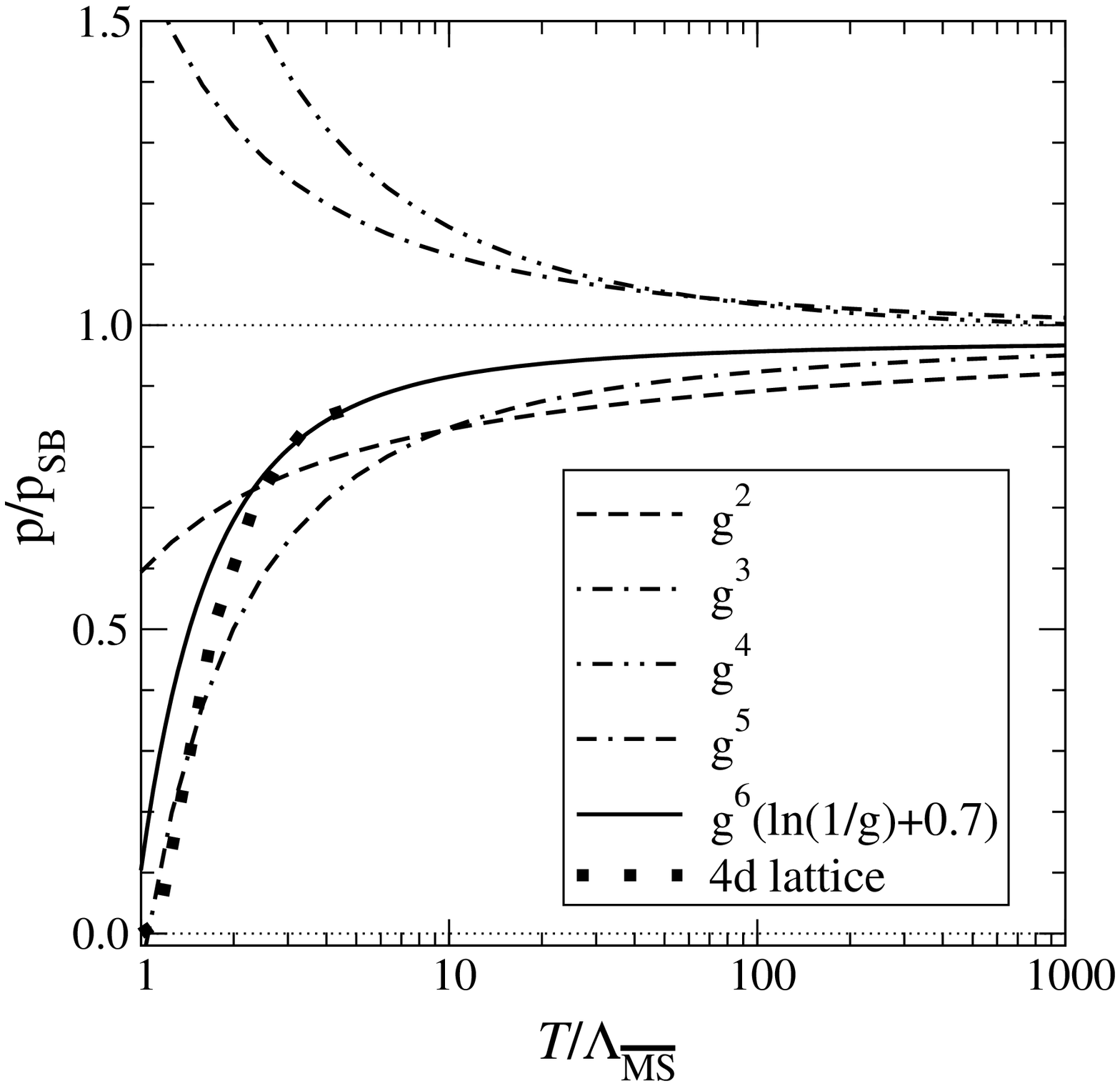}}
\caption[a]{Left: the order $g^6$ perturbative result for the $n_f=0$ pressure plotted for different values of the
$\mc{O}(g^6)$ coefficient. Right: the result plotted to different perturbative orders. The graphs are from Ref.~\cite{klry}
and the lattice data from Ref.~\cite{boyd}.}
\end{figure}

\section{The pressure at vanishing density}
At zero chemical potential the order $g^6\ln\,g$ expression for the QCD pressure was first derived in Ref.~\cite{klry}. The
analysis of this result is complicated by the undetermined $\mathcal{O}(g^6)$ term in the expansion, which
induces an arbitrariness to the numerical factors that appear inside the logarithms of the $g^6\ln\,g$ part. This has a
considerable effect on the behavior of the result, and until the full $\mc{O}(g^6)$ contribution has been computed, the
best we can do is to treat the coefficient of this term as a free parameter. This has been done in Fig.~4.1 \cite{klry},
where the pressure has been plotted as a function of temperature at $n_f=0$ using the value\footnote{This value has been
obtained by optimizing the convergence of the one-loop (NLO) expansion of $g_\rmi{E}^2$. See Ref.~\cite{klrs} for details.}
$\bar{\Lambda}_\rmi{opt}=6.742T$ for the scale parameter. It is seen that by optimizing the value of the unknown $g^6$
part, the perturbative results can be made to cleanly approach the lattice datapoints even at temperatures surprisingly
close to $T_c$, but that the effect of varying the coefficient of the $\mc{O}(g^6)$ term is so large that the predictive
power of perturbation theory remains regrettably weak at present. As has been conjectured in Ref.~\cite{lainesewm}, there is nevertheless reason to believe that once the full order $g^6$ computation has been finished, the situation will improve significantly. At this order the magnetostatic sector namely enters the expansion of the pressure non-perturbatively, after which one has accounted for the dynamics of all the relevant energy scales --- $2\pi T$, $gT$ and $g^2T$.

\begin{figure}[t]
\centerline{\epsfxsize=7.4cm\epsfysize=6.8cm\epsfbox{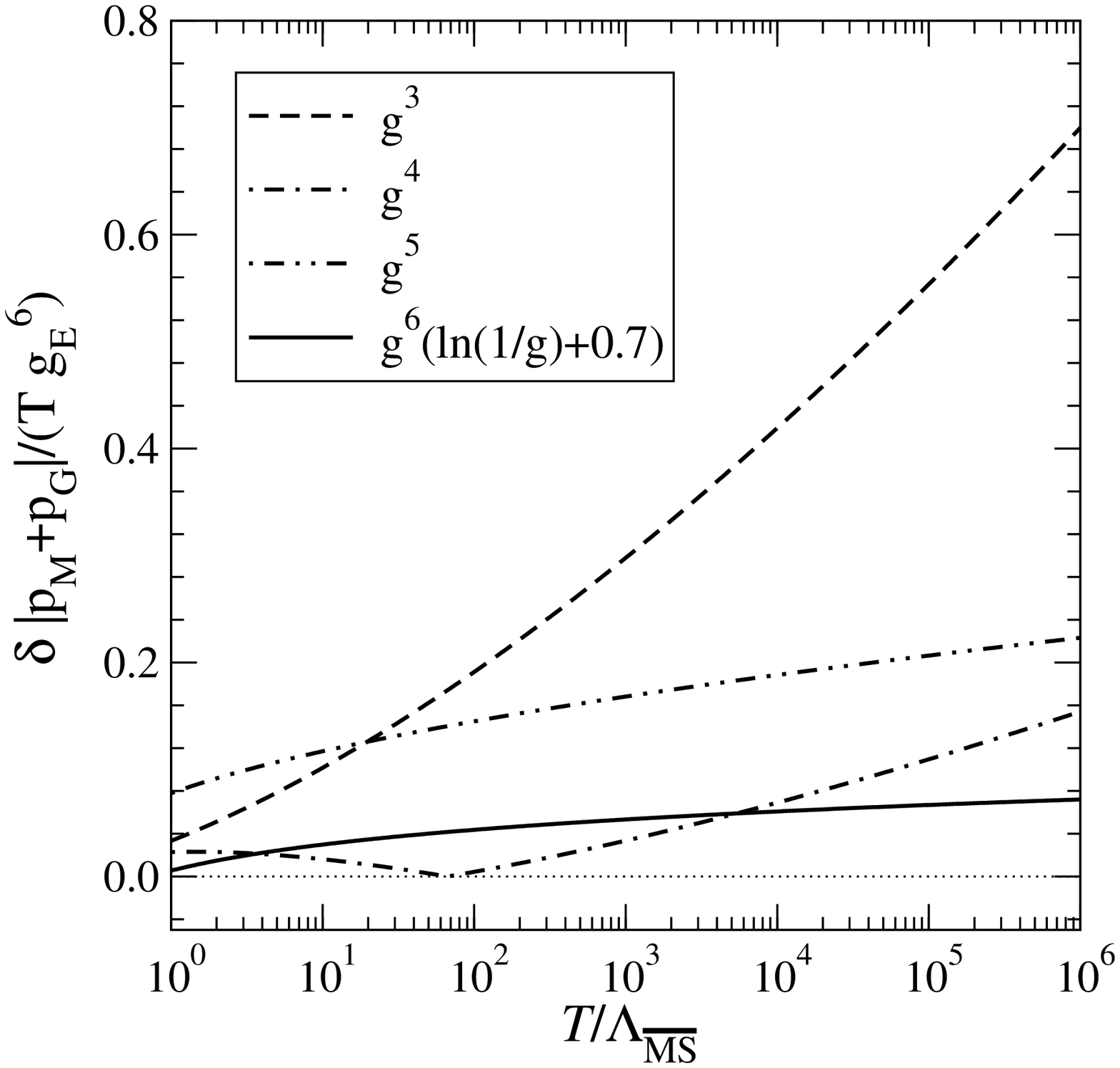}\;\;\;\;\;\;\epsfxsize=7.4cm\epsfysize=6.8cm\epsfbox{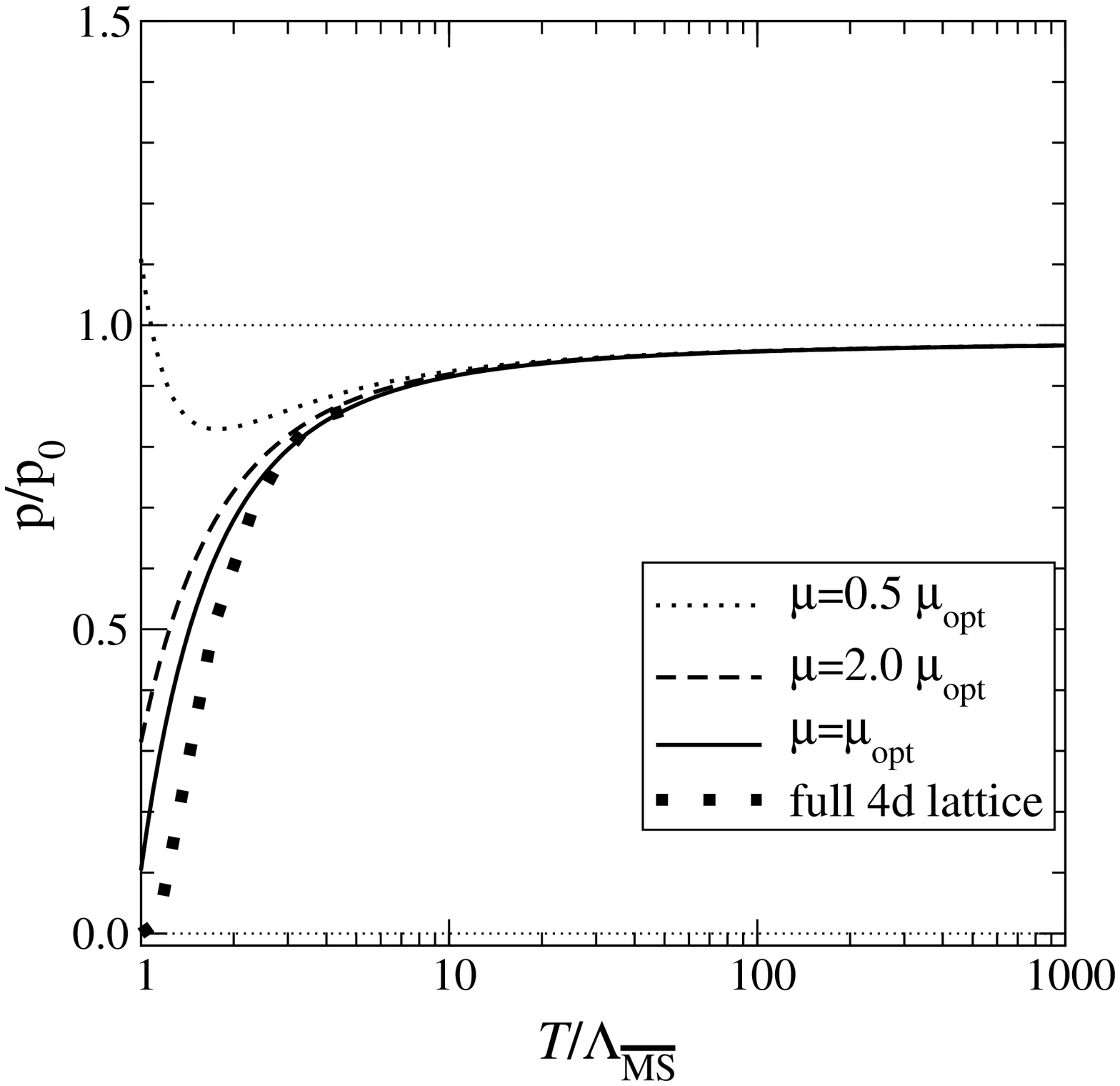}}
\caption[a]{Left: the different terms of the perturbative expansion of $p_\rmi{M}+p_\rmi{G}$ at $n_f=0$ normalized by
$Tg_\rmi{E}^6=T^4(g^6+\mc{O}(g^8))$ \cite{klry}. Right: the scale dependence of the $\mu=0$ pressure (the $g^6(\ln(1/g)+0.7)$ case) at $n_f=0$ \cite{lainesewm}. Lattice data is again from Ref.~\cite{boyd}.}
\end{figure}

The convergence properties of the $\mu=0$ expansion have been analyzed in detail in Refs.~\cite{bn1,klry,bir3} with
the result that the slowest convergence has been traced back to the contributions of the soft momentum scales entering
through the effective theories. This is due to the simple fact that while for hard momenta of order $2\pi T$ the expansion parameter is $g^2/\pi^2$, the situation is completely different for the soft scales: for the color-electric modes the expansion is in powers of $g/\pi$ and for color-magnetic modes there is no perturbative series at all. This part of the pressure, or more specifically $p_\rmi{M}+p_\rmi{G}$, is studied in
Fig.~4.2 a, where the absolute values of the different terms of the function are plotted appropriately normalized. The
general tendency in the plot is clearly favorable in the sense that at sufficiently large energies the magnitudes of
the different terms keep decreasing as we move on to higher orders. This, however, only happens at
very high temperatures, and the convergence of the perturbative expansion close to $T_c$ is far more questionable.

The scale dependence of Eq.~(\ref{pres}) is an equally important issue to address, as it induces an ambiguity to the results, which in principle is at least as serious as that originating from the unknown order $g^6$ term. There are, however, several methods we may try to use to reduce this effect, of which the most widely used is due to Brodsky, Lepage and Mackenzie \cite{blm}. The BLM criterion states that the scale should be chosen in such a way that in the highest perturbative order it makes the term proportional to the highest power of $n_f$ vanish. The scheme has been applied to the expansions of the parameters of the effective theories in Ref.~\cite{bn1} with the result that the optimal scale has been observed to vary roughly between $\pi T$ and $4\pi T$. This seems physically very reasonable in light of the fact that the lowest non-zero bosonic Matsubara mode $2\pi T$ falls within this interval and that with this choice the magnitudes of the logarithms appearing inside the matching coefficients remain small.

In Fig.~4.2 b the renormalization scale dependence of the pure gauge pressure (at $\mu=0$) is analyzed by plotting the function at $\bar{\Lambda}=\bar{\Lambda}_\rmi{opt}/2$ and $\bar{\Lambda}=2\bar{\Lambda}_\rmi{opt}$. It is clearly seen that while at high temperatures the effect of the scale variation rapidly decreases, it does produce a sizable ambiguity to the results at temperatures close to $T_c$. This implies that even after the determination of the order $g^6$ contribution to the pressure, the perturbative result will always contain one free parameter that can in principle be adjusted to match the lattice data at low temperatures. There have been attempts to improve the situation through different resummation schemes, where the contributions of the soft scales have been treated without truncations \cite{bir3}, but a detailed analysis of this work is outside the scope of our present treatment.

\section{Effects of small but finite chemical potentials}
Moving on to study the chemical potential dependence of the high-temperature pressure \cite{avpres}, we have a large selection of
lattice data available for comparison. The pressure itself has been computed at small values of $\mu$ in Refs.~\cite{karsch2,fodor,gup4}, and at $\mu=0$ its higher derivatives with respect to the chemical potentials, i.e.~the quark number
susceptibilities, are available from Refs.~\cite{karsch2,gup4,gup2,gup3}. The compatibility of the perturbative results
and the mainly quenched lattice data of Gupta \textit{et al.} \cite{gup4,gup3,gup2} has been analyzed in detail in the
papers \cite{avsusc,avpres}, but here we will extend the analysis by comparing the behavior of the perturbative expansion also with the $n_f=2$ work of Allton \textit{et al.} \cite{karsch2}.

\begin{figure}[t]

\centerline{\epsfxsize=7.4cm\epsfysize=6.8cm\epsfbox{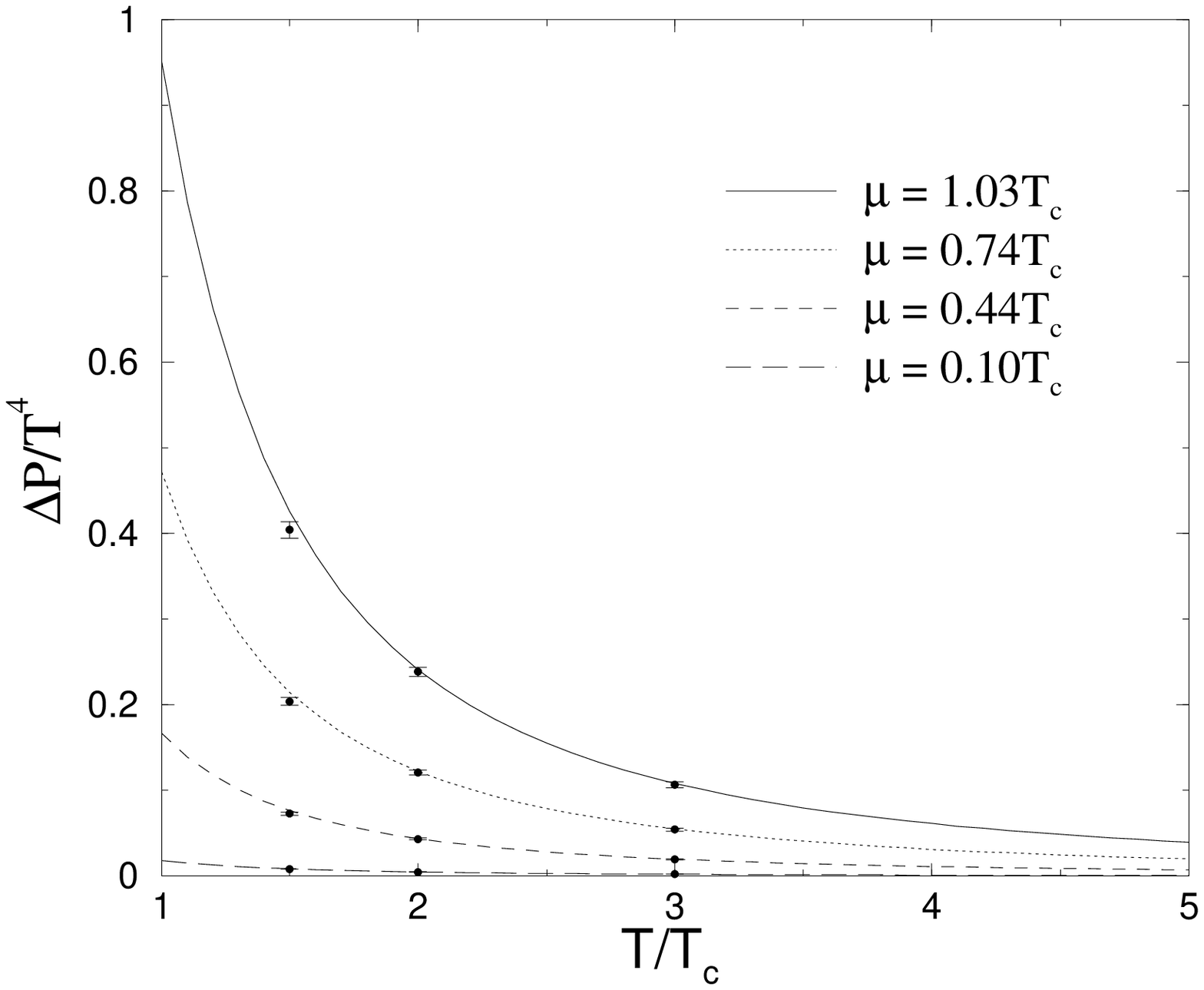}\;\;\;\;\;\;\epsfxsize=7.4cm\epsfysize=6.8cm\epsfbox{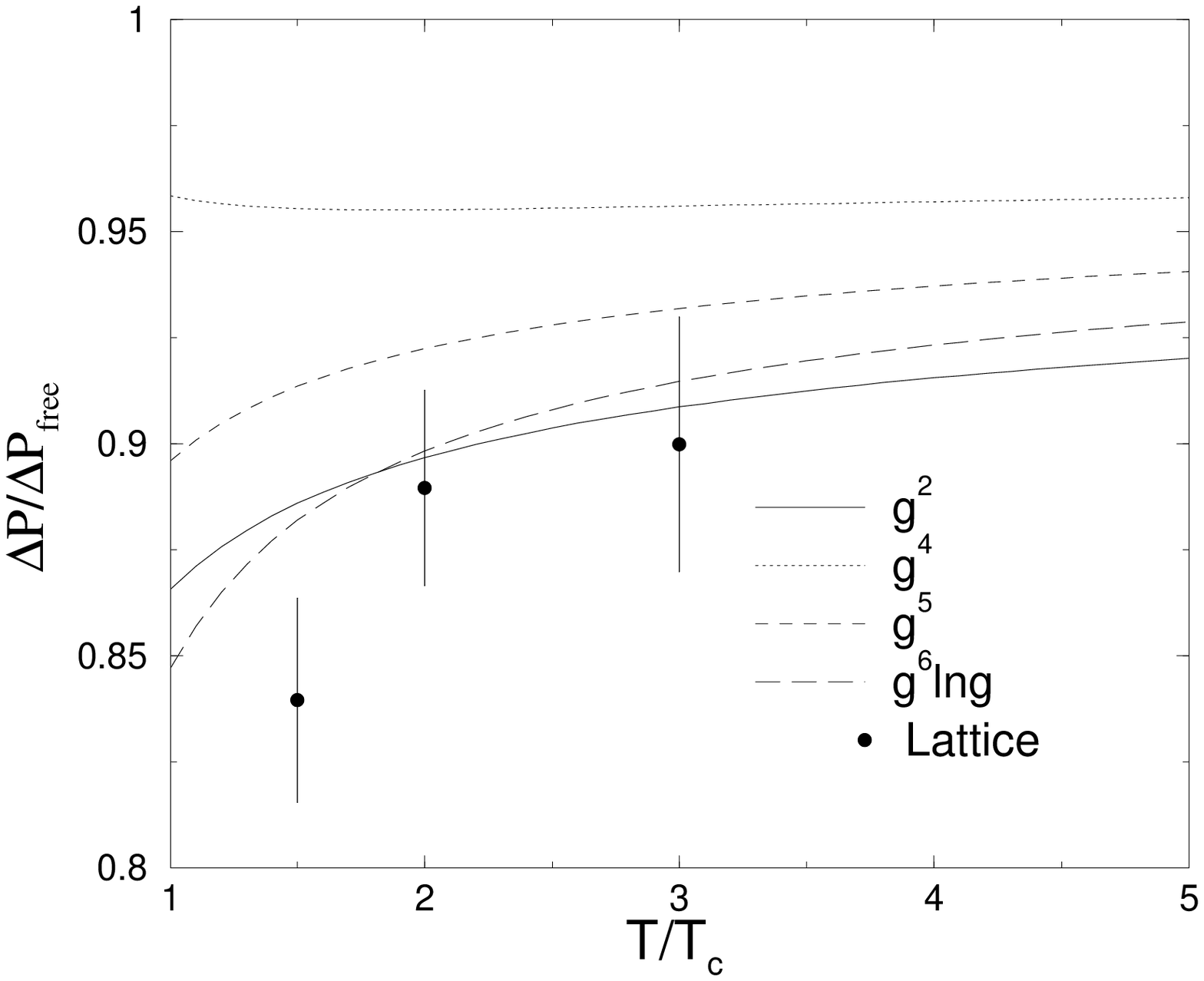}}

\caption[a]{Left: $\Delta P|_{n_f=0}$ plotted for different values of $\mu$. Right: the different
perturbative orders of $\Delta P(\mu=0.44T_c)|_{n_f=0}$ normalized by the free theory result. The lattice data is from Ref.~\cite{gup4},
and the result $T_c/\Lambda_{{\overline{\rm MS}}}|_{n_f=0}=1.15$ from Ref.~\cite{gup1} has been applied.}
\end{figure}

\begin{figure}[t]

\centerline{\epsfxsize=7.4cm\epsfysize=6.8cm\epsfbox{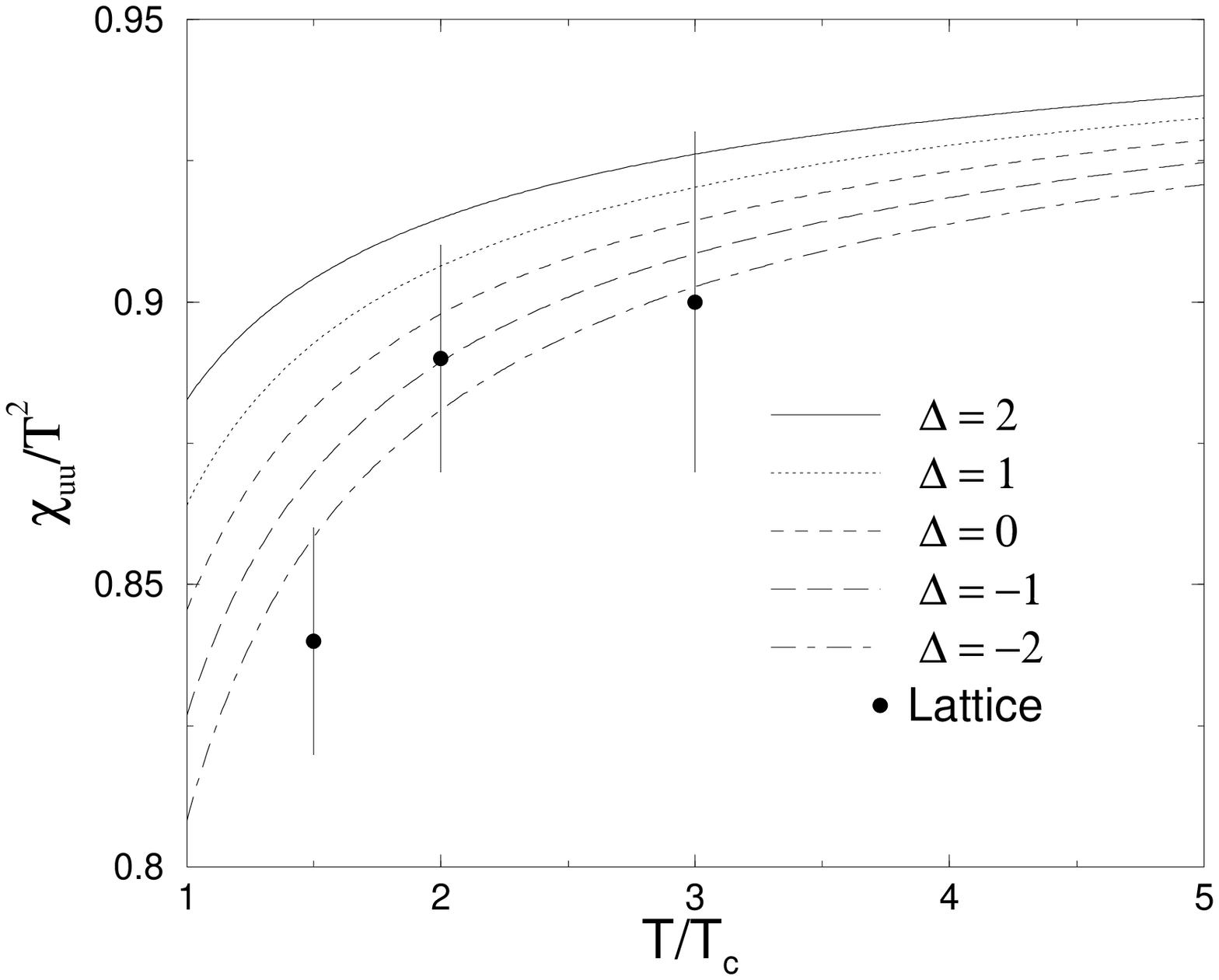}\;\;\;\;\;\;\;\epsfxsize=7.4cm\epsfysize=6.8cm\epsfbox{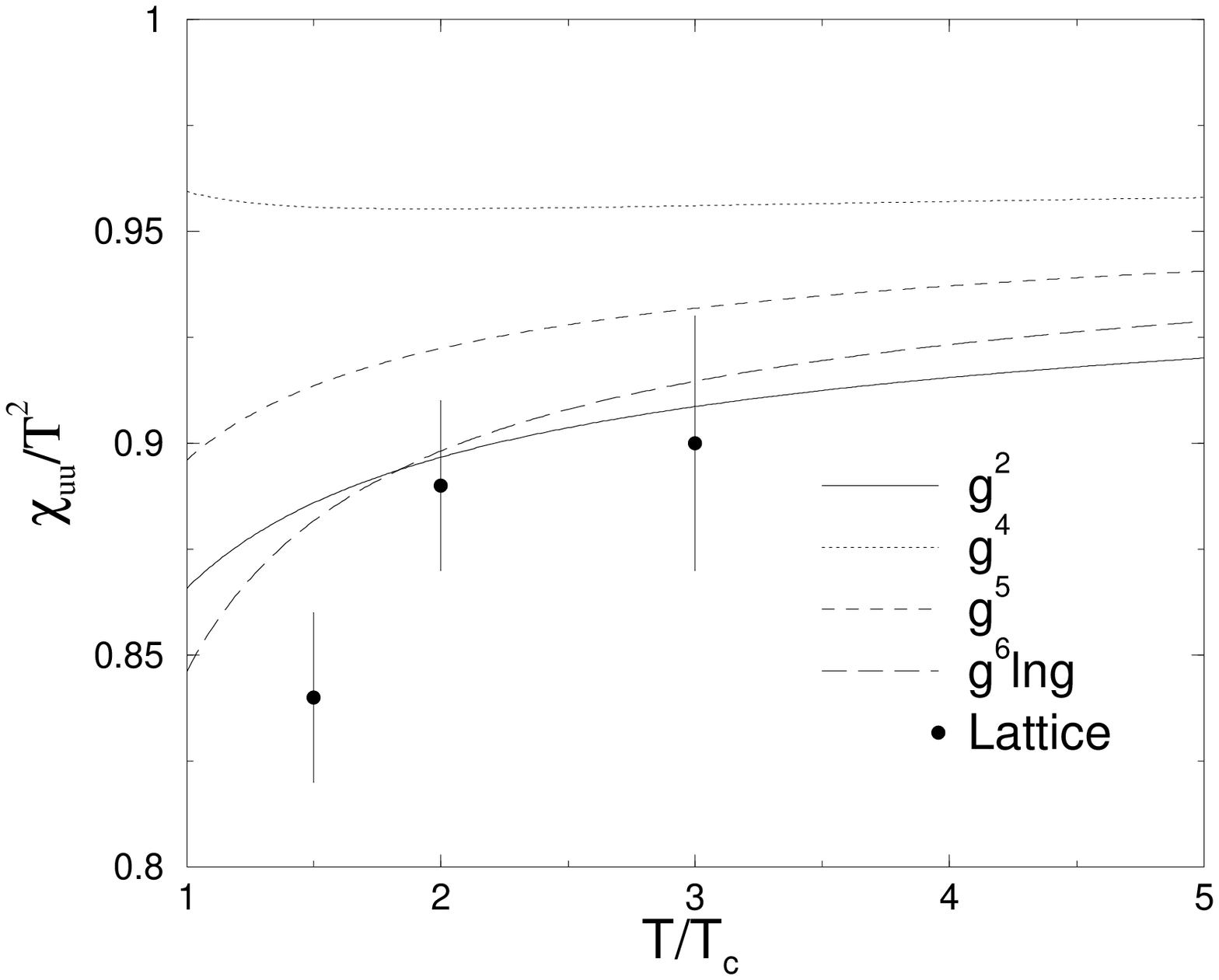}}

\caption[a]{Left: $\chi_{uu}/T^2|_{n_f=0}$ plotted for different values of $\Delta$ corresponding to the expected effect
of the $\mc{O}(g^6)$ term. Right: the expansion of the quantity at $\Delta=0$ to different perturbative orders. The lattice data is from Ref.~\cite{gup2}.}
\end{figure}
\begin{figure}[t!]

\centerline{\epsfxsize=7.4cm\epsfysize=6.8cm\epsfbox{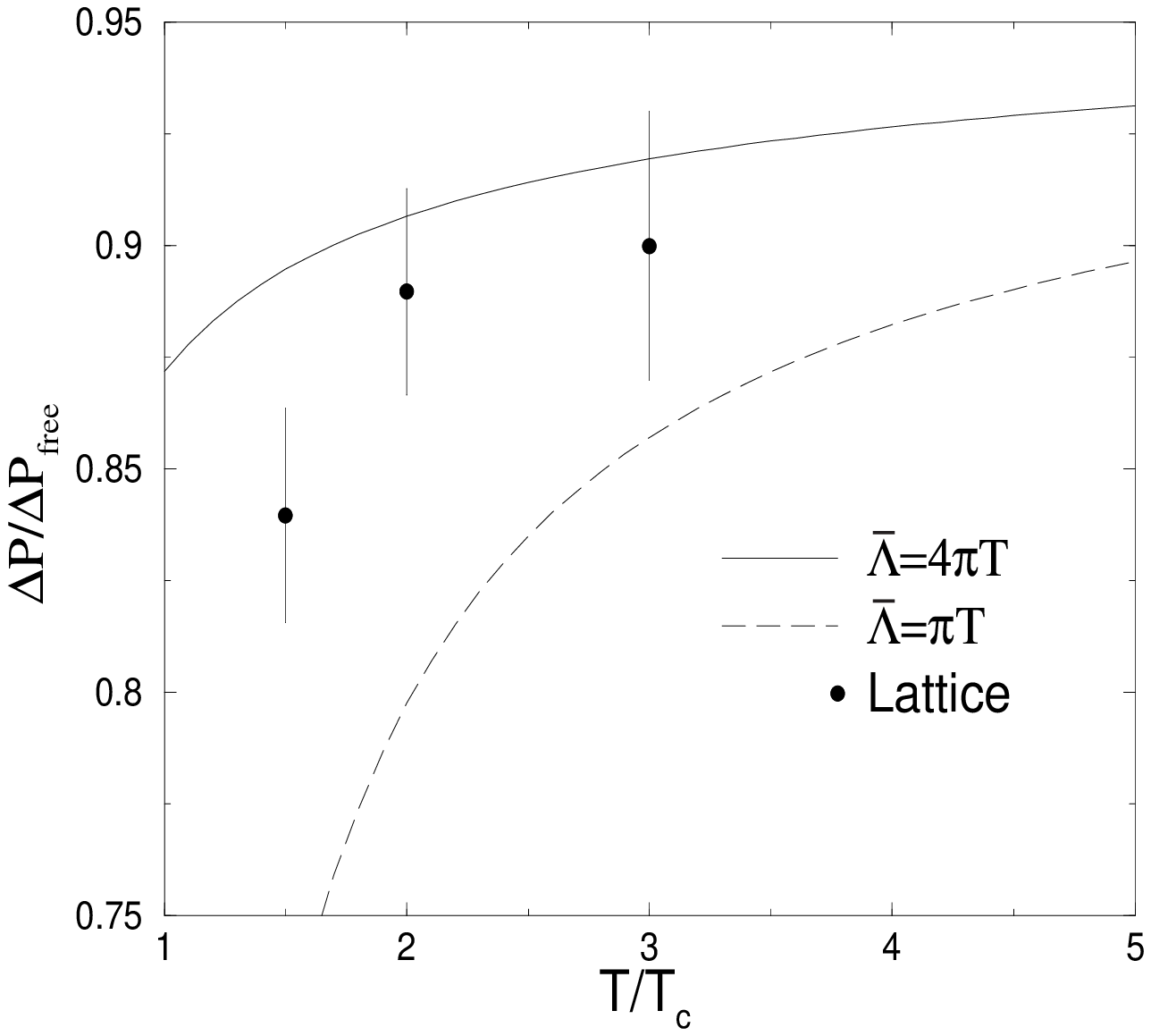}\;\;\;\;\;\;\;\epsfxsize=7.4cm\epsfysize=6.8cm\epsfbox{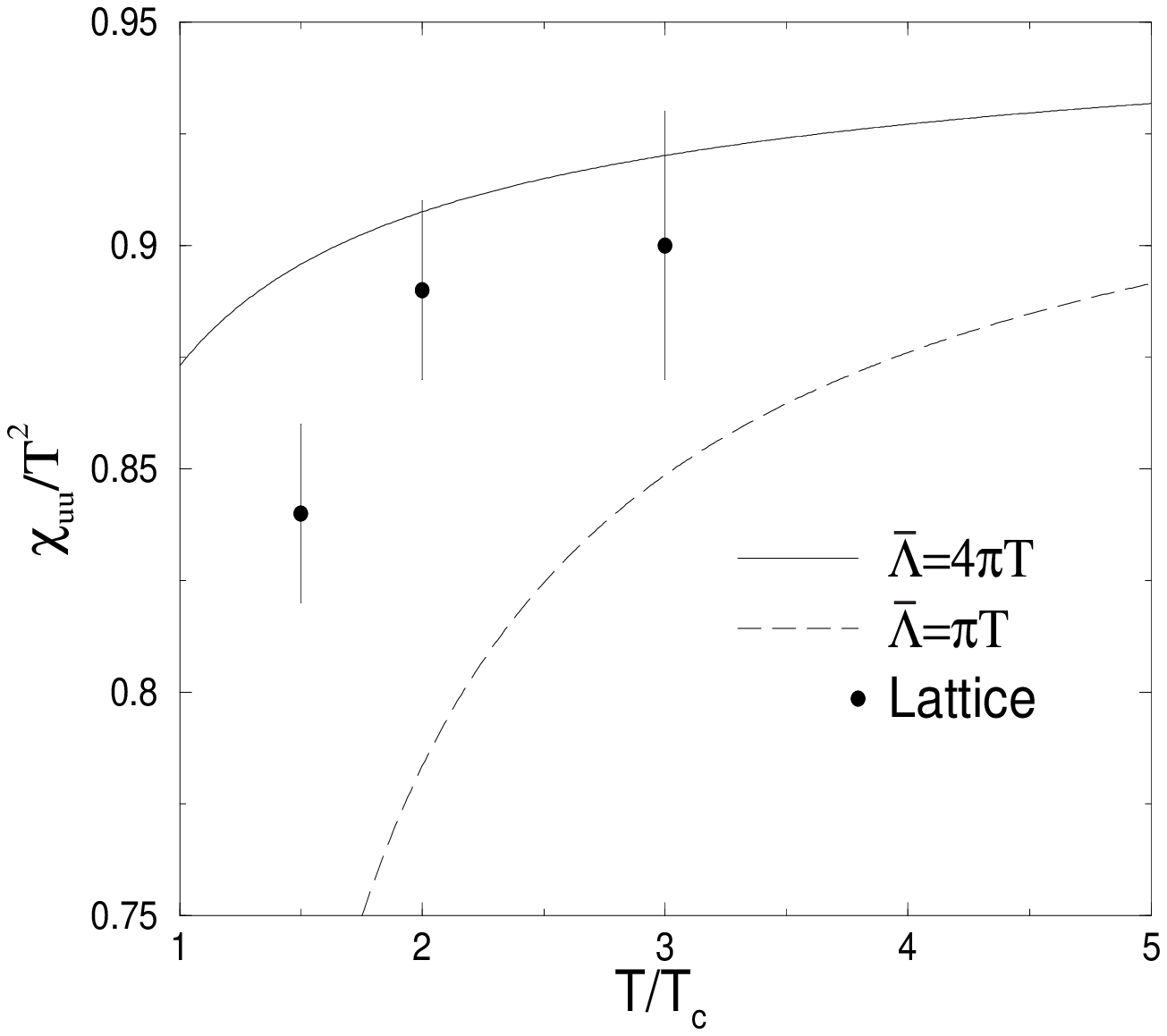}}

\caption[a]{Left: the scale dependence of $\Delta P|_{n_f=0}$ at $\mu=0.44T_c$. Right: the scale dependence of
$\chi_{uu}/T^2|_{n_f=0}$. The lattice results are from Refs.~\cite{gup4,gup2}.}
\end{figure}

\begin{figure}[t]

\centerline{\epsfxsize=7.4cm\epsfysize=6.8cm\epsfbox{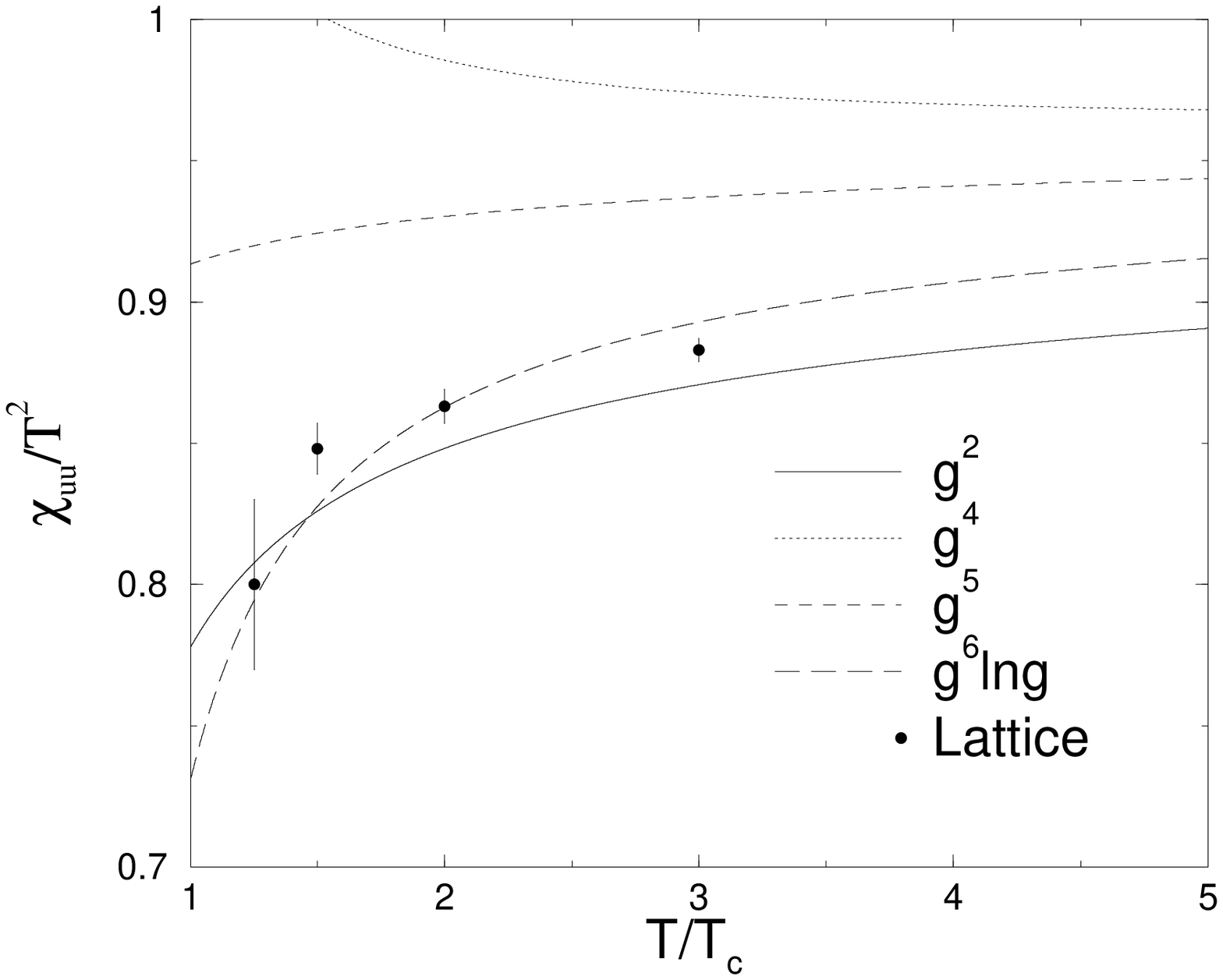}\;\;\;\;\;\;\epsfig{file=qns_0423.eps, width=6.8cm, height=6.7cm}}

\caption[a]{Left: $\chi_{uu}/T^2|_{n_f=2}$ plotted to different orders together with lattice data \cite{gup3}. Right: lattice results for different susceptibilities \cite{karsch2} (for definitions, see the original paper). In the first plot $\bar{\Lambda}=8.112T$ \cite{klrs} and $T_c/\Lambda_{{\overline{\rm MS}}}|_{n_f=2}=0.49$ \cite{gup1} have been used; in the second $T_0\equiv T_c$.}
\end{figure}

\begin{figure}[t!]

\centerline{\epsfxsize=7.4cm\epsfysize=6.8cm\epsfbox{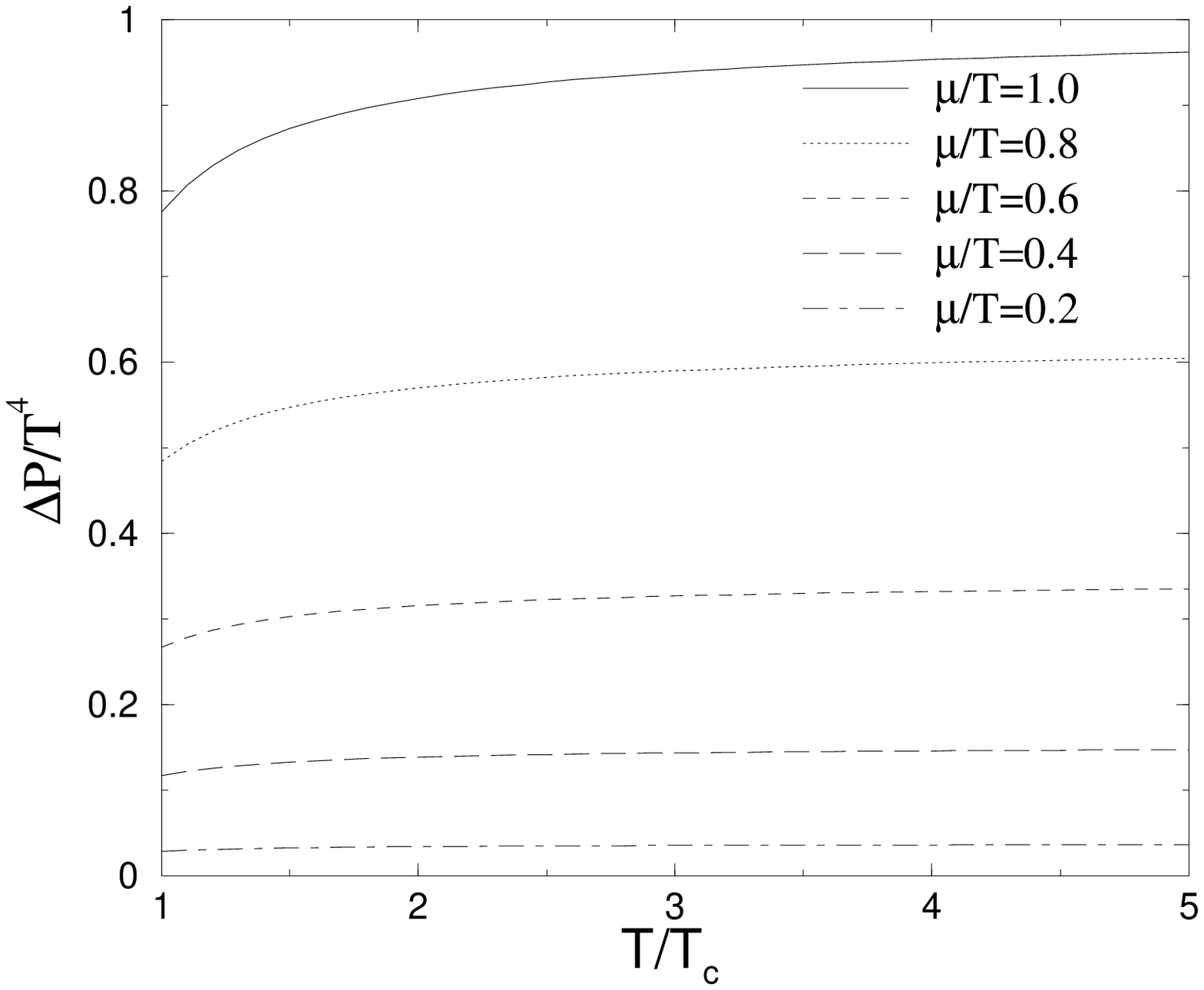}\;\;\;\;\;\;\epsfig{file=eosmu4T_nf2_0424C.eps, width=6.8cm, height=6.7cm}}

\caption[a]{Left: $\Delta P/T^4|_{n_f=2}$ plotted for different ratios of $\mu/T$, i.e.~not for fixed $\mu/T_c$. Right: lattice results for the same
quantity from Ref.~\cite{karsch2}.}
\end{figure}

The quantities we analyze in this section are the deviation of the pressure from its $\mu=0$ value,
\ba
\Delta P(T,\mu)&=&p_\rmi{QCD}(T,\mu)-p_\rmi{QCD}(T,0), \label{deltap}
\ea
as well as the quark number susceptibilities
\ba
\chi_{ijk...}&\equiv&\fr{\partial^n p_\rmi{QCD}}{\partial\mu_i \partial\mu_j \partial\mu_k ...}.
\ea
In Ref.~\cite{gup4} these functions were computed on the lattice in the quenched approximation, where no sea quarks are
present. From the perturbative expression of the pressure, Eq.~(\ref{pres}), the corresponding results can on the other hand
be obtained by first computing
the susceptibilities at arbitrary $n_f$, then taking the formal limit $n_f=0$ and in the end integrating the result back
to obtain $\Delta P$ (see the discussion in Ref.~\cite{avpres}). We have compared these results for
$\Delta P$ and $\chi_{uu}$, the linear (second-order) diagonal susceptibility, in Figs.~4.3 and 4.4, which show remarkable agreement even at small values
of $T$. In addition to this, we observe a considerable improvement in the convergence properties of the perturbative expansions in
comparison with the zero chemical potential case. This time also the scale dependence of the results, shown in Fig.~4.5,
remains almost entirely within the error bars of the lattice data already at $T=3T_c$. In studying the above figures it should in particular be noted that unlike in the $\mu=0$ case we have here in no way optimized the coefficient of the $g^6$ term in the perturbative result, but instead simply used the $\mc{O}(g^6\ln\,g)$ part of Eq.~(\ref{pres}).

Figs.~4.3 to 4.5 reveal also another very interesting feature of the quantities $\Delta P$ and $\chi_{uu}$: when
normalized by their respective free theory values, the corresponding curves practically fall on top of each
other. This observation, which at first instant seems curious, is explained by the fact that at small values of
$\mu$ the behavior of the quantity $\Delta P$ is almost solely determined by the leading order term of its Taylor
expansion in $\mubar^2$, the linear diagonal susceptibility. The
linear non-diagonal susceptibility $\chi_{ud}$ is of negligible magnitude in comparison with $\chi_{uu}$, and the
contribution of the non-linear susceptibilities to $\Delta P$ is suppressed by a factor $\mubar^2$,
which for $\mu=0.44T_c$, $T > T_c$ is very small. In the present work we will therefore not consider the non-linear and
non-diagonal susceptibilities at all, but refer the reader to Refs.~\cite{avpres,bir}.

\begin{figure}[t]

\centerline{\epsfxsize=15cm\epsfysize=11.5cm \epsfbox{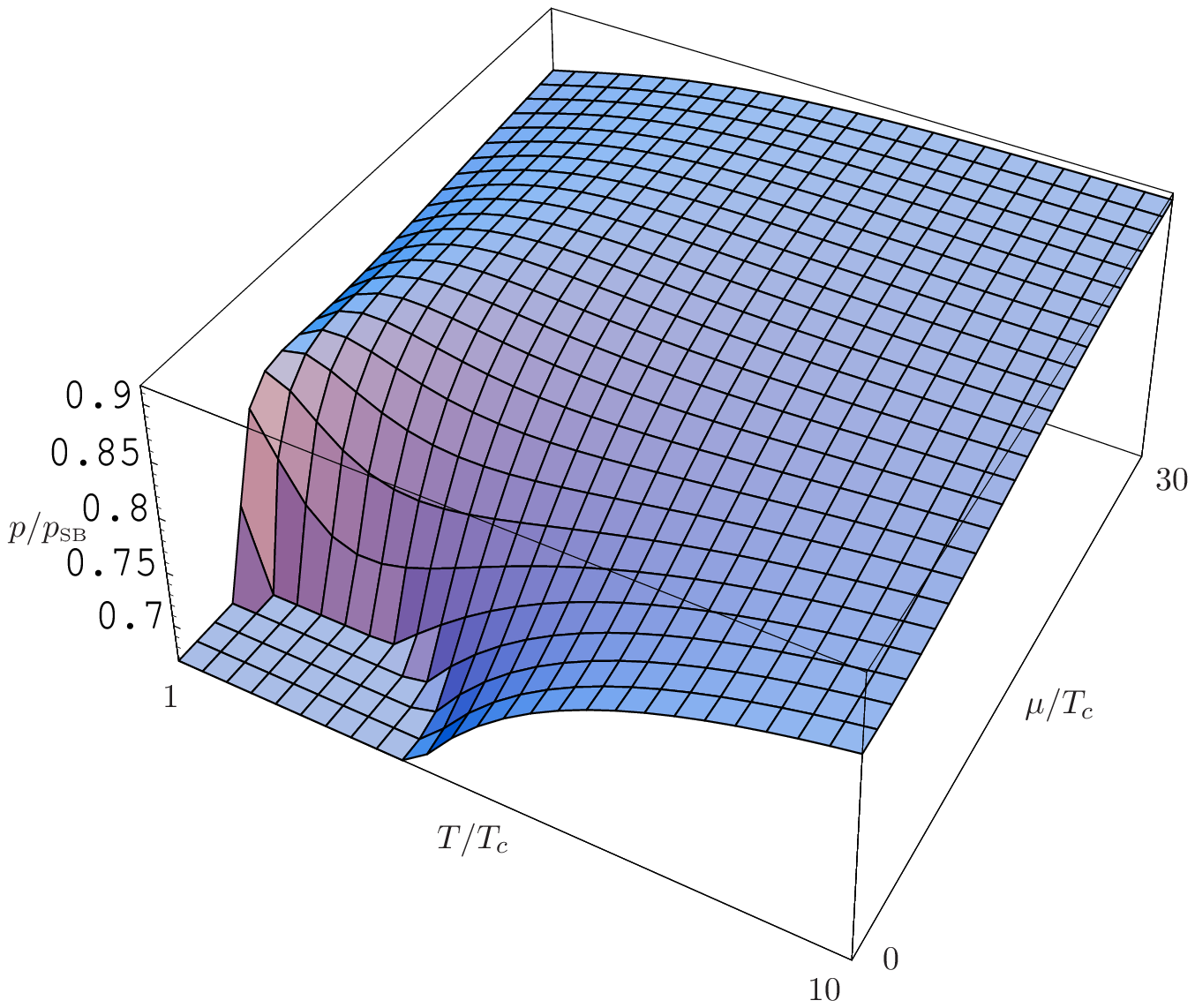}}

\caption[a]{$p/p_{\mbox{\tiny SB}}$ plotted on the ($\mu$-$T$)-plane for $n_f=2$.}
\end{figure}

The physically more interesting case of two quark flavors with $\mu_u=\mu_d\equiv \mu$ has been considered on the lattice in Refs.~\cite{karsch2,gup3}. For the linear diagonal susceptibilities the compatibility of these results and their perturbative counterparts is analyzed in Fig.~4.6, which shows good agreement at least with the data of Ref.~\cite{gup3}. Taking into account the relation
\ba
\chi_{uu}&=&\fr{1}{4}\(\chi_\rmi{q}+4\chi_\rmi{I}\)
\ea
between the definitions of Refs.~\cite{avpres} and \cite{karsch2}, the outcome of Ref.~\cite{karsch2} can, however, also be seen to
be fit in very nicely with the perturbative result, with the exception of temperatures very close to
$T_c$. A similar behavior is observed for the pressure in Fig.~4.7, where the quantity $\Delta P$ is plotted for
different fixed values of $\mu/T$. The good agreement of our predictions and the lattice results of Ref.~\cite{karsch2}, obtained using an unphysically high value $0.4T$ for the fermion masses, implies that the dependence of the results on quark masses should be of negligible magnitude.

\section{The zero-temperature limit}
While we have just witnessed the dimensionally reduced perturbative expansion of the pressure do exceptionally well in all
lattice and convergence tests especially at finite chemical potentials, it is --- as we pointed out already in chapter 3
--- \textit{a priori} not known how far away from the line $\mu=0$ we may continue to use it. This question can again be
ultimately answered only through comparison with experimental data or with the results of a genuinely non-perturbative
first-principles calculation. In the absence of both alternatives we must here resort to other means and choose
to follow and extend the treatment of Ref.~\cite{avpres}.

\begin{figure}[t]

\centerline{\epsfxsize=15cm\epsfysize=11.5cm \epsfbox{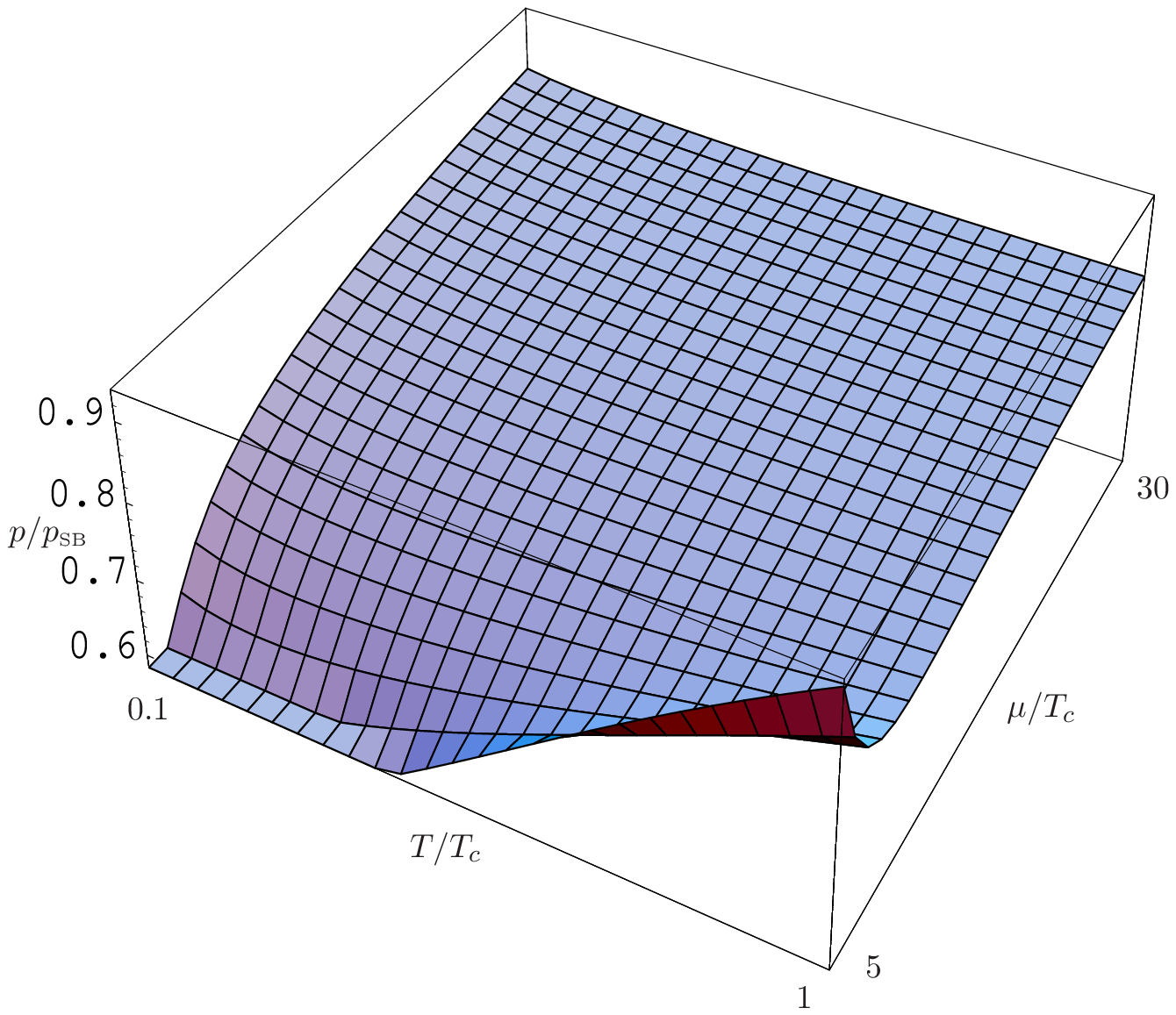}}
\caption[a]{The $\mc{O}(g^4)$ result for $p/p_{\mbox{\tiny SB}}|_{n_f=2}$ plotted for temperatures below $T_c$.}
\end{figure}

Let us first consider a strongly interacting system at a temperature well above the critical one, which ensures that
we are working in the deconfined phase. As has been argued in Ref.~\cite{hlp}, we expect dimensional reduction and hence
Eq.~(\ref{pres}) to be applicable at least as long as the chemical potentials satisfy $\mu_f\lesssim 4T$ for all flavors.
There is, however, no specific reason why we should expect the method to break down even at this point\footnote{This will
happen inevitably only when the chemical potentials reach values $\mu \sim 2\pi T/g$ making the three-dimensional mass
parameter $m_\rmi{E}$ to be of the same order of magnitude as the hard scales.}, and it is
therefore worthwhile to investigate the behavior of our result also at relatively high $\mu/T$. This has been done in
Fig.~4.8, where $p/p_{\mbox{\tiny SB}}$ has been plotted as a function of $\mu$ and $T$ in the case $n_f=2$,
$\mu_u=\mu_d\equiv \mu$ with the scale parameter (somewhat arbitrarily) reading
\ba
\bar{\Lambda}&=& 2 \pi T \sqrt{1+2\mubar^2}.\label{scale1}
\ea
We observe that the quantity stays below its free theory result 1 everywhere and only slowly approaches it when the values
of the temperature and the chemical potential increase. In the region of low $T$ and $\mu$ the result on the other hand
rapidly decreases reaching a value close to zero when $T=T_c$ and $\mu=0$. There the applicability of perturbation theory
is, however, anyway highly questionable due to the large value of the coupling constant.

\begin{figure}[t]

\centerline{\epsfxsize=15cm\epsfysize=11.5cm \epsfbox{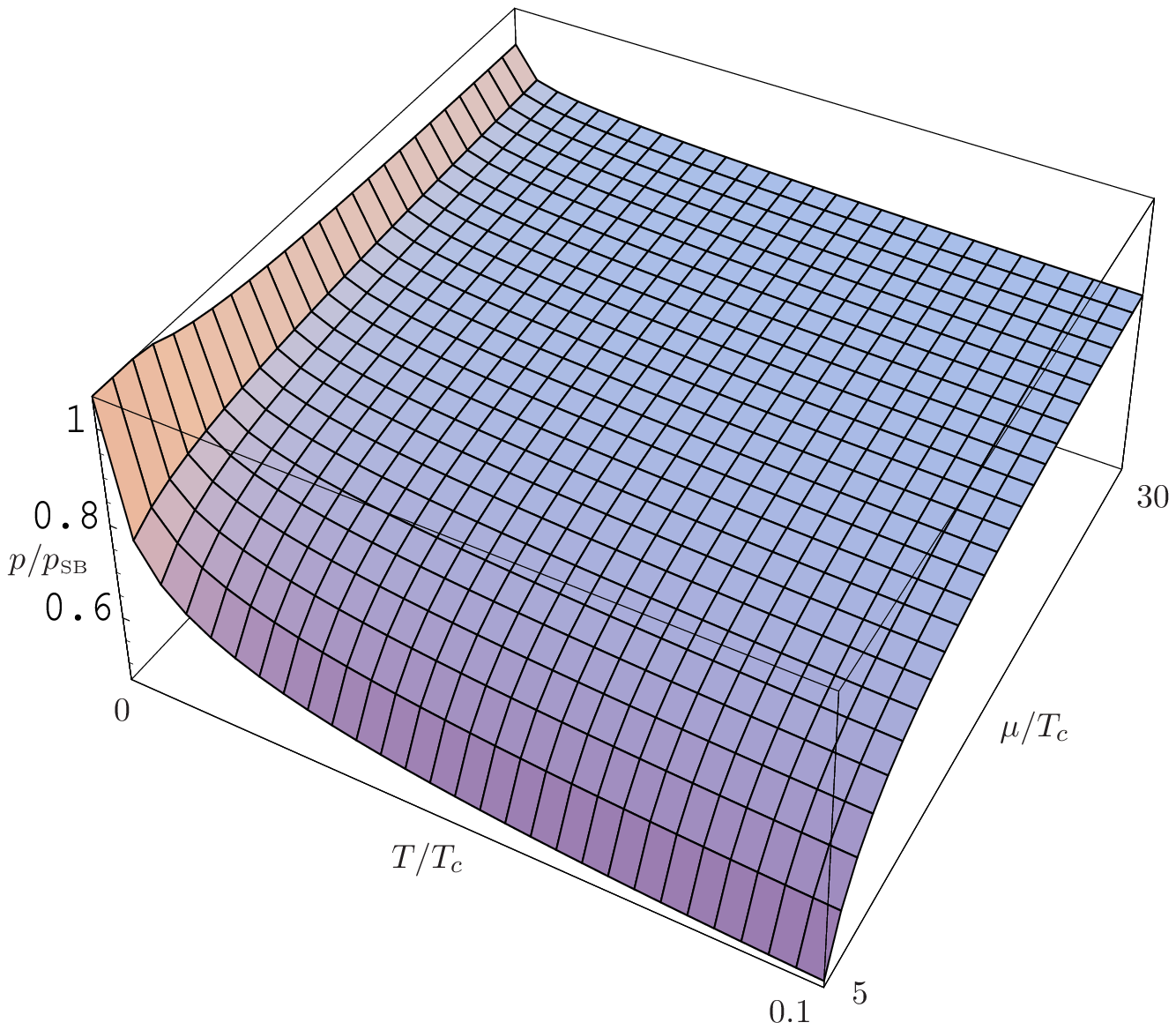}}

\caption[a]{The behavior of $p/p_{\mbox{\tiny SB}}|_{n_f=2}$ at $\mc{O}(g^4)$ near the $T=0$ line.}
\end{figure}

Moving down along the temperature axis, it is clear that the dimensionally reduced result ceases to be applicable at some
point. If we let the chemical potentials to be high enough to keep us in the deconfined phase, we may, however, keep
inspecting the behavior of the function and especially compare its order $g^4$
part with the zero-temperature result of Eq.~(\ref{pt0res}). Continuing to
work at $n_f=2$, we observe from Fig.~4.9, where the pressure has been plotted to $\mc{O}(g^4)$ at temperatures below
$T_c$, that the perturbative result slowly decreases as a function of $T$ all the way down to $T/\mu\sim 1/40$. At even lower temperatures than this the unphysical nature of the result\footnote{In Ref.~\cite{avippreb} the breakdown of dimensional reduction has been studied more closely in the limit of large $n_f$ with the result that its onset seems to take place already, when the ratio $\mu/T$ becomes of order 10.}, however, manifests itself as a divergence proportional to $\ln\fr{\mu}{T}$, as can easily be verified using the results of Appendix B.1.3. This behavior is demonstrated in Fig.~4.10.

The order $g^4$ perturbative expression for the zero-temperature pressure, Eq.~(\ref{pt0res}), is analyzed in Fig.~4.11,
where it is plotted as a function of $\mu$. Comparing its behavior with that
of the dimensionally reduced $\mc{O}(g^4)$ result in Fig 4.9, we see that as the temperature decreases the
high-$T$ expansion very smoothly approaches the
zero-temperature one, until the unphysical logarithms become dominant at small $T$. This indicates that our order
$g^6\ln\,g$ result may actually be applicable even at very low temperatures and high values of
$\mu/T$. Remembering on the other hand that the critical temperature for the meltdown of the color superconductor
is of the order $0.1T_c$, it would be extremely tempting to argue that the dimensionally reduced result is valid almost
throughout the quark-gluon plasma phase. We must, however, exercise caution in drawing these conclusions until the order
$g^4$ result applicable throughout the whole phase diagram is available. This computation obviously needs to
be performed for the sake of completeness alone, even though it now seems that its numerical effect on our present
results may well turn out to be of negligible magnitude in most parts of the ($\mu$-$T$)-plane.

\begin{figure}[t]
\centerline{\epsfxsize=7.4cm\epsfysize=6.8cm\epsfbox{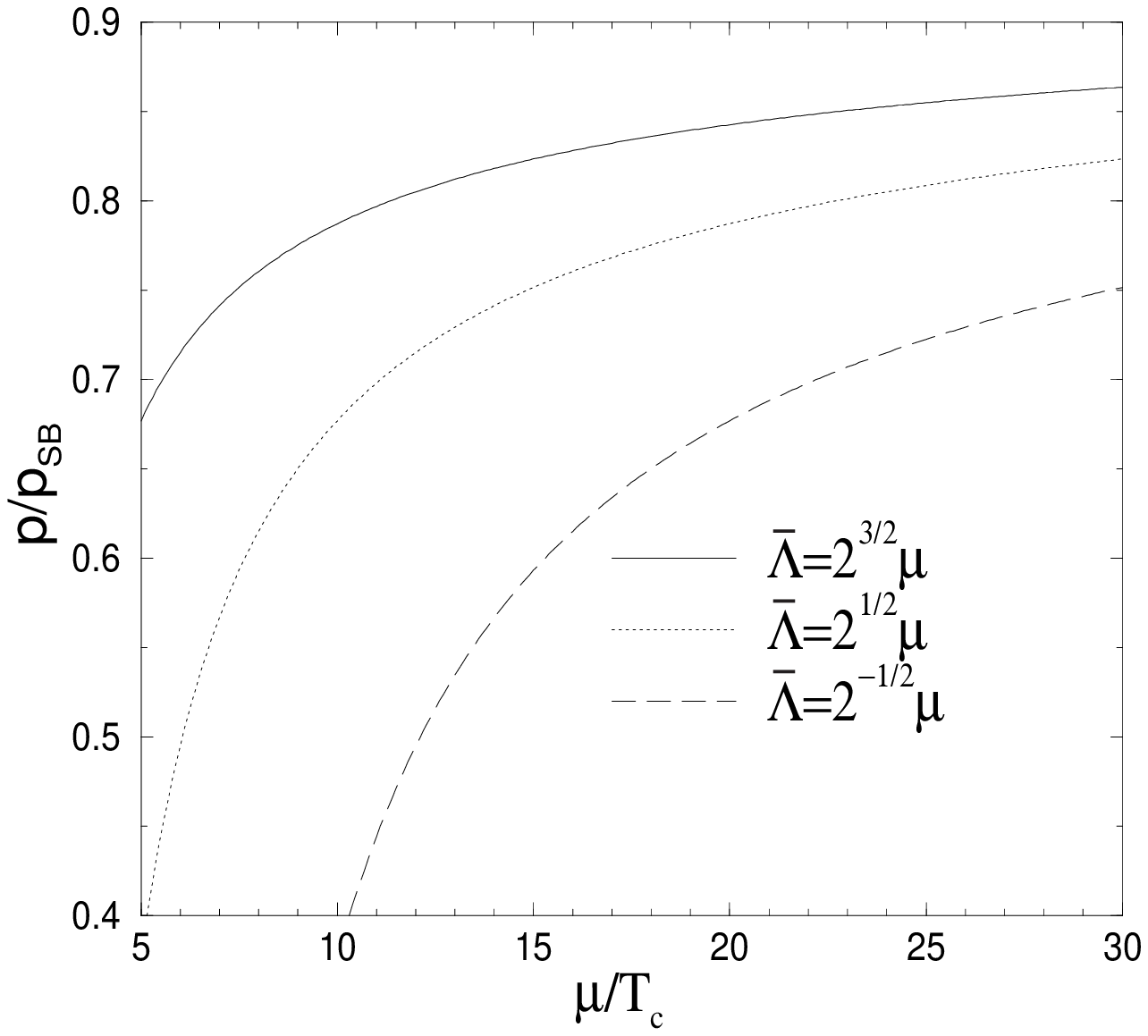}\;\;\;\;\;\;\epsfxsize=7.4cm\epsfysize=6.8cm \epsfbox{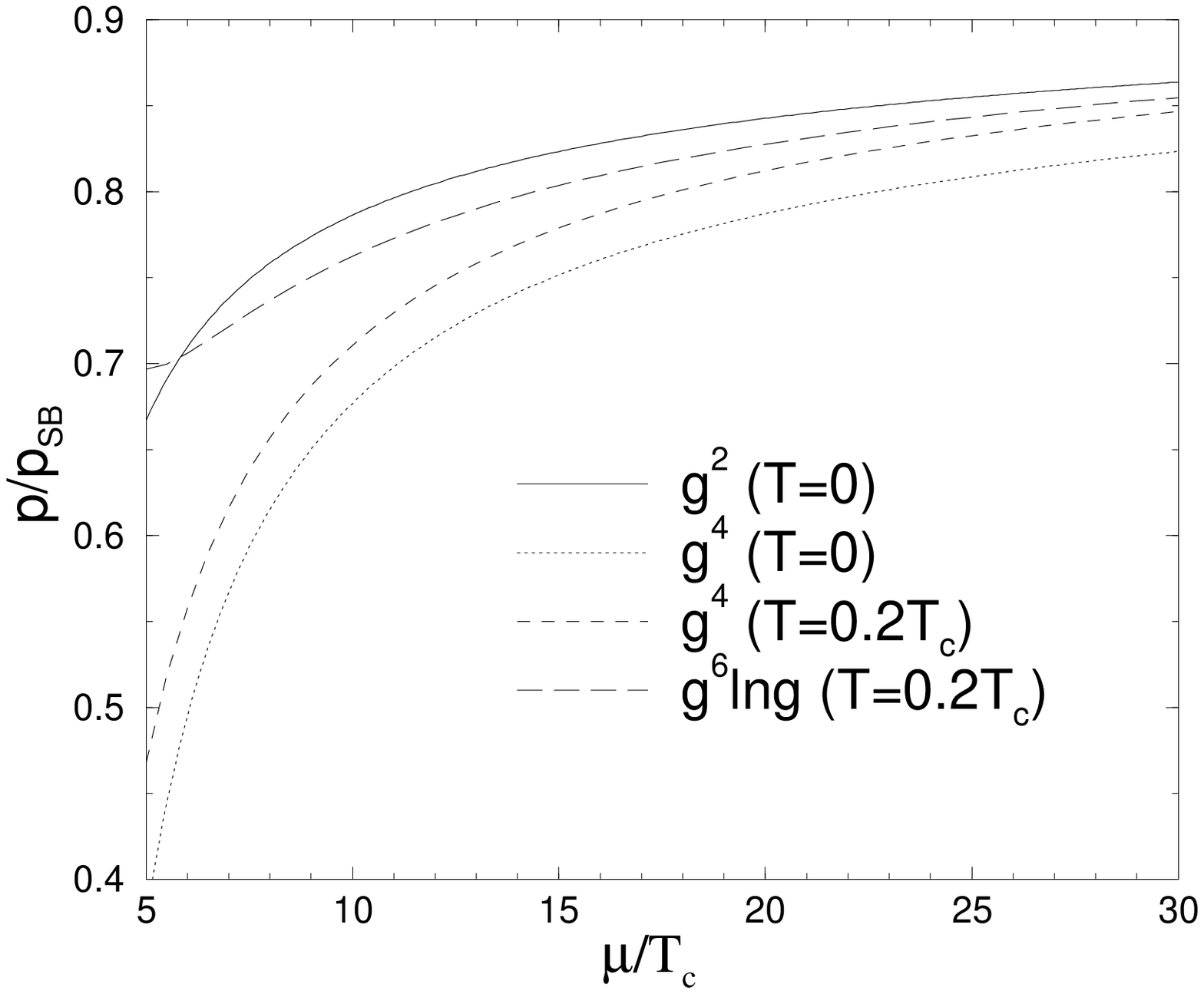}}
\caption[a]{Left: the scale dependence of the zero-temperature result of Eq.~(\ref{pt0res}). Right: the behavior of the order $g^2$ and $g^4$ perturbative results at $T=0$ and the order $g^6\ln\,g$ result at $T=0.2T_c$. The scale parameter is chosen as in Eq.~(\ref{scale1}).}
\end{figure}

To finish the chapter, let us still briefly dwell upon the properties of the zero-temperature expansion of the pressure.
As we only have the first three perturbative orders available, it is
somewhat difficult to quantitatively analyze the convergence properties of the series. A comparison of Fig.~4.11 with the
corresponding $\mu=0$ plots of Figs.~4.1 and 4.2 however indicates that the situation at $T=0$ is considerably more
favorable than at zero chemical potentials --- with the possible exception of scale dependence. The magnitude of the $\mc{O}(g^4)$ term in the expansion
is nevertheless considerable, and the extension of the result to further perturbative orders is therefore certainly
well-motivated. At the same time we must of course remember that in these calculations we are always developing
perturbation theory around a physically incorrect ground state, which includes no Cooper pairs.

\chapter{Conclusions and acknowledgements}
QCD is the quantum field theory that describes one of the four basic interactions of nature, the strong nuclear force. At
extremely high temperatures and densities a strongly interacting system undergoes a phase transition from the hadronic
phase into a new form of matter called the quark-gluon plasma, where the constituents of nucleons and other hadrons are
liberated from their confinement and act almost freely like an ideal gas. In view of the on-going experimental efforts to
detect and study this exotic phase of QCD it is essential that we continuously thrive to learn more about its properties.
Analytic calculations in the theory should therefore be pushed forward, as they provide reliable and easily tractable
results free from the practical and theoretical restrictions of numerical studies, such as lattice simulations.
High-order perturbative computations, which provide a conceptually simple and efficient means to access the properties of
quark-gluon plasma at very high temperatures and chemical potentials, certainly fall into this category.

The work described in the present paper has dealt with the perturbative determination of the most fundamental quantity describing the equilibrium thermodynamics of quark-gluon plasma, the pressure of QCD at finite temperature and quark number density. After reviewing the necessary theoretical and mathematical background we proceeded to analyze the derivation of the most recent perturbative results for the quantity, and then moved on to compare them with up-to-date lattice data. As expected,
our studies showed that the convergence properties of high-temperature perturbation theory improve
considerably as we move away from the line $\mu=0$. After analyzing the behavior of the different perturbative results
for the pressure at small temperatures we then came to a somewhat surprising conclusion: the $\mc{O}(g^6\ln\,g)$
expression derived for the quantity using three-dimensional effective theories in Ref.~\cite{avpres} appears to
be applicable almost throughout the quark-gluon plasma phase. At present this statement is, however, still of highly
speculative nature at finite values of $n_f$, and the existing order $g^4$ result for the zero-temperature pressure~\cite{fmcl} should therefore
definitely be generalized to cover the region of small but finite temperatures as well.

Taking into account the behavior of the pressure's high-temperature expansion at vanishing chemical potentials, it
nevertheless seems to us that the most pressing challenge in the field is the determination of the $\mc{O}(g^6)$
correction to this quantity. At this order perturbative QCD, however, runs into serious infrared problems, for which the
only cure comes from non-perturbative methods, such as three-dimensional lattice simulations. As the results of
four-dimensional lattice studies are on the other hand at present still a long way from covering the region
of large $T$ and $\mu$, our conclusion is that in the near future the most fruitful approach to high-temperature QCD may
well be a combination of perturbative and non-perturbative techniques. First steps in this direction have already been
taken on both fronts \cite{avyork,klry,parma}.

\vspace{0.5cm}

\noindent I want to thank my supervisor Keijo Kajantie as well as Mikko Laine and York Schr\"oder for their invaluable help and guidance in all of my research --- including the writing of the present manuscript. In addition, I wish to express my gratitude towards Christofer Cronstr\"om, Kari~J.~Eskola, Antti Gynther, Andreas Ipp and Anton Rebhan, who have offered valuable and important comments on this paper. The work has been financially supported by the Magnus Ehrnrooth Foundation, the V\"ais\"al\"a Foundation and the Academy of Finland, Contract no. 77744.

\appendix

\chapter{The matching coefficients}
Due to small notational differences between our treatment and that of Ref.~\cite{avpres} regarding the definition of the matching coefficients $\alpha_\rmi{M1}$, $\alpha_\rmi{M2}$ and $\alpha_\rmi{G}$, we include here a list of all coefficients required to obtain the pressure to $\mc{O}(g^6\ln\,g)$. In terms of the $\aleph$ functions introduced at the end of chapter 2 their explicit expressions read
\ba
\aE{1} &=& \fr{\pi^2}{45}\fr{1}{n_f}\!\sum_f\bigg\{d_A+\bigg(\fr{7}{4} + 30\mubar^2 + 60\mubar^4\bigg)d_F\bigg\},
\label{ae1} \\
\aE{2} &=& -\fr{d_A}{144}\fr{1}{n_f}\!\sum_f\bigg\{C_A + \fr{T_F}{2}\(1+12\mubar^2\)\(5+12\mubar^2\)\!\bigg\}, \\
\aE{3} &=& \fr{d_A}{144}\Bigg[\fr{1}{n_f}\!\sum_f\bigg\{C_A^2\bigg(\fr{12}{\e}
+\fr{194}{3}\ln\fr{\bar{\Lambda}}{4\pi T} + \fr{116}{5} + 4\gamma -\fr{38}{3}\fr{\zeta'(-3)}{\zeta(-3)} +
\fr{220}{3}\fr{\zeta'(-1)}{\zeta(-1)}\bigg) \nn
&+& C_A T_F\bigg( \!12\(1+12\mubar^2\)\fr{1}{\e} +
\bigg(\fr{169}{3}+600\mubar^2-528\mubar^4\bigg)\ln\fr{\bar{\Lambda}}{4\pi T} +\fr{1121}{60} + 8\gamma \nn
&+& 2\(127+48\gamma\)\mubar^2 - 644\mubar^4
+ \fr{268}{15}\fr{\zeta '(-3)}{\zeta(-3)} + \fr{4}{3}\(11+156\mubar^2\)\fr{\zeta'(-1)}{\zeta(-1)} \nn
&+& 24\Big[52\,\aleph(3,z)
+ 144\imathb\mubar\,\aleph(2,z)+\(17-92\mubar^2\)\aleph(1,z)+4\imathb\mubar\,\aleph(0,z)\Big]\bigg) \nn
&+& C_F T_F \bigg(\fr{3}{4}\(1+4\mubar^2\)\(35+332\mubar^2\)-24\(1-4\mubar^2\)\fr{\zeta'(-1)}{\zeta(-1)} \nn
&-& 144\Big[12\imathb\mubar\,\aleph(2,z)-2\(1+8\mubar^2\)\aleph(1,z)
-\imathb\mubar\(1+4\mubar^2\)\aleph(0,z)\Big]\bigg) \nn
&+& T_F^2 \bigg(\fr{4}{3}\(1+12\mubar^2\)\(5 + 12\mubar^2\)\ln\fr{\bar{\Lambda}}{4\pi T}
+ \fr{1}{3}+4\gamma + 8\(7+12\gamma\)\mubar^2 + 112\mubar^4 \nn
&-& \fr{64}{15}\fr{\zeta'(-3)}{\zeta(-3)}
- \fr{32}{3}\(1+12\mubar^2\)\fr{\zeta'(-1)}{\zeta(-1)} \nn
&-& 96\Big[8\,\aleph(3,z) + 12\imathb\mubar\,\aleph(2,z) - 2\(1+2\mubar^2\)\aleph(1,z)
- \imathb\mubar\,\aleph(0,z)\Big]\bigg)\bigg\} \nn
&+& 288\,T_F^2\fr{1}{n_f^2}\sum_{f\,g}\bigg\{2\(1+\gamma\)\mubar_f^2\mubar_g^2
-\Big[\aleph(3,z_f+z_g)+\aleph(3,z_f+z_g^*) \nn
&+&4\imathb\mubar_f\Big(\aleph(2,z_f+z_g) + \aleph(2,z_f+z_g^*)\Big)
- 4\mubar_g^2\,\aleph(1,z_f) -\(\mubar_f+\mubar_g\)^2\aleph(1,z_f+z_g) \nonumber
\ea
\ba
&-&\(\mubar_f-\mubar_g\)^2\aleph(1,z_f+z_g^*)
-4\imathb\mubar_f\mubar_g^2\,\aleph(0,z_f) \Big]\bigg\}\Bigg], \label{alphae3}\\
\aE{4} &=& \fr{1}{3}\fr{1}{n_f}\!\sum_f\Big\{C_A+T_F\(1+12\mubar^2\)\Big\}, \\
\aE{5} &=& \fr{1}{3}\fr{1}{n_f}\!\sum_{f}
\bigg\{2\,C_A\bigg(\ln\fr{\bar{\Lambda}}{4\pi T} + \fr{\zeta'(-1)}{\zeta(-1)}\bigg) \nn
&+& T_F \bigg(\(1+12\mubar^2\)\(2\,\ln\fr{\bar{\Lambda}}{4\pi T}+1\) + 24\,\aleph(1,z)\bigg)\bigg\}, \\
\aE{6} &=& \fr{1}{9}\fr{1}{n_f}\!\sum_f\bigg\{C_A^2\bigg(22\,\ln\fr{e^{\gamma}\bar{\Lambda}}{4\pi T}+5\bigg) \nn
&+&C_AT_F\bigg(2\(7+132\mubar^2\)\ln\fr{e^{\gamma}\bar{\Lambda}}{4\pi T}+9+132\mubar^2+8\gamma + 4\,\aleph(z)\bigg) \nn
&-& 18\,C_F T_F\Big(1+12\mubar^2\Big) - 4\,T_F^2\Big(1+12\mubar^2\Big)\bigg(2\,\ln\fr{\bar{\Lambda}}{4\pi T} - 1 - \aleph(z) \bigg)\bigg\}, \\
\aE{7} &=& \fr{1}{3}\fr{1}{n_f}\sum_{f}\bigg\{C_A\bigg(22\,\ln\fr{e^{\gamma}\bar{\Lambda}}{4\pi T}
+1\bigg) - 4\,T_F\(2\,\ln\fr{\bar{\Lambda}}{4\pi T}-\aleph(z)\)\bigg\}, \\
\alpha_\rmi{M1}&=&\fr{43}{32}-\fr{491}{6144}\pi^2, \\
\alpha_\rmi{M2}&=&-\fr{4}{3}, \\
\alpha_\rmi{G}&=&\fr{43}{96}-\fr{157}{6144}\pi^2.
\ea

\chapter{Diagrammatic methods}
Up to this point our approach to perturbative QCD has been somewhat limited, as
the most complicated computational aspect, the evaluation of multi-loop Feynman diagrams, has been left completely
outside the treatment. In this appendix we will introduce to the reader the analytic techniques used in the articles
\cite{avsusc,avpres,avyork} by presenting a few simple, but very detailed examples of our computations. In this process we will try to avoid a repetition of the material of the original papers, and rather concentrate on new and non-trivial
aspects of the computations, such as the analyticity properties of the special functions encountered.

\section{Three-loop diagrams in the full theory}
To illustrate the computational methods used in Ref.~\cite{avpres} in determining the strict perturbation expansion of the
QCD pressure up to three-loop order, let us consider as an example the evaluation of the last diagram of Fig.~3.1 c, the
fermionic `Mercedes'. We begin the calculation by first reducing the diagram into a sum of master sum-integrals and then
proceed to evaluate these functions. The computation will be performed in the Feynman gauge, where the value of the gauge
parameter has been set to $\xi=1$.

\subsection{Reduction to master sum-integrals}
Applying the finite-temperature Feynman rules of section 2 to the Mercedes graph, denoted here by $I_M$, we
straightforwardly obtain
\ba
I_M &=& -\sumint_{\{PQR\}} {\rm Tr} \Big[\fr{1}{\slashed{P}}\(g\gamma _{\mu}T_{ij}^{a}\)
\fr{1}{\slashed{Q}}\(g\gamma _{\nu}T_{jk}^{b}\)\fr{1}{\slashed{R}}\(g\gamma _{\rho}T_{ki}^{c}\)\Big] \imathb g f^{abc} \nn
&&\times \fr{\delta_{\mu\rho}\(2P-Q-R\)_{\nu}+\delta_{\rho\nu}\(2R-P-Q\)_{\mu}+
\delta_{\nu\mu}\(2Q-R-P\)_{\rho}}{\(P-Q\)^2\(Q-R\)^2\(R-P\)^2}, \label{ib1}
\ea
where a sum over flavors has been suppressed and the sum-integration measure is as defined in Eq.~(\ref{asdfg}). The color
sum is easily performed here by exploiting the antisymmetricity property of the structure coefficients $f^{abc}$,
\ba
T_{ij}^{a}T_{jk}^{b}T_{ki}^{c}f^{abc}&=&\fr{1}{2}\tr\big([T^a,T^b]T^c\big)f^{abc}\;\;\,=\;\;\,
\fr{i}{2}f^{abc}f^{abd}\,\tr\big(T^cT^d\big)\nn
&=&\fr{i}2T_Ff^{abc}f^{abc}\;\;\,=\;\;\,\fr{i}{2}C_AT_F\delta^{cc}\nn
&=&\fr{i}{2}d_AC_AT_F,
\ea
and simplifying Eq.~(\ref{ib1}) further we eventually get
\ba
I_M\;\;\,=\;\;\,\fr{3}{2}d_A C_A T_F g^4 \sumint_{\{PQR\}} \fr{P_{\alpha}Q_{\beta}R_{\gamma}\(2P-Q-R\)_{\nu}}
{P^2Q^2R^2\(P-Q\)^2\(Q-R\)^2\(R-P\)^2}{\rm Tr}\big[\gamma_{\alpha}\gamma_{\mu}\gamma_{\beta}
\gamma_{\nu}\gamma_{\gamma}\gamma_{\mu}\big].
\ea
Here we can at once take the trace over the gamma matrices by applying standard formulae (not specific to 4 dimensions)
and get as the result a combination of metric tensors, which in our case are Kronecker $\delta$-symbols. The subsequent
tensorial contractions then become trivial and produce
\ba
I_M&=& 48 \(1-\e\)d_A C_A T_F g^4 \sumint_{\{PQR\}} \fr{P\cdot\(P-Q\) Q \cdot R}{P^2Q^2R^2\(P-Q\)^2\(Q-R\)^2\(R-P\)^2},
\ea
where the explicit appearance of $\e$ signals that we are working in $3-2\e$ space dimensions. Combining the scalar products
of the momenta into squares by using trivial identities such as
\ba
P \cdot Q &=&-\fr{1}{2}\((P-Q)^2-P^2-Q^2\),
\ea
we can then cancel two propagators in each term and finally get
\ba
I_M&=& 12 \(1-\e\)d_A C_A T_F g^4 \Big[\({\cal I}_{1}^0-\widetilde{\cal I}_{1}^0\)\widetilde{\mLARGE{\tau}}+
\fr{1}{2} \widetilde{\cal M}_{0,0}\Big],
\label{diaga}
\ea
where we have in analogy with Refs.~\cite{avpres,bn1} defined
\ba
{\cal I}_{1}^0 &\equiv& \sumint_P \fr{1}{P^2}, \\
\widetilde{\cal I}_{1}^0 &\equiv& \sumint_{\{P\}} \fr{1}{P^2}, \\
\widetilde{\mLARGE{\tau}} &\equiv& \sumint_{\{PQ\}}
\fr{1}{P^2 Q^2 \(P-Q\)^2}, \label{tautde}\\
\widetilde{\cal M}_{0,0} &\equiv& \sumint_{\{PQR\}}
\fr{1}{P^2Q^2\(P-R\)^2\(Q-R\)^2}.
\ea

Next we move on to consider the evaluation of these functions, which form an illustrative subset of the class of
all one-, two- and three-loop master sum-integrals appearing in the covariant gauges of QCD. All sum-integrals required to determine the QCD pressure up to order $g^6\ln\,g$ have been evaluated in Ref.~\cite{az} for $\mu=0$ and in Ref.~\cite{avpres} for $\mu\neq 0$. In the present work we will keep the chemical potentials arbitrary, and start our treatment from the fermionic one-loop case $\widetilde{\cal I}_{1}^0$. Our approach has been significantly motivated by the $\mu=0$ work of Ref. \cite{az}.

\subsection{Evaluation of the sum-integrals}
Written out using the integration measure chosen, the definition of $\widetilde{\cal I}_{1}^0$ reads
\ba
\widetilde{\cal I}_{1}^0 &\equiv& \sumint_{\{P\}} \fr{1}{P^2}\;\;\,=\;\;\, \Lambda^{2\e}T \sum_{n=-\infty}^{\infty}\int
\fr{{\rm d}^{3-2\e} p}{(2\pi)^{3-2\e}}\fr{1}{p^2+((2n+1)T-i\mu)^2}.
\ea
After the trivial evaluation of the momentum integral (see e.g.~the appendix of Ref.~\cite{ramond}) we get
\ba
\widetilde{\cal I}_{1}^0
&=& \fr{\Gamma(-1/2+\e)}{\(4\pi\)^{3/2-\e}}\Lambda^{2\e}T \sum_{n=-\infty}^{\infty}
\Big[\(\(2n+1\)\pi T-i\mu\)^2\Big]^{1/2-\e},
\ea
where the sum bears a close resemblance to the definition of the generalized Riemann $\zeta$-function,
\ba
\zeta(z,q) &=& \sum_{n=0}^{\infty} \fr{1}{\(n+q\)^z}.
\ea
The sum-integral can therefore clearly be written in the form
\ba
\widetilde{\cal I}_{1}^0&=& 2^{-2}\pi^{-1/2}T^{2}\(\fr{\Lambda^2}{\pi T^2}\)^{\e} \Gamma(-1/2+\e) \nn
&&\times \Big(\zeta(-1+2\e,1/2-\imathb \bar{\mu})+\zeta(-1+2\e,1/2+\imathb \bar{\mu})\Big),
\ea
which after an expansion in $\e$ --- performed using the standard formulae of Ref.~\cite{gr} --- gives
\ba
\widetilde{\cal I}_{1}^0 &=& -\fr{T^2}{24}\bigg\{1+12\mubar^2
+2\e\bigg[\(1+12\mubar^2\)\(1+\ln\,\fr{\bar{\Lambda}}{4\pi T}\)
+12\,\aleph(1,z)\bigg]\bigg\}.
\ea
Here we have switched to using the $\msbar$ scale parameter $\bar{\Lambda}$ and now encounter the $\aleph$-functions for
the first time. The bosonic sum-integral ${\cal I}_{1}^0$ is evaluated in an analogous manner.

As we move on to consider the two-fold sum-integral $\widetilde{\mLARGE{\tau}}$ --- first solved in Ref.~\cite{antti} --- the computational methods required get
somewhat more complicated. The initial step in the calculation
is to notice that the definition of the sum-integral, Eq.~(\ref{tautde}), can be written in a simpler form, if we shift the integration
variables by
\ba
P&\rightarrow &P'\;\;\equiv\;\;P-Q,
\ea
and define a `polarization' function $\Pi_{\rm f}$ as
\ba
\Pi_{\rm f} \!\(P\) &\equiv& \sumint_{\{Q\}} \fr{1}{Q^2(P+Q)^2}. \label{piff}
\ea
We then get for the sum-integral
\ba
\widetilde{\mLARGE{\tau}} &\equiv& \sumint_{P\{Q\}}
\fr{1}{P^2 Q^2 \(P+Q\)^2}\;\;=\;\;\sumint_P \fr{1}{P^2}\Pi_{\rm f} \!\(P\),\label{tauti}
\ea
and begin its evaluation by transforming the function $\Pi_{\rm f}$ into coordinate space at $\e=0$.

As can be easily verified, a Fourier-transform in three dimensions gives for the fermionic propagator
\ba
\int\!\fr{{\rm d}^{3}q}{(2\pi)^{3}}\fr{e^{\imathb \mathbf{q}\cdot \mathbf{r}}}{q^2+\(\(2n+1\)\pi T-\imathb \mu\)^2}
&=& \fr{e^{-\(|2n+1|\pi T-\imathb\mu\,{\rm sign}(2n+1)\)r}}{4\pi T}.
\ea
Inserting the inverse transform into the definition of $\Pi_{\rm f}$ we then obtain ($p_0\equiv 2\pi mT$)
\ba
\Pi_{\rm f} \!\(P\) &=&\fr{1}{(4\pi)^2T}\int\!\fr{{\rm d^{3}}q}{(2\pi)^{3}}\sum_{n=-\infty}^{\infty}\int {\rm d}^3 r_1\int {\rm d}^3 r_2\, e^{\imathb (\mathbf{q} \cdot \mathbf{r}_1+(\mathbf{q}+\mathbf{p})\cdot \mathbf{r}_2)}\nn
&\times&{\rm exp}\big[-\(|2n+1|\pi T-i\mu\,{\rm sign}(2n+1)\)r_1\big]\nn
&\times&{\rm exp}\big[-\(|2n+2m+1|\pi T-i\mu\,{\rm sign}(2n+2m+1)\)r_2\big],
\ea
where the momentum integral can be easily performed and produces a Dirac $\delta$-function. This, on the other hand,
helps us to remove one of the $r$-integrals and gives
\ba
\Pi_{\rm f} \!\(P\) &=&\fr{1}{(4\pi)^2T}\sum_{n=-\infty}^{\infty}\int {\rm d}^3 r\, e^{\imathb \mathbf{p}\cdot \mathbf{r}}
{\rm exp}\big[-\(|2n+1|\pi T-i\mu\,{\rm sign}(2n+1)\)r\big]\nn
&\times&{\rm exp}\big[-\(|2n+2m+1|\pi T-i\mu\,{\rm sign}(2n+2m+1)\)r\big],
\ea
in which the infinite sum over $n$ is quickly performed with the help of the standard formula
\ba
\sum_{n=n_0}^{n_1}e^{-\alpha n}&=&\fr{{\rm csch}(\alpha/2)}{2}\(e^{-\alpha(n_0-1/2)}-e^{-\alpha(n_1+1/2)}\). \label{summ1}
\ea
Our result for the polarization function in three dimensions then finally reads
\ba
\Pi_{\rm f} \!\(P\)&=& \fr{T}{\(4\pi\)^2} \int {\rm d}^3 r\,\fr{e^{\imathb \mathbf{p} \cdot \mathbf{r}}}{r^2}
\( | \bar{p}_0 | + \cos(2\bar{\mu} \bar{r}) \,{\rm csch} \,\bar{r} \) e^{-|p_0| r}, \label{pifresu}
\ea
with the dimensionless variables $\bar{p}_0$ and $\bar{r}$ defined by
\ba
\bar{p}_0 &\equiv& \fr{p_0}{2\pi T}, \\
\bar{r} &\equiv& 2\pi T r.
\ea

The insertion of Eq.~(\ref{pifresu}) into the definition of $\widetilde{\mLARGE{\tau}}$ provides us a
straightforward way to compute the sum-integral. Before doing this we however must find a way to regulate
its obvious UV divergence, which otherwise would render the value of the graph infinite. The most straightforward
way to proceed is to analyze the behavior of $\Pi_{\rm f}$ at large external momenta and then subtract the leading order
terms from the sum-integral. The subtracted part namely becomes finite and can be evaluated at $d=3$, while the rest is
easily dealt with while keeping $\e$ nonzero.

The large-$P$ behavior of the polarization function is most easily obtained through the standard procedure of converting
the $q_0$ sum in Eq.~(\ref{piff}) into a contour integral (see e.g.~Refs.~\cite{az,kap}) using the formula
\ba
2\pi iT\sum_{n=-\infty}^{\infty } f(i\omega^\rmi{f}_n)&=&\int_{-i\infty}^{i\infty}{\rm d}p_0 f(p_0)+\oint_C{\rm d}p_0f(p_0) \label{intsum1}\\
&-&\bigg\{\int_{-i\infty+\mu+\e}^{i\infty+\mu+\e}{\rm d}p_0 \,n(p_0-\mu)
+\int_{-i\infty+\mu-\e}^{i\infty+\mu-\e}{\rm d}p_0\,n(\mu-p_0)\bigg\}f(p_0), \nonumber
\ea
where we have denoted
\ba
\omega^\rmi{f}_n&\equiv&(2n+1)\pi T-i\mu, \\
n(x)&\equiv&\fr{1}{e^{\beta x}+1}
\ea
and the integration contour $C$ is defined by
\ba
C&\equiv&\pics{\Line(-5,30)(65,30)\Line(10,60)(10,30)\Line(10,30)(10,0)\Lsc(10,0)(50,0)\Line(50,0)(50,60)\Lsc(50,60)(10,60)
\Text(5,35)[c]{0}\Text(56,35)[c]{$\mu$}\Text(-6,0)[c]{$-i\infty$}\Text(-2,60)[c]{$i\infty$}\Text(65,60)[c]{$p_0$}}
\;\;\;\;\;\;\;\;\;\;\;\; .
\ea
Concentrating first on the divergent vacuum ($T=\mu=0$) part of $\Pi_\rmi{f}$, represented by the first term in
Eq.~(\ref{intsum1}), we straightforwardly obtain
\ba
\Pi^{(0)}\!\(P\)&=&\Lambda^{2\e}\int\fr{{\rm d}^{4-2\e} q}{(2\pi)^{4-2\e}}\fr{1}{q^2(q+P)^2}\nn
&=&\Lambda^{2\e}\int_0^1{\rm d}x \int\fr{{\rm d}^{4-2\e} q}{(2\pi)^{4-2\e}}\fr{1}{(q^2+xP^2+2x\,\mathbf{q}\cdot\mathbf{P})^2} \nn
&=&\fr{\Lambda^{2\e}\Gamma(\e)}{(4\pi)^{2-\e}}\int_0^1{\rm d}x \fr{1}{(x(1-x)P^2)^{\e}}\nn
&=&\fr{1}{\(4\pi\)^{2-\e}}\fr{\Gamma (\e)\Gamma^2(1-\e)}{\Gamma (2-2\e)}\(\fr{\Lambda^2}{P^2}\)^{\!\!\e}
\;\;\,\equiv\;\;\, \beta_0 \(\fr{T^2}{P^2}\)^{\!\!\e},\label{rtyui}
\ea
where $\beta_0$ clearly contains an $1/\e$ divergence. We have suppressed the subscript f, as the
vacuum part of the polarization function is independent of the bosonic or fermionic nature of $Q$.

Let us then move on to consider the matter part of $\Pi_\rmi{f}$ and define
\ba
f(iq_0)&\equiv&\fr{1}{(i(q_0+p_0))^2-\mathbf{(q+p)}^2}
\fr{1}{(iq_0)^2-\mathbf{q}^2}.
\ea
Following the notation of Eq.~(\ref{intsum1}), we now get
\ba
\Pi_{\rm f}^{(T)}\!\(P\)&\equiv&\Pi_{\rm f} \!\(P\)-\Pi^{(0)}\!\(P\) \nn
&=&\bigg\{\Lambda^{2\e} T \int\fr{{\rm d}^{3-2\e} q}{(2\pi)^{3-2\e}}\sum_{\{q_0\}}
f(iq_0)\bigg\}^{(T)} \nn
&=&\fr{\Lambda^{2\e}}{2\pi i}\int\fr{{\rm d}^{3-2\e} q}{(2\pi)^{3-2\e}}\bigg(\oint_C{\rm d}q_0f(q_0)\nn
&-&\bigg\{\int_{-i\infty+\mu+\e}^{i\infty+\mu+\e}{\rm d}q_0 \,n(q_0-\mu)
+\int_{-i\infty+\mu-\e}^{i\infty+\mu-\e}{\rm d}q_0\,n(\mu-q_0)\bigg\}f(q_0)\bigg),
\ea
in which the $q_0$-integrals are trivial to perform using the Residue theorem. After some straightforward
simplifications this produces
\ba
\Pi_{\rm f}^{(T)}\!\(P\)&=&\Lambda^{2\e}\int\fr{{\rm d}^{3-2\e} q}{(2\pi)^{3-2\e}}\fr{1}{2q} \Big(\theta(\mu-q)+n(q+\mu)+{\rm sign}(q-\mu)n(|q-\mu|)\Big)\nn
&\times& \(\fr{1}{(q-ip_0)^2-(\mathbf{q+p})^2}+\fr{1}{(q+ip_0)^2-(\mathbf{q+p})^2}\) \nn
&=&\Lambda^{2\e}\int\fr{{\rm d}^{3-2\e} q}{(2\pi)^{3-2\e}}\fr{1}{2q} \Big(n(q+\mu)+n(q-\mu)\Big)\nn
&\times& \(\fr{1}{(q-ip_0)^2-(\mathbf{q+p})^2}+\fr{1}{(q+ip_0)^2-(\mathbf{q+p})^2}\), \label{pifint}
\ea
where we have used the trivial identities
\ba
n(q-ip_0)&=&n(q), \\
n(q-\mu)&=&\theta(\mu-q)+{\rm sign}(q-\mu)n(|q-\mu|).
\ea

The leading large-$P$ behavior of $\Pi^{(T)}_\rmi{f}$ is now available through a simple power series expansion of the integrand
of Eq.~(\ref{pifint}), as the functions $n$ ensure that only the region of small $q$ contributes to the integral. This
on the other hand helps us to immediately take care of the angular integrals and leads to the result
\ba
\Pi_{\rm f}^{(T)}\!\(P\)&=&-\fr{\Lambda^{2\e}}{(4\pi)^{3/2-\e}\Gamma(3/2-\e)}\int_0^{\infty}{\rm d}q q^{1-2\e}\Big(n(q+\mu)+n(q-\mu)\Big) \nn &\times&\(\fr{2}{P^2}+\fr{8}{3-2\e}\bigg[\fr{1}{P^4}-(4-2\e)\fr{p_0^2}{P^6}\bigg]q^2+\mc{O}\Big(\fr{1}{P^6}\Big)\),
\ea
where the remaining $q$-integrals are expressible in terms of the polylogarithm function
\ba
{\rm Li}_{\nu}(z)&=&\sum_{k=1}^{\infty}\fr{z^k}{k^{\nu}}\;\;\,=\;\;\,\fr{z}{\Gamma(\nu)}\int_0^{\infty}{\rm d}t\fr{t^{\nu-1}}{e^{t}-z}.
\ea
Our final result for the leading large-$P$ behavior of the matter part of the polarization function then reads
\ba
\label{piflp}
\Pi^{(T)}_{\rm f}\!\(P\) &=& \beta_1\fr{1}{P^2}+\beta_2 T^4 \(\fr{1}{P^4}-\(4-2\e\)
\fr{p_0^2}{P^6}\)+\mc{O}\Big(\fr{1}{P^6}\Big)\nn
&\equiv&\Pi^{(T)}_{\rm f,UV}\!\(P\)+\mc{O}\Big(\fr{1}{P^6}\Big),
\ea
with the coefficients $\beta$ defined by
\ba
\beta_1&=& 2^{-2+2\e}\pi^{-3/2+\e}\fr{\Gamma (2-2\e )}{\Gamma (3/2-\e )}\(\fr{\Lambda}{T}\)^{\!2\e}
\( {\rm Li}_{2-2\e}(\!-e^{\mu /T})+{\rm Li}_{2-2\e}(\!-e^{-\mu /T})\),\\
\beta_2 &=& 2^{2\e}\pi^{-3/2+\e}\fr{\Gamma (3-2\e )}{\Gamma (3/2-\e )}\(\fr{\Lambda}{T}\)^{\!2\e}
\( {\rm Li}_{4-2\e}(\!-e^{\mu /T})+{\rm Li}_{4-2\e}(\!-e^{-\mu /T})\).
\ea
As can be easily verified starting from the definition of $\Pi_\rmi{f}$, we could have immediately written down the first coefficient
in the form $\beta_1=2\widetilde{\cal I}_{1}^0$, which actually is enough for our present purposes. We, however, wanted
to present here the general scheme used in analyzing the large-$P$ behavior of different polarization functions and
therefore extended the treatment to the next order. The higher order terms are frequently needed at three-loop level.

Returning back to the computation of $\widetilde{\mLARGE{\tau}}$, we now see that the explicit subtraction of the leading
order terms of $\Pi_\rmi{f}$ from Eq.~(\ref{tauti}) is enough to render the sum-integral finite. We, however, still need
a coordinate space representation for the regularized polarization function at $\e=0$, which is straightforwardly
obtained and reads
\ba
&&\Pi_\rmi{f}(P)-\Pi^{(0)}\!\(P\)-\Pi^{(T)}_{\rm f,UV}\!\(P\)\;\;\,= \\
&& \fr{T}{\(4\pi\)^2} \!\int\! {\rm d}^3 r\,\fr{e^{\imathb \mathbf{p} \cdot \mathbf{r}}}{r^2}\!
\bigg[\cos(2\bar{\mu} \bar{r}) {\rm csch} \,\bar{r} -\fr{1}{\bar{r}}+\(\fr{1}{6}+2\bar{\mu}^2\)\bar{r}-
\(\fr{7}{360}+\fr{\bar{\mu}^2}{3}+\fr{2\bar{\mu}^4}{3}\)\bar{r}^3\bigg]e^{-|p_0| r}. \nonumber
\ea
In deriving this result, we have employed e.g.~the identities
\ba
{\rm Li}_{2}(-e^{x}) + {\rm Li}_{2}(-e^{-x}) = -\fr{\pi^2}{6}-\fr{1}{2}x^2,\\
{\rm Li}_{4}(-e^{x}) + {\rm Li}_{4}(-e^{-x}) = -\fr{7\pi^4}{360}-\fr{\pi^2}{12}x^2-\fr{1}{24}x^4,
\ea
which are special cases of a more general relation between the polylogarithm function and the Bernoulli polynomials (see
Ref.~\cite{wolfram}). As a curiosity, it should be noted here that despite the pole in the coefficient $\beta_0$, even the Fourier transform of $\Pi^{(0)}$ is finite at $d=3$ due to the appearance of an additional factor of $1/\Gamma(\e)$ in the transformation formula of $1/(P^2)^{\e}$.

We are now finally ready to evaluate the initial sum-integral. Plugging the above results into Eq.~(\ref{tauti}), we get
\ba
\widetilde{\mLARGE{\tau}} &=& \sumint_P \fr{1}{P^2}
\bigg( \beta_0 \(\fr{T^2}{P^2}\)^{\!\!\e}+2\widetilde{\cal I}_{1}^0\fr{1}{P^2} \nn
&+& \fr{T}{\(4\pi\)^2} \!\int\! {\rm d}^3 r\,\fr{e^{\imathb \mathbf{p} \cdot \mathbf{r}}}{r^2}
\bigg[\cos(2\bar{\mu} \bar{r}) {\rm csch} \,\bar{r} -\fr{1}{\bar{r}}
+ \(\fr{1}{6}+2\bar{\mu}^2\)\bar{r}\bigg]e^{-|p_0| r}\bigg) + \mathcal{O}\(\e\),
\label{tau1}
\ea
where the two first terms are easily computable by methods analogous to those used when considering
$\widetilde{\cal I}_{1}^0$. For the finite third term the evaluation of the momentum space integral and the
subsequent $p_0$ sum on the other hand produce
\ba
&& \fr{T}{\(4\pi\)^2} \sumint_P \fr{1}{P^2}\!\int\! {\rm d}^3 r\,\fr{e^{\imathb \mathbf{p} \cdot \mathbf{r}}}{r^2}
\bigg[\cos(2\bar{\mu} \bar{r}) {\rm csch} \,\bar{r} -\fr{1}{\bar{r}}
+ \(\fr{1}{6}+2\bar{\mu}^2\)\bar{r}\bigg]e^{-|p_0| r} \nn
&=& \fr{T^2}{\(4\pi\)^3}\!\int\! {\rm d}^3 r \sum_{p_0} \fr{e^{-2|p_0|r}}{r^3}
\bigg[\cos(2\bar{\mu} \bar{r}) {\rm csch} \,\bar{r} -\fr{1}{\bar{r}}
+ \(\fr{1}{6}+2\bar{\mu}^2\)\bar{r}\bigg] \nn
&=& \fr{T^2}{\(4\pi\)^2}\int_0^{\infty} {\rm d} r \,\fr{{\rm coth }\,r}{r}
\bigg[\cos(2\bar{\mu} r) {\rm csch} \,r -\fr{1}{r} + \(\fr{1}{6}+2\bar{\mu}^2\)r\bigg],
\label{tau2}
\ea
where the final coordinate space integral is UV convergent but has an IR divergence due to the zero mode in the bosonic
frequency sum. We could easily remove this by separating the zeroth Matsubara mode from the sum, but
need not actually do that; the singular part namely originates from a term proportional to
$\int_0^{\infty}{\rm d}r \, r^{z}$, which automatically
vanishes under dimensional regularization. We can therefore instantly proceed to evaluate the integral, which is most
easily accomplished by regulating each term of the integrand separately with a factor $r^{\alpha}$. In the end we will
sum the different pieces together and proceed to the limit $\alpha\rightarrow 0+$.

The last two terms of Eq.~(\ref{tau2}) are easily dealt with by applying the relation
\ba
\int_0^{\infty}{\rm d}x\,x^z {\rm coth}\,x&=&2^{-z}\Gamma(1+z)\zeta(1+z),
\ea
while with the first one we first need to perform a partial integration and then use the straightforwardly derivable identity
\ba
\int_0^{\infty}{\rm d}x \, x^z e^{\imathb \beta x}{\rm csch}\,x &=& 2^{-z}\Gamma(1+z)\zeta(1+z,1/2-\imathb\beta /2).
\ea
When adding the three parts of Eq.~(\ref{tau2}) together, all terms singular in $\alpha$ cancel and we are left with an
expression, where we may easily proceed to the limit $\alpha=0$. The inclusion of the first two terms of Eq.~(\ref{tau1})
to the sum then finally leads us to the result
\ba
\widetilde{\mLARGE{\tau}} &=& -\fr{T^2\mubar^2}{\(4\pi\)^2}\(\fr{1}{\e} +
2\(1+2\,\ln\,\fr{\bar{\Lambda}}{4\pi T}\) -
\fr{2\imathb}{\mubar} \aleph(0,z)\)
\ea
for the entire sum-integral.

For three-fold sum-integrals the calculations become still somewhat more evolved \cite{avpres,az}, but the general features of the above coordinate space method still apply; in particular, we always end up evaluating one-dimensional hyperbolic integrals, which lead to results containing different combinations of the $\aleph$-functions. We will
therefore not present a detailed account of the computation of the sum-integral $\widetilde{\cal M}_{0,0}$ here, but merely refer the
reader to Ref.~\cite{avpres}. In the following we instead concentrate on analyzing the properties of the special functions
encountered.

\subsection{Properties of the $\aleph$-functions}
As can be seen from the explicit form of the matching coefficients listed in the appendix A, the only special functions that
appear in the perturbative expansion of the QCD pressure up to order $g^6\ln\,g$ are the $\aleph$-functions. These
were defined in chapter 2 by
\ba
\aleph(n,w) &\equiv& \zeta'(-n,w)+\(-1\)^{n+1}\zeta'(-n,w^{*}), \\
\aleph(w) &\equiv& \Psi(w)+\Psi(w^*),
\ea
where $n\in\{0,1,2,3\}$ and the arguments of the functions read either
\ba
w&=&z\;\;\, =\;\;\, 1/2-i\mubar\;\;\;\;{\rm or} \\
w&=&z_f+z_g\;\;\,=\,\;\;1-i(\mubar_f+\mubar_g).
\ea
We will now examine the behavior of these functions and in particular show how one can derive generalized power series expansions for them in the limits of small and large $\mubar$. These results have played a crucial role in the derivation of analytic expressions for the quark number susceptibilities in Ref.~\cite{avsusc} and for the $T=0$ pressure in Ref.~\cite{avpres}.

Our first task is to find convenient (integral) representations for the $\zeta$- and $\Psi$-functions. The generalized Riemann $\zeta$-function is usually defined by the infinite sum
\ba
\zeta(x,w) &=& \sum_{n=0}^{\infty} \fr{1}{\(n+w\)^x}
\ea
or the integral \cite{gr}
\ba
\zeta(x,w)&=&\fr{1}{\Gamma(x)}\int_0^{\infty}{\rm d}t\,t^{x-1}\fr{e^{-wt}}{1-e^{-t}}. \label{defint1}
\ea
Using the latter it is especially straightforward to analytically continue\footnote{Recalling the analytic continuation of the $\Gamma$ function
onto $\mathbb{C}\backslash(-\mathbb{N})$ using its integral representation.} the function to the whole $x$-plane
excluding the point $x=1$, if we assume the parameter $w$ to satisfy the condition Re($w$)$>0$. For the digamma function
$\Psi$ we will use a similar representation \cite{gr}
\ba
\Psi(w)&=&-\gamma+\int_0^{\infty}{\rm d}t\,\fr{e^{-t}-e^{-wt}}{1-e^{-t}}, \label{defint2}
\ea
where Re($w$)$>0$ is again implicitly assumed.

\begin{figure}[t]

\centerline{\epsfxsize=13cm \epsfbox{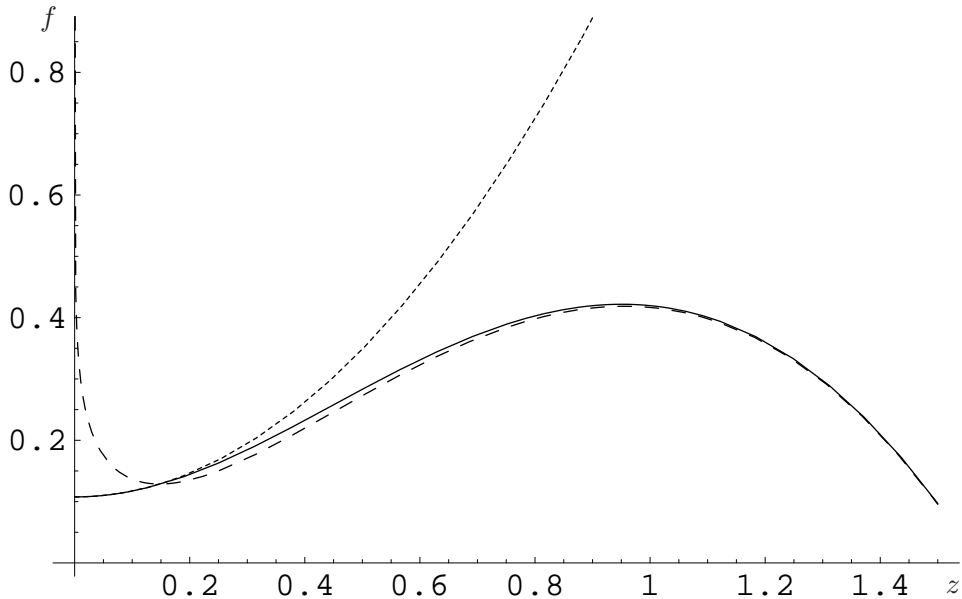}}

\caption[a]{The behavior of $\aleph(1,z)$ (solid line) and its asymptotic limits, Eqs.~(\ref{smallmu1}) (dotted line) and
(\ref{largemu1}) (dashed line), as functions of $\mubar$.}
\end{figure}

\subsubsection{Small $\mubar$}
Let us begin our analysis from the limit of small $\mubar$, where we must expand the integrals of Eqs.~(\ref{defint1}) and (\ref{defint2}) in powers of Im($w$). For the $\zeta$ function this is easily achieved via an
expansion of the denominator of the integrand in powers of $e^{-t}$, which assumes that the factor $t^{x-1}$
regulates the divergence at small $t$. If we denote ${\rm Re}(w)\equiv w_0,\;{\rm Im}(w)\equiv w_1$, we get
\ba
\zeta(x,w)&=&\fr{1}{\Gamma(x)}\sum_{n=0}^{\infty}\int_0^{\infty}{\rm d}t\,t^{x-1}e^{-(w+n)t}\nn
&=&\fr{1}{\Gamma(x)}\sum_{n=0}^{\infty}\sum_{k=0}^{\infty}\fr{(-iw_1)^k}{k!}\int_0^{\infty}{\rm d}t\,t^{k+x-1}e^{-(w_0+n)t}
\ea
\ba
&=&\fr{1}{\Gamma(x)}\sum_{n=0}^{\infty}\sum_{k=0}^{\infty}\fr{(-iw_1)^k}{(n+w_0)^{k+x}k!}\Gamma(k+x) \nn
&=&\sum_{k=0}^{\infty}{k+x-1\choose k}\zeta(k+x,w_0)(-iw_1)^k,
\ea
where the binomial coefficient in the final expression reveals that the result could as well have been derived starting
from the sum representation of $\zeta$. A derivative of the above expression evaluated at negative integer values of $x$
then gives us the desired results for $\aleph(n,w)$. Just to demonstrate this, for $n=1$ the first few coefficients of the expansion read
\ba
\aleph(1,z) &=& -\fr{1}{12}\Big(\ln\,2-\fr{\zeta'(-1)}{\zeta(-1)}\Big) - \(1-2\,\ln\,2-\gamma\)\mubar^2
+\mathcal{O}(\mubar^4). \label{smallmu1}
\ea

For the functions $\aleph(w)$ and $\Psi$ the calculation becomes somewhat more complicated. Introducing an IR regulator $t^{\alpha}$ to Eq.~(\ref{defint2}) and expanding the denominator as before, the integrals corresponding to the two terms of the integrand can, however, be analytically computed and produce a finite result when added together. We are then able to expand also the function $\aleph(z)$ to an arbitrary order in $\mubar$, and obtain
\ba
\aleph(z) &=& -2\gamma -4\,\ln\,2 + 14\,\zeta(3)\,\mubar^2+\mathcal{O}(\mubar^4).\label{smallmu2}
\ea

\begin{figure}[t]

\centerline{\epsfxsize=13cm \epsfbox{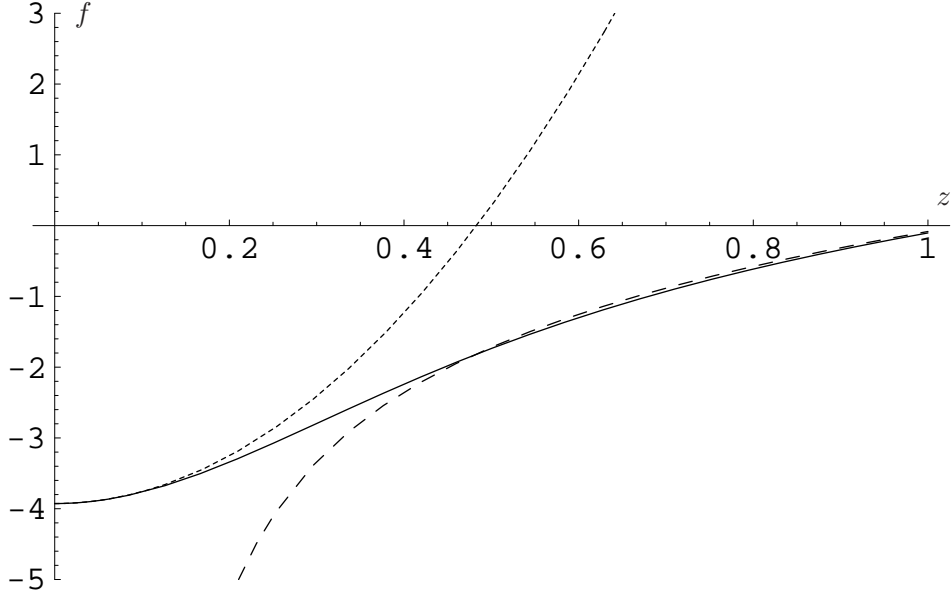}}

\caption[a]{The behavior of $\aleph(z)$ (solid line) and its asymptotic limits, Eqs.~(\ref{smallmu2}) (dotted line) and
(\ref{largemu2}) (dashed line), as functions of $\mubar$.}
\end{figure}

\subsubsection{Large $\mubar$}
Moving on to the limit of large $\mubar$, we want to expand the $\aleph$-functions in powers of $1/\mubar$, or
equivalently the integral representations of Eqs.~(\ref{defint1}) and (\ref{defint2}) in powers of $1/{\rm Im}(w)$. This
is most easily accomplished by noticing that a large value of $\mubar$ causes the dominant contribution to the
integrals to originate from the region of small $t$, which means we can expand the integrands in powers of $t$
with the exception of the factor $e^{-iw t}$. We thereby get
\ba
\zeta(x,w)&=&\fr{1}{\Gamma(x)}\int_0^{\infty}{\rm d}t\,t^{x-1}e^{-iwt}\(t^{-1}+1/2+t/12+\mc{O}(t^2)\) \nn
&=&\fr{(iw)^{-x+1}}{x-1}\(1+\fr{x-1}{2iw}+\fr{x(x-1)}{12(iw)^2}+\mc{O}(1/w^3)\),
\ea
which after some trivial algebra leads to
\ba
\aleph(1,z) &=& -\mubar^2\Big(\ln\,\mubar-\fr{1}{2}\Big)-\fr{1}{12}\Big(\ln\,\mubar+1\Big) +
\mathcal{O}\(\fr{\ln\,\mubar}{\mubar^2}\) \label{largemu1}
\ea
for $\aleph(1,z)$, and similarly for the other $\aleph(n,z)$-functions.

Analogously to the limit of small $\mubar$, the determination of the behavior of the functions $\aleph(w)$ and $\Psi$ at
large $\mubar$ requires the introduction of an additional factor $t^{\alpha}$ to the integrand of Eq.~(\ref{defint2}) to
regulate the divergent behavior of the two terms at $t=0$. In other respects the computation however proceeds in a manner
highly analogous to that presented above and in the end gives us
\ba
\aleph(z) &=& 2\,\ln\,\mubar -\fr{1}{12}\fr{1}{\mubar^2}+\mathcal{O}(\fr{\ln\,\mubar}{\mubar^4}).\label{largemu2}
\ea
The general behavior of the functions $\aleph(1,z)$ and $\aleph(z)$ is depicted in Figs.~B.1. and B.2.

\section{Four-loop diagrams in the effective theories}
In Ref.~\cite{avyork} it is described how all four-loop vacuum diagrams
appearing in three-dimensional massive field theories can be numerically computed to a high order in $\e=(3-d)/2$.
For these purposes we have in addition derived previously unknown analytic results for a small class of four-loop master diagrams, whose
structure is simple enough to allow for their analytic evaluation. As this analytic work is only briefly quoted in Ref.~\cite{avyork}, we will here explore the subject in more detail and introduce the methods used in the computations. An even more detailed treatment of some of this material can be found in Ref.~\cite{avgradu}.

The diagrams, whose analytic evaluation turns out to be possible, fall into two general categories: graphs with a small number of vertices
can be evaluated by coordinate space methods (see e.g.~Refs.~\cite{gkp,arttu}), while the ones containing only simple one-loop
sub-diagrams are most easily dealt with directly in momentum space. Our treatment is therefore too divided into two parts, of
which the first deals with coordinate space computations and the latter with momentum space ones. The diagrams to be covered
are shown in Fig.~B.3, where also the numbering of the propagator masses can be found. We will keep the
values of these masses arbitrary throughout our treatment, which enables us to afterwards easily obtain results for
diagrams with some propagators raised to higher powers through differentiation. The momentum integration measure we will
be using in our calculations is the $\msbar$ one defined at the end of chapter 2.

\def\diaaa(#1,#2,#3,#4,#5,#6){\pic{#1(15,15)(15,90,210)%
 #2(15,15)(15,210,330) #3(15,15)(15,-30,90) #4(2,7.5)(15,30)%
 #6(28,7.5)(2,7.5) #5(15,30)(28,7.5)
 \Text(1,26)[c]{$_1$}\Text(11,17)[c]{$_2$}\Text(15,-3)[c]{$_3$}\Text(15,10.5)[c]{$_4$}
 \Text(29,26)[c]{$_5$}\Text(19,17)[c]{$_6$}}}

\def\diabb(#1,#2,#3,#4,#5,#6,#7){\pic{#3(15,15)(15,0,90)%
 #2(15,15)(15,90,180) #4(15,15)(15,180,270) #1(15,15)(15,270,360)%
 #6(0,15)(15,30) #7(15,0)(0,15) #5(30,15)(15,0)
 \Text(2,28)[c]{$_1$}\Text(10,20)[c]{$_2$}\Text(2,2)[c]{$_3$}\Text(10,10)[c]{$_4$}
  \Text(28,2)[c]{$_5$}\Text(20,10)[c]{$_6$}\Text(28,28)[c]{$_7$}}}

\def\diacc(#1,#2,#3,#4,#5,#6,#7){\picb{#1(15,15)(15,90,270)%
 #2(30,15)(15,-90,90) #4(30,30)(15,30) #3(15,0)(30,0)%
 #5(15,0)(15,30) #6(30,30)(30,0) #7(15,30)(30,0)
 \Text(-2,15)[c]{$_1$}\Text(13,15)[c]{$_2$}
 \Text(48,15)[c]{$_3$}\Text(33,15)[c]{$_4$}
 \Text(23,-3)[c]{$_5$}\Text(23,33.5)[c]{$_6$}\Text(20.5,12)[c]{$_7$}}}

\def\diadd(#1,#2,#3,#4,#5){\picb{#1(30,15)(15,-120,120)%
 #2(30,15)(15,120,240) #3(15,15)(15,60,300) #4(15,15)(15,-60,60)%
 #5(22.5,3)(22.5,27)
 \Text(-2.5,15)[c]{$_1$}\Text(12.5,15)[c]{$_2$}
 \Text(20.5,15)[c]{$_3$}\Text(33,15)[c]{$_4$} \Text(48,15)[c]{$_5$}}}

\def\diaee(#1,#2,#3,#4,#5,#6){\!\!\picb{#2(26.25,15)(15.5,256,76)%
 #3(30,30)(15,30) #1(18.75,15)(15.5,104,284) #4(15,30)(22.5,0)%
 #5(30,30)(22.5,0) #6(15,17.8)(19.3,292.8,39.1)%
 \Text(23,33.5)[c]{$_1$} \Text(1,15)[c]{$_2$}
 \Text(16.5,15)[c]{$_3$} \Text(29,15)[c]{$_4$}
 \Text(37,15)[c]{$_5$} \Text(44.5,15)[c]{$_6$}}}

\begin{figure}[t]
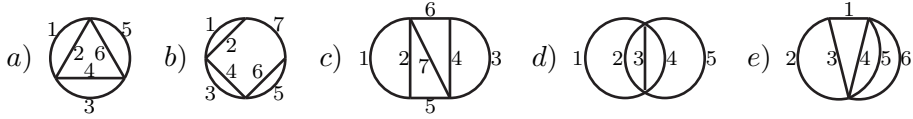

\centering
\ba \nonumber
\begin{array}{rrrrr}
a)~
\diaaa(\CArc,\CArc,\CArc,\Line,\Line,\Line)
& b)~
\diabb(\CArc,\CArc,\CArc,\CArc,\Line,\Line,\Line)
& c)~
\!\!\!\diacc(\CArc,\CArc,\Line,\Line,\Line,\Line,\Line)\;\;\;
& d)~
\!\!\!\diadd(\CArc,\CArc,\CArc,\CArc,\Line)\;\;\;
& e)~
\!\!\!\diaee(\CArc,\CArc,\Line,\Line,\Line,\CArc)
\nn
\end{array}
\ea
\caption[a]{Diagrams evaluated in the present paper.}
\end{figure}

\subsection{Coordinate space calculations}
When the number of vertices in a diagram is small, it often proves useful to convert the momentum space integrals into coordinate
space ones by taking Fourier transforms of the propagators. The success of this method is based on the obvious fact
that the number of coordinate space integrations needed in the end will always be one less than the number of vertices in the graph. The
coordinate representation of the scalar propagator is, however, fairly complicated in an arbitrary dimension, and angular
integrals furthermore often prove to be surprisingly difficult to perform. This makes the application of the
method increasingly difficult when the number of vertices arises.

\subsubsection{The basic setup}
The most fundamental quantity required in coordinate space calculations is the Fourier transform of a massive
propagator, for which a straightforward computation produces
\ba
{\rm D}\(x,m\)&\equiv&\bar{\Lambda}^{2\e}\int\fr{{\rm d}^{3-2\e}p}{\(2\pi\)^{3-2\e}} ~\fr{e^{-\imath p\cdot x}}{p^2+m^2}\;\;=\;\;
\fr{\bar{\Lambda}^{2\e}}{\(2\pi\)^{3/2-\e}}\(\fr{m}{x}\)^{1/2-\e}{\rm K_{1/2-\e}}\(mx\),
\ea
with K denoting a modified Bessel function. An expansion of this result in powers of $x$ then gives
\ba
{\rm D}\(x,m\)&=& \fr{\bar{\Lambda}^{2\e}\Gamma\(1/2-\e\)}{4\pi^{1-\e}\Gamma\(1/2\)}x^{-1+2\e}\bigg(1+\fr{\(m x\)^2}{2\(1+2\e\)}+
{\cal O}\(\(mx\)^4\)\bigg)\nn
&-&\fr{\bar{\Lambda}^{2\e}\Gamma\(-1/2+\e\)}{\(4\pi\)^{1-\e}\Gamma\(-1/2\)}m^{1-2\e}\(1+{\cal O}\(\(mx\)^2\)\)
\ea
\ba
&\equiv& m^{1-2\e}\bigg[\alpha\(mx\)^{-1+2\e}\bigg(1+\fr{\(m x\)^2}{2\(1+2\e\)}+{\cal O}\(\(mx\)^4\)\bigg)\nn
&-&\beta\(1+{\cal O}\(\(mx\)^2\)\)\bigg],
\label{corp}
\ea
where the coefficients $\alpha$ and $\beta$ read
\ba
\alpha &=& \fr{\bar{\Lambda}^{2\e}\Gamma\(1/2-\e\)}{4\pi^{1-\e}\Gamma\(1/2\)},\\
\beta &=& \fr{\bar{\Lambda}^{2\e}\Gamma\(-1/2+\e\)}{\(4\pi\)^{1-\e}\Gamma\(-1/2\)}.
\ea
Correspondingly, an $\e$-expansion of D produces
\ba
{\rm D}\(x,m\)&=&\fr{e^{-m x}}{4\pi x}\bigg[1+\e\bigg(\ln\fr{2\pi \bar{\Lambda}^2 x}{m}+e^{2 m x}{\rm Ei}\(-2m x\)\bigg)
+{\cal O}\(\e^2\)\bigg],
\label{propexp}
\ea
where ${\rm Ei}$ denotes the exponential integral function
\ba
{\rm Ei}(x)&=&-\int_{-x}^{\infty}{\rm d}t\,\fr{e^{-t}}{t}
\ea
and the relations \cite{gr}
\ba
{\rm K}_{1/2}\(x\) &=& \sqrt{\fr{\pi}{2x}}e^{-x} \\
\lk\fr{\partial {\rm K}_\nu\(x\)}{\partial \nu}\rk_{\nu = 1/2} &=& -\(\fr{\pi}{2x}\)^{1/2}e^{x}~{\rm Ei}\(-2x\)
\ea
have been used. These results will be frequently used in the following.

\subsubsection{Diagram {\textit d}}
The basketball-type diagrams containing only two vertices are particularly well suited for integration in coordinate space
due to their trivial angular structure. Writing the propagators in terms of their Fourier transforms we easily obtain for
the graph $d$,
\ba
I_d &=& \bar{\Lambda}^{-10\e}\(\fr{ \bar{\Lambda}^{2\e}}{\(4\pi e^{-\gamma}\)^\e}\)^4 \int{\rm
d^{3-2\e}}x \prod_{i=1}^{5}{\rm D}\(x,m_i\) \nn
&=& \(\fr{ \bar{\Lambda}^{-\e /2}}{\(4\pi e^{-\gamma}\)^\e}\)^4 \fr{2
\pi^{3/2-\e}}{\Gamma\(3/2-\e\)} \int_0^\infty{\rm d}x ~x^{2-2\e} \prod_{i=1}^{5}{\rm D}\(x,m_i\) \nn
&\equiv&\(\fr{\bar{\Lambda}^{-1}}{\(4\pi e^{-\gamma}\)^2}\)^{2\e} \fr{2 \pi^{3/2-\e}}{\Gamma\(3/2-\e\)}{\cal I},
\ea
where the integral ${\cal I}$ clearly contains a singularity at the origin signalling of the ultraviolet divergence of the
diagram. In order to separate this divergent part we introduce a small positive parameter $r$ to divide the radial
integration path into two subintervals $[0,r]$ and $[r,\infty [$ while keeping in mind that in the end we will sum the
two contributions together and proceed to the limit $r\rightarrow 0$. On the first interval one may use the small-$x$
expansion of the propagators, whereas on the latter we can simply set $\e=0$.

Starting from the first interval we get by applying Eq.~(\ref{corp}),
\ba
{\cal I}_1 &\equiv&\int_0^r{\rm d}x ~x^{2-2\e} \prod_{i=1}^{5}{\rm D}\(x,m_i\)\nn
&=& \int_0^r{\rm d}x ~x^{2-2\e}
\prod_{i=1}^{5} \bigg\{ m_i^{1-2\e}\bigg[\alpha\(m_i x\)^{-1+2\e}\bigg(1+\fr{\(m_i x\)^2}{2\(1+2\e\)}
+{\cal O}\(\(m_ix\)^4\)\bigg)\nn
&-&\beta\(1+{\cal O}\(\(m_i x\)^2\)\)\bigg] \bigg\} + \mathcal{O}(r) \nn
&=& \int_0^r{\rm d}x ~x^{2-2\e}\bigg\{\alpha^5 ~x^{-5+10\e}-\alpha^4 \beta \sum_{i=1}^5 m_i^{1-2\e}
~x^{-4+8\e} +\fr{\alpha^5}{2\(1+2\e\)} \sum_{i=1}^5 m_i^2 ~x^{-3+10\e} \nn
&+&\fr{\alpha^3\beta^2}{2}\sum_{i\neq j}\(m_i m_j\)^{1-2\e}~x^{-3+6\e} \bigg\}+\mathcal{O}(r),
\ea
where the final integrations are trivial to perform and lead to
\ba
{\cal I}_1 &=& -\fr{\alpha^5}{2}r^{-2}+\alpha^4 \beta \sum_{i=1}^5 m_i ~r^{-1}+
\fr{\alpha^5 r^{8\e}}{16\e\(1+2\e\)}\sum_{i=1}^5 m_i ^2\nn
&+&\fr{\alpha^3\beta^2 r^{4\e}}{8\e}\sum_{i\neq j}\(m_i m_j\)^{1-2\e}+\mathcal{O}(r).
\ea
On the second interval the calculation is even more straightforward, as at $d=3$ we only need to evaluate the integral
\ba
{\cal I}_2 &=& \fr{1}{\(4\pi\)^5} \int_r^\infty{\rm d}x~\fr{e^{-M x}}{x^3},
\ea
with $M$ defined by
\ba
M&\equiv&\sum_{i=1}^5 m_i.
\ea
After several partial integrations and the dropping of terms that vanish in the limit $r\rightarrow 0$ we eventually obtain
\ba
{\cal I}_2 &=& \fr{1}{\(4\pi\)^5}\(\fr{e^{-M r}}{2r^2}-\fr{Me^{-M r}}{2r}-\fr{M^2 e^{-M r}\ln \(M r\)}{2}+
\fr{M^2}{2}\int_0^\infty {\rm d}x~\ln x ~e^{-x} \) \nn
&=&\fr{1}{\(4\pi\)^5} \(\fr{1}{2r^2}-\fr{M}{r}+\fr{1}{2}\(\fr{3}{2}-\gamma\)M^2-\fr{M^2\ln \(M r\)}{2}\) +\mathcal{O}(r).
\ea

When adding up ${\cal I}_1$ and ${\cal I}_2$, all terms divergent in the limit $r\rightarrow 0$ cancel. An expansion in $\e$ then gives for the whole integral the simple result
\ba
I_d &=&\fr{M^2}{2(4\pi)^4}\bigg\{\fr{1}{8\e}\bigg[ 1+\sum_{i\neq j} \fr{m_i m_j}{M^2}\bigg]+ \fr{3}{2} +
\ln \fr{\bar{\Lambda}}{M}\nn
&+&\sum_{i\neq j}\fr{m_i m_j}{M^2}\(\ln \( \fr{\bar{\Lambda}}{m_i} \)+\fr{3}{2}-\ln\,2\)\bigg\},
\ea
where we have finally proceeded to the limit $r=0$. When all masses are equal, $m_i=m\,\forall i$, we in particular get
\ba
I_d\(m_i=m\) &=& \fr{45m^2}{(8\pi)^4}\(\fr{1}{\e}+8\,\ln\fr{\bar{\Lambda}}{m}+12-\fr{32}{9}\ln\,2-\fr{40}{9}\ln\,5\).
\ea

\def\fig1dia(#1,#2,#3,#4,#5,#6){\!\!\picb{#2(26.25,15)(15.5,256,76)%
 #3(30,30)(15,30) #1(18.75,15)(15.5,104,284) #4(15,30)(22.5,0)%
 #5(30,30)(22.5,0) #6(15,17.8)(19.3,292.8,39.1)%
 \Text(23,34)[c]{$_1$} \Text(1,15)[c]{$_2$}
 \Text(16,15)[c]{$_3$} \Text(29,15)[c]{$_4$}
 \Text(37,15)[c]{$_5$} \Text(44.5,15)[c]{$_6$}
 \Text(23,-5)[c]{$0$}\Text(15,34.6)[c]{$x$} \Text(31,34.6)[c]{$y$}}}

\begin{figure}[t]
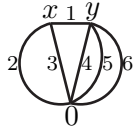

\centering
\ba
\fig1dia(\CArc,\CArc,\Line,\Line,\Line,\CArc)\nonumber
\ea
\caption{The structure of the diagram $e$ in coordinate space.}
\end{figure}

\subsubsection{Diagram {\textit e}}
The coordinate space evaluation of diagrams containing more than two vertices is generally very difficult to approach,
but may anyhow prove successful in certain special cases. One such case is the four-loop graph $e$, which after a
Fourier-transform reads
\ba
\label{intse}
I_e &=& \(\fr{\bar{\Lambda}^{-\e}}{\(4\pi e^{-\gamma}\)^\e}\)^4\int{\rm d^{3-2\e}}x
\int{\rm d^{3-2\e}}y ~ {\rm D}\(|x-y|,m_1\) \nn
&\times & {\rm D}\(x,m_2\){\rm D}\(x,m_3\){\rm D}\(y,m_4\){\rm D}\(y,m_5\){\rm D}\(y,m_6\),
\ea
as portrayed in Fig.~B.4. Here we can at once perform the angular integrations by applying the identity \cite{gkp}
\ba
\int{\rm d}\Omega_\rho \fr{{\rm K_\lambda}\(|r-\rho |\)} {|r-\rho| ^\lambda}
&=& \(2\pi\)^{\lambda+1} \fr{{\rm I_\lambda} \(\rho\)}{\rho^{\lambda}}\fr{{\rm K_\lambda}\(r\)}{r^{\lambda}},
\label{ang}
\ea
where $\lambda=(d-2)/2$ and $r>\rho$. This on the other hand gives us
\ba
I_e &=& \(\fr{1}{\(4\pi e^{-\gamma}\)^\e}\)^4 \int_0^\infty{\rm d}x\int_0^\infty{\rm d}y ~x^{2-2\e} y^{2-2\e}
\prod_{i=2}^{3}{\rm D}\(x,m_i\)\prod_{i=4}^{6} {\rm D}\(y,m_i\)\nn
& \times & \({\rm D}\(x,m_1\){\rm G}\(m_1y\)\theta\(x-y\)+{\rm D}\(y,m_1\){\rm G}\(m_1x\)\theta\(y-x\)\) \nn
&\equiv &\(\fr{1} {4\pi e^{-\gamma}}\)^{4\e}{\cal I}, \label{intse2}
\ea
in which we have defined the function G by
\ba
{\rm G}\(x\)&\equiv&\fr{2^{5/2-\e} \pi ^{3-2\e}\bar{\Lambda}^{-4\e}}{\Gamma\(3/2-\e\)}
\fr{{\rm I_{1/2-\e}}\(x\)}{x^{1/2-\e}}.
\ea
For future purposes we note that at small values of its argument this function has the expansion
\ba
{\rm G}\(x\)&=& \fr{4\pi^{3-2\e}\bar{\Lambda}^{-4\e}}{\Gamma\(3/2-\e\)^2 }\(1+{\cal O}\(x^2\)\)\;\;\,\equiv\;\;\,\kappa\(1+{\cal O}\(x^2\)\),
\ea
and with $\e$ set to zero it reads
\ba
{\rm G}\(x\)&=& \(4\pi\)^2 \fr{\sinh\,x}{x}.
\ea

\newcommand{\kuva}[1]{\;\parbox[c]{120pt}{\begin{picture}(120,120)(0,0)
\SetWidth{1.0}\SetScale{1.0} #1 \end{picture}}\;}

\def\XYDIA{\kuva{\Line(0,0)(120,120)\Line(0,0)(120,0)\Line(0,0)(0,120)\Line(35,0)(35,120)\Line(0,35)(120,35)
\Text(112.5,-5)[c]{$x$}\Text(-5,112.5)[c]{$y$}\Text(24,12)[c]{$A$}\Text(12,24)[c]{$B$}\Text(80,17)[c]{$C$}
\Text(17,80)[c]{$D$}\Text(92,60)[c]{$E$}\Text(60,92)[c]{$F$}\Text(35,-5)[c]{$r$}\Text(-5,35)[c]{$r$}
\Line(117,-3)(120,0)\Line(117,3)(120,0)\Line(-3,117)(0,120)\Line(3,117)(0,120)}}

\begin{figure}[t]
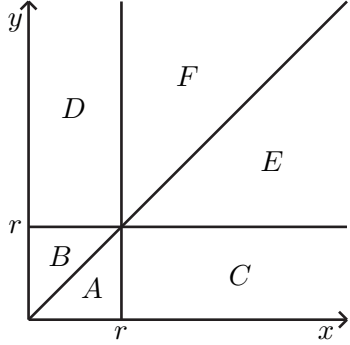

\centering
\ba
\XYDIA \nonumber
\ea
\caption{The division of the $xy$-plane into 6 subregions in the evaluation of the integral of Eq.~(\ref{intse2}).}
\end{figure}

Let us then move on to evaluate the integral ${\cal I}$. In analogy with the computation of the basketball
diagram we again introduce a small positive parameter $r$ to divide the integration plane into different regions in
order to isolate the singularities of the graph. This division is depicted in Fig.~B.5, and it is obvious that at least
the two regions E and F will be safe from UV divergences meaning that we can there set $\e=0$. In the end we will again sum the different contributions together and proceed to the limit $r\rightarrow 0$.

In the computations to follow we will frequently apply the notation
\ba
m_i+m_j+m_k+... &\equiv & M_{ijk...}, \\
m^i\times m^j\times m^k\times ... &\equiv & M^{ijk...}.
\ea

\subsubsection*{Regions A and B}

In the regions A and B the coordinates $x$ and $y$ are confined near the origin as $x,y\leq r$. Denoting $y=tx$ and using the small-$x$ expansions of
the different functions we easily obtain
\ba
{\cal I}_A &=&\int_0^r{\rm d}x\int_0^x{\rm d}y ~x^{2-2\e} y^{2-2\e}{\rm G}
\(m_1y\)\prod_{i=1}^{3}{\rm D}\(x,m_i\)\prod_{i=4}^{6}{\rm D}\(y,m_i\) \nn
&=& \kappa \alpha^{6}\int_0^r{\rm d}x\int_0^1{\rm d}t~x^{-1+8\e}t^{-1+4\e} + \mathcal{O}(r)
\;\;=\;\;\kappa \alpha^{6} \fr{r^{8\e}}{32\e^2} + \mathcal{O}(r),
\ea
and similarly
\ba
{\cal I}_B &=&\int_0^r{\rm d}y\int_0^y{\rm d}x ~x^{2-2\e} y^{2-2\e}{\rm G}\(m_1x\){\rm D}\(y,m_1\)
\prod_{i=2}^{3}{\rm D}\(x,m_i\)
\prod_{i=4}^{6}{\rm D}\(y,m_i\) \nn
&=& \kappa \alpha^{6} \fr{r^{8\e}}{8\e\(1+2\e\)}+ \mathcal{O}(r).
\ea
As we will in the end take the limit of $r=0$, this accuracy suffices to us at present.

\subsubsection*{Region C}
In the region C --- which turns out to be the most problematic one --- the integration variable $y$ in confined near the origin, while $x$ on the other hand ranges from $r$ ro $\infty$. Proceeding in the same fashion as before, we get
\ba
{\cal I}_C &=& \int_r^\infty{\rm d}x\int_0^r{\rm d}y ~x^{2-2\e} y^{2-2\e}{\rm G}\(m_1y\)\prod_{i=1}^{3}{\rm D}\(x,m_i\)
\prod_{i=4}^{6}{\rm D}\(y,m_i\)\nn
&=& \kappa\alpha^3\int_r^\infty{\rm d}x ~x^{2-2\e}\prod_{i=1}^{3}{\rm D}\(x,m_i\)\int_0^r{\rm d}y~y^{-1+4\e}
+ \mathcal{O}(r) \nn
&=&\kappa\alpha^3 \fr{r^{4\e}}{4\e}\int_r^\infty{\rm d}x ~x^{2-2\e}\prod_{i=1}^{3}{\rm D}\(x,m_i\) + \mathcal{O}(r),
\label{ic}
\ea
where the remaining $x$-integral must be computed to $\mathcal{O}(\e)$ in order to obtain also the finite part of
${\cal I}_C$. This can on the other hand be achieved through an application of the $\e$-expansion of D, Eq.~(\ref{propexp}), which gives us
\ba
\label{intd}
&&\int_r^\infty{\rm d}x ~x^{2-2\e}\prod_{i=1}^{3}{\rm D}\(x,m_i\) \\
&& =\;\; \fr{1}{\(4\pi\)^3}\int_r^\infty{\rm d}x~\fr{e^{-M_{123}x}}{x} \bigg[ 1 + \e\bigg(\ln x +
 \ln\fr{(2\pi m^2)^3}{M^{123}}+\sum_{i=1}^{3}e^{2m_i x}{\rm Ei}\(-2m_i x\)\bigg)
+\mc{O}(\e^2)\bigg], \nonumber
\ea
but leaves a number of non-trivial one-dimensional integrals to be computed. Performing several partial integrations
all of these can nevertheless be evaluated analytically, and we finally obtain
\ba
{\cal I}_C &=& -\fr{1}{4\(4\pi\)^4}\bigg\{\Big(\gamma+\ln \(M_{123}r\)\Big)\bigg(\fr{1}{\e}+4+\gamma+\ln\,2
+\ln \fr{\(2\pi \bar{\Lambda}^2r\)^4}{M_{123}M^{123}}\bigg)
+ \fr{\pi ^2}{6} + \gamma ^2\\
&+&\fr{1}{2}\(\ln \(M_{123}r\)\)^2+\sum_{i=1}^{3} \Big[\fr{1}{2}\(\ln \(2m_i r\)\)^2
+\gamma \ln \(2m_i r\)+{\rm Li_2}\(1-\fr{M_{123}}{2m_i}\)\Big] \bigg\}+\mathcal{O}(r) \nonumber
\ea
as the contribution of the region C to the integral $\mc{I}$.

\subsubsection*{Region D}

It can be easily verified that the region D produces no divergences, so we may set $\e =0$ in
the beginning. Then all the required integrals reduce to familiar types and we easily obtain, following the above treatment,
\ba
{\cal I}_D &=& \int_0^r{\rm d}x\int_r^\infty{\rm d}y ~x^2 y^2{\rm G}\(m_1x\){\rm D}\(y,m_1\)\prod_{i=2}^{3}{\rm D}\(x,m_i\)
\prod_{i=4}^{6}{\rm D}\(y,m_i\) \nn
&=&\fr{1}{\(4\pi\)^4 m_1}\int_0^r{\rm d}x~\fr{e^{-M_{23}x}}{x}\sinh \(m_1 x\)
\int_r^\infty{\rm d}y ~\fr{e^{-M_{1456}y}}{y^2} \nn
&=&\fr{1}{\(4\pi\)^4}+\mathcal{O}(r).
\ea

\subsubsection*{Regions E and F}
The regions E and F contain no points near the origin and will therefore also give only finite contributions to the
integral ${\cal I}$. Proceeding as before we get for E,
\ba
{\cal I}_E &=& \int_r^\infty{\rm d}y\int_y^\infty{\rm d}x ~x^2 y^2 {\rm G}\(m_1y\)\prod_{i=1}^{3}{\rm D}\(x,m_i\)
\prod_{i=4}^{6}{\rm D}\(y,m_i\)
\ea
\ba
&=&\fr{1}{\(4\pi\)^4 m_1}\int_r^\infty{\rm d}y ~\fr{e^{-M_{456}y}}{y^2}\sinh \(m_1 y\)
\int_y^\infty{\rm d}x \fr{e^{-M_{123}x}}{x}\nn
&=&-\fr{1}{\(4\pi\)^4 m_1}\int_r^\infty{\rm d}y ~\fr{e^{-M_{456}y}}{y^2}\sinh \(m_1 y\){\rm Ei}\(-M_{123}y\),
\ea
and similarly for F,
\ba
{\cal I}_F &=& \int_r^\infty{\rm d}x\int_x^\infty{\rm d}y ~x^2 y^2{\rm G}\(m_1x\){\rm D}\(y,m_1\)
\prod_{i=2}^{3}{\rm D}\(x,m_i\)
\prod_{i=4}^{6}{\rm D}\(y,m_i\) \nn
&=&\fr{1}{\(4\pi\)^4 m_1}\int_r^\infty{\rm d}x ~\fr{e^{-M_{23}x}}{x}\sinh \(m_1 x\)
\int_x^\infty{\rm d}y \fr{e^{-M_{1456}y}}{y^2} \nn
&=&\fr{1}{\(4\pi\)^4 m_1}\int_r^\infty{\rm d}x ~\fr{e^{-M_{23}x}}{x}\sinh \(m_1 x\) \(\fr{e^{-M_{1456}x}}{x}+
M_{1456}{\rm Ei}\(-M_{1456}x\)\).
\ea
Here the remaining integrals are straightforward to evaluate by employing partial integrations and in the end produce the results
\ba
{\cal I}_E &=& \fr{1}{2\(4\pi\)^4}\bigg\{-2+\fr{\pi^2}{6}+\gamma^2+\fr{M_{1123456}}{m_1}\ln\fr{M_{1123456}}{M_{123}}
-\fr{M_{23456}}{m_1}\ln \fr{M_{23456}}{M_{123}}\nn
&+&\fr{M_{1456}}{m_1}{\rm Li_2}\(-\fr{M_{1456}}{M_{123}}\)
+\(1-\fr{M_{456}}{m_1}\){\rm Li_2}\(-\fr{-m_1+M_{456}}{M_{123}}\)
+2\gamma\,\ln \(M_{123}r\)\nn
&+&\(\ln \(M_{123}r\)\)^2\bigg\}+\mathcal{O}(r), \\
{\cal I}_F&=&\fr{1}{2\(4\pi\)^4}\bigg\{2\(1-\gamma\)+\fr{M_{23456}}{m_1}\ln \(M_{23456}r\)
-\fr{M_{1123456}}{m_1}\ln \(M_{1123456}r\)\nn
&+&\fr{M_{1456}}{m_1}\bigg({\rm Li_2}\(-\fr{M_{123}}{M_{1456}}\)-{\rm Li_2}\(-\fr{-m_1+M_{23}}{M_{1456}}\)\bigg)\bigg\}+\mathcal{O}(r).
\ea

\subsubsection*{Final result}
All that is left to do is to add up the contributions of the different regions to the initial integral and proceed to the limit $r\rightarrow 0+$. This finally gives us as the result for the whole diagram
\ba
I_e &=& \fr{1}{2(8\pi)^4}\Bigg\{\fr{1}{\e^2}
+\fr{8}{\e}\bigg[1+\ln\fr{\bar{\Lambda}}{M_{123}}\bigg]+4\bigg[13+\fr{7\pi^2}{12}-3\(\ln\,2\)^2 +16\,\ln
\fr{\bar{\Lambda}}{M_{123}}+8\(\ln \fr{\bar{\Lambda}}{M_{123}}\)^2 \nn
&+&2\,\ln\,2\,\ln \fr{\(M_{123}\)^3}{M^{123}}+4\(1-\fr{M_{456}}{m_1}\){\rm Li_2}\(-\fr{-m_1+M_{456}}{M_{123}}\)\
+4\fr{M_{1456}}{m_1}{\rm Li_2}\(-\fr{M_{1456}}{M_{123}}\)\nn
&+&4\fr{M_{1456}}{m_1}\bigg({\rm
Li_2}\(-\fr{M_{123}}{M_{1456}}\)-{\rm Li_2}\(-\fr{-m_1+M_{23}}{M_{1456}}\)\bigg)\nn
&+& -\sum_{i=1}^{3}\bigg\{\(\ln \fr{m_i}{M_{123}}
\)^2+2{\rm Li_2}\(1-\fr{M_{123}}{2m_i}\)\bigg\} \bigg]\Bigg\}.
\ea

\subsubsection*{Special mass configurations}
Important special mass configurations include the identical mass case, $m_i=m\;\forall i$, as well as the one with
$m_3=0,~ m_1=m_2=m_4=m_5=m_6=m$. Using the easily verifiable dilogarithmic relations
\ba
{\rm Li_2}\(-x\)&=&-\fr{\pi^2}{6}-\fr{1}{2}\(\ln \(1+x\)\)^2+\int_x^\infty{\rm d}t~ \fr{\ln \(1+t\)}{t\(1+t\)},\nn
{\rm Li_2}\(-x\)+{\rm Li_2}\(-\fr{1}{x}\)&=&-\fr{1}{2}\(\ln \(-x\)\)^2-\fr{\pi ^2}{6},
\ea
we straightforwardly get
\ba
I_e\(m_i=m\) &=& \fr{1}{2(8\pi)^4}\bigg\{\fr{1}{\e^2}+\fr{8}{\e}\bigg[1+\ln \fr{\bar{\Lambda}}{3m} \bigg]
+4\bigg[ 13+\fr{7\pi^2}{12} +16\,\ln \fr{\bar{\Lambda}}{3m}+8\(\ln \fr{\bar{\Lambda}}{3m}\)^2\nn
&-&3\(\ln\fr{2}{3}\)^2+16\({\rm Li_2}\(-\fr{3}{4}\)+{\rm Li_2}\(-\fr{4}{3}\)\)-8\,{\rm Li_2}\(-\fr{2}{3}\)\nn
&-&6\,{\rm Li_2}\(-\fr{1}{2}\)
-16\,{\rm Li_2}\(-\fr{1}{4}\) \bigg] \bigg\}, \\
I_e\(m_3=0\) &=& \fr{1}{2(8\pi)^4}\bigg\{\fr{1}{\e^2}+\fr{8}{\e}\bigg[1+\ln \fr{\bar{\Lambda}}{2m} \bigg] +
4\bigg[13-\fr{13\pi^2}{12}-8\(\ln\,2\)^2+16\,\ln\fr{\bar{\Lambda}}{2m}\nn
&+&8\(\ln\fr{\bar{\Lambda}}{2m}\)^2\bigg]\bigg\},
\ea
where especially the remarkably simple form of the latter result is worthy of some attention.

\subsection{Momentum space calculations}
Diagrams containing arbitrarily many vertices but only trivial one-loop subdiagrams are most easily handled directly in momentum space. The number of such graphs is, however, small and analytic results are so far available only in a few special cases. We will now briefly address the computation of three four-loop diagrams, for which at least simple integral representations can be derived using this method.

\subsubsection{Diagram {\textit a}}
The triangle diagram has previously been studied both in three and four dimensions \cite{bn2,kast1} with the result that its
divergent part has been successfully obtained. Here our objective is, however, to compute its finite term as well, and we
start this by defining a scalar two-point function
\ba
\tilde{\Pi}(p,m_1,m_2)&\equiv& \int_q \fr{1}{q^2+m_1^2}\fr{1}{\(q-p\)^2+m_2^2}
\ea
analogous to the polarization function of appendix B.1. This allows us to write the diagram in the form
\ba
I_a&=&\int_p \tilde{\Pi}(p,m_1,m_2)\tilde{\Pi}(p,m_3,m_4)\tilde{\Pi}(p,m_5,m_6),
\label{ia}
\ea
which serves as a useful starting point for our analysis.

The momentum space integral appearing in the definition of $\tilde{\Pi}$ can be easily performed by introducing a Feynman
parameter $x$ to combine the two propagators in the integrand (see also Eq.~(\ref{rtyui})). We thereby get
\ba
\tilde{\Pi}(p,m_1,m_2) &=& \fr{e^{\gamma\e}\bar{\Lambda}^{2\e}\Gamma(1/2+\e)}{8\pi^{3/2}\(p^2\)^{1/2+\e}}
\int_0^1{\rm d}x \fr1{\(x(1-x)+\(xm_1^2+(1-x)m_2^2\)/p^2\)^ {1/2+\e}}\nn
&\equiv&  \fr{e^{\gamma\e}\bar{\Lambda}^{2\e}\Gamma(1/2+\e)}{8\pi^{3/2}\(p^2\)^{1/2+\e}}
{\rm B }\(m_1^2/p^2,m_2^2/p^2,\e\),
\label{subd}
\ea
where the function B possesses the easily verifiable properties
\ba
\label{b1}
\lim_{x_1,x_2 \to 0} {\rm B }\(x_1,x_2,\e\) &=& \fr{2^{2\e}\sqrt{\pi}\Gamma(1/2-\e)}{\Gamma(1-\e)},  \\
{\rm B }\(x_1,x_2,0\) &=& 2\,{\rm arccot}(x_1^{1/2} + x_2^{1/2}).
\label{b2}
\ea
If we now define $p\equiv\bar{\Lambda} u$, Eq.~(\ref{ia}) becomes
\ba
\label{intrea}
I_a &=&\fr{2e^{4\gamma\e}}{(4\pi)^6}
\fr{\Gamma(1/2+\e)^{3}}{\Gamma(3/2-\e)} \int_0^\infty{\rm d}u ~u^{-1-8\e}\nn
&\times&{\rm B }\(\fr{m_1^2}{u^2\bar{\Lambda}^2},\fr{m_2^2}{u^2\bar{\Lambda}^2},\e\){\rm B }\(\fr{m_3^2}{u^2\bar{\Lambda}^2},\fr{m_4^2}{u^2\bar{\Lambda}^2},\e\)
{\rm B }\(\fr{m_5^2}{u^2\bar{\Lambda}^2},\fr{m_6^2}{u^2\bar{\Lambda}^2},\e\),
\ea
which after a partial integration in $u$ gives
\ba
I_a &=& \fr{2e^{4\gamma\e}}{(4\pi)^6}
\fr{\Gamma(1/2+\e)^{3}}{\Gamma(3/2-\e)}\fr{1}{8\e}
\int_0^\infty{\rm d}u ~u^{-8\e}\nn
&\times& \fr{d}{du}
\lk {\rm B }\(\fr{m_1^2}{u^2\bar{\Lambda}^2},\fr{m_2^2}{u^2\bar{\Lambda}^2},\e\){\rm B }\(\fr{m_3^2}{u^2\bar{\Lambda}^2},\fr{m_4^2}{u^2\bar{\Lambda}^2},\e\)
{\rm B }\(\fr{m_5^2}{u^2\bar{\Lambda}^2},\fr{m_6^2}{u^2\bar{\Lambda}^2},\e\)\rk.
\ea
Here the $1/\e$ part of the integral has been separated, which means we can expand the factor $u^{-8\e}$ in $\e$. Using
Eqs.~(\ref{b1}) and (\ref{b2}) this finally leads us to the result
\ba
I_a &=& \fr{2e^{4\gamma\e}}{(4\pi)^6}\fr{\Gamma(1/2+\e)^{3}}{\Gamma(3/2-\e)}\bigg\{\fr{1}{8\e}
\(\fr{2^{2\e}\sqrt{\pi}\Gamma(1/2-\e)}{\Gamma(1-\e)}\)^{3} \label{aaerd} \\
&-&8\int_0^\infty{\rm d}u \,\ln\,u\fr{d}{du}\lk\arctan\(\fr{u\bar{\Lambda}}{m_1+m_2}\)\arctan\(\fr{u\bar{\Lambda}}{m_3+m_4}\)
\arctan\(\fr{u\bar{\Lambda}}{m_5+m_6}\)\rk \bigg\}, \nonumber
\ea
where terms of order $\e$ have been dropped.

The final result for the triangle diagram is now available through a simple $\e$-expansion of Eq.~(\ref{aaerd}), which
gives us
\ba
I_a &=& 2^{-13}\pi^{-2}\Bigg\{\fr{1}{\e}+4\,\ln\,2+2
-\fr{64}{\pi^{3}}\int_0^\infty{\rm d}u\, \ln\,u \nn
&\times&\fr{d}{du}\lk\arctan\(\fr{u\bar{\Lambda}}{m_1+m_2}\)\arctan\(\fr{u\bar{\Lambda}}{m_3+m_4}\)
\arctan\(\fr{u\bar{\Lambda}}{m_5+m_6}\)\rk\Bigg\}.
\ea
Here the remaining integral can be analytically performed only in certain special cases, which fortunately include the one with identical propagator masses ($m_i = m ~ \forall i$),
\ba
I_a\(m_i=m\) &=& 2^{-13}\pi^{-2}\bigg\{\fr{1}{\e}+4\,\ln\fr{\bar{\Lambda}^2}{2m^2}+2-\fr{84}{\pi^2}\zeta\(3\)\bigg\}.
\ea

\subsubsection{Diagrams $b$ and $c$}

The graphs $b$ and $c$ are clearly finite in three dimensions. Using the above results for the function $\tilde{\Pi}$ we
may almost instantly write for them the integral representations
\ba
I_b &=& 2^{-7}\pi^{-5}\int_0^\infty{\rm d}p
\,p^{-1}\(p^2+m_7^2\)^{-1} \nn
&\times&\arctan\(\fr{p}{m_1+m_2}\)\arctan\(\fr{p}{m_3+m_4}\)\arctan\(\fr{p}{m_5+m_6}\), \\
I_c &=& 2^{-8}\pi^{-6}\int_0^\infty{\rm d}p\int_0^\infty{\rm d}q
\arctan\(\fr{p}{m_1+m_2}\)\arctan\(\fr{q}{m_3+m_4}\)\nn
&\times&\fr{\ln ((p+q)^2+m_7^2)
/((p-q)^2+m_7^2)}{(p^2+m_5^2)(q^2+m_6^2)}.
\ea
The analytic evaluation of these integrals has, however, not been possible even in the identical mass case, so at
present these expressions only serve as sources for accurate numerical results.




\begin{thebibliography}{99}

\addcontentsline{toc}{chapter}{The bibliography}

\bibitem{avsusc}
A.~Vuorinen,
Phys.\ Rev.\ D {\bf 67} (2003) 074032
[hep-ph/0212283].

\bibitem{avpres}
A.~Vuorinen,
Phys.\ Rev.\ D {\bf 68} (2003) 054017
[hep-ph/0305183].


\bibitem{avyork}
Y.~Schr\"oder and A.~Vuorinen,
``High-precision evaluation of four-loop vacuum bubbles in three dimensions,''
hep-ph/0311323


\bibitem{fmcl}
B.~A.~Freedman and L.~D.~McLerran,
Phys.\ Rev.\ D {\bf 16} (1977) 1169.




\bibitem{az}
P.~Arnold and C.~X.~Zhai,
Phys.\ Rev.\ D {\bf 50} (1994) 7603
[hep-ph/9408276];
Phys.\ Rev.\ D {\bf 51} (1995) 1906
[hep-ph/9410360].

\bibitem{lapo}
S.~Laporta,
Int.\ J.\ Mod.\ Phys.\ A {\bf 15} (2000) 5087
[hep-ph/0102033].

\bibitem{karsch2}
C.~R.~Allton, S.~Ejiri, S.~J.~Hands, O.~Kaczmarek, F.~Karsch, E.~Laermann and C.~Schmidt,
``The equation of state for two flavor QCD at non-zero chemical  potential,''
hep-lat/0305007.

\bibitem{kap}
J.~Kapusta,``Finite-temperature field theory", Cambridge University Press, Cambridge (1989).

\bibitem{leb}
M.~Le~Bellac, ``Thermal field theory", Cambridge University Press, Cambridge (1996).


\bibitem{gpy}
D.J.~Gross, R.D.~Pisarski and L.G.~Yaffe,
Rev.\ Mod.\ Phys.\ {53} (1981) 43.



\bibitem{tt2}
T.~Toimela,
Int.\ J.\ Theor.\ Phys.\  {\bf 24} (1985) 901
[Erratum-ibid.\  {\bf 26} (1987) 1021].

\bibitem{QCD1}
H.~Fritzsch, M.~Gell-Mann and H.~Leutwyler,
Phys.\ Lett.\ B {\bf 47} (1973) 365.


\bibitem{growil}
D.~J.~Gross and F.~Wilczek,
Phys.\ Rev.\ Lett.\  {\bf 30} (1973) 1343.

\bibitem{pol}
H.~D.~Politzer,
Phys.\ Rev.\ Lett.\  {\bf 30} (1973) 1346.

\bibitem{dris}
D.~H.~Rischke,
``The quark-gluon plasma in equilibrium,''
nucl-th/0305030.

\bibitem{laeph}
E.~Laermann and O.~Philipsen,
``Status of lattice QCD at finite temperature,''
hep-ph/0303042.


\bibitem{rajwil}
M.~G.~Alford, K.~Rajagopal and F.~Wilczek,
Phys.\ Lett.\ B {\bf 422} (1998) 247
[hep-ph/9711395].

\bibitem{rajwil2}
K.~Rajagopal and F.~Wilczek,
``The condensed matter physics of QCD,''
hep-ph/0011333.

\bibitem{rothe}
Heinz~J.~Rothe,
``Lattice gauge theories - an introduction", World Scientific (1997).

\bibitem{bl}
D.~Bailin and A.~Love,
``Introduction to gauge field theory", IOP publishing, London (1993).

\bibitem{taylor}
J.~C.~Taylor,
``Gauge theories of weak interactions", Cambridge University Press, Cambridge (1975).

\bibitem{ippdis}
A.~Ipp,
``Quantum Corrections to Thermodynamic Properties in the Large Nf limit of the Quark Gluon Plasma", Academic dissertation, Technical University of Vienna (2003).


\bibitem{antti}
A.~Gynther,
Phys.\ Rev.\ D {\bf 68} (2003) 016001
[hep-ph/0303019].

\bibitem{cutsiv}
R.~Cutler and D.~W.~Sivers,
Phys.\ Rev.\ D {\bf 17} (1978) 196.



\bibitem{es}
E.~V.~Shuryak,
Sov.\ Phys.\ JETP {\bf 47} (1978) 212
[Zh.\ Eksp.\ Teor.\ Fiz.\  {\bf 74} (1978) 408];
S.~A.~Chin,
Phys.\ Lett.\ B {\bf 78} (1978) 552.



\bibitem{jk}
J.~I.~Kapusta,
Nucl.\ Phys.\ B {\bf 148} (1979) 461.



\bibitem{tt}
T.~Toimela,
Phys.\ Lett.\ B {\bf 124} (1983) 407.



\bibitem{zk}
C.~X.~Zhai and B.~Kastening,
Phys.\ Rev.\ D {\bf 52} (1995) 7232
[hep-ph/9507380].



\bibitem{bn1}
E.~Braaten and A.~Nieto,
Phys.\ Rev.\ D {\bf 53} (1996) 3421
[hep-ph/9510408].

\bibitem{dr}
P. Ginsparg,
Nucl.\ Phys.\ B 170 (1980) 388;
%
T. Appelquist and R.D. Pisarski,
Phys.\ Rev.\ D 23 (1981) 2305.


\bibitem{klrs2}
K.~Kajantie, M.~Laine, K.~Rummukainen and M.~E.~Shaposhnikov,
Nucl.\ Phys.\ B {\bf 458} (1996) 90
[hep-ph/9508379].

\bibitem{klry}
K.~Kajantie, M.~Laine, K.~Rummukainen and Y.~Schr\"oder,
Phys.\ Rev.\ D {\bf 67} (2003) 105008 [hep-ph/0211321].



\bibitem{klry2}
K.~Kajantie, M.~Laine, K.~Rummukainen and Y.~Schr\"oder,
JHEP {\bf 0304} (2003) 036
[hep-ph/0304048].

\bibitem{yorkll}
Y.~Schroder,
Nucl.\ Phys.\ Proc.\ Suppl.\  {\bf 116} (2003) 402
[hep-ph/0211288].


\bibitem{form}
J.~A.~Vermaseren,
``New features of FORM,''
math-ph/0010025.


\bibitem{linde}
A.~D.~Linde,
Phys.\ Lett.\ B {\bf 96} (1980) 289.



\bibitem{ikrv}
A.~Ipp, K.~Kajantie, A.~Rebhan and A.~Vuorinen,
in preparation.


\bibitem{karsch1}
G.~Boyd, J.~Engels, F.~Karsch, E.~Laermann, C.~Legeland, M.~Lutgemeier and B.~Petersson,
Nucl.\ Phys.\ B {\bf 469} (1996) 419
[hep-lat/9602007].



\bibitem{fodor}
Z.~Fodor, S.~D.~Katz and K.~K.~Szabo,
``The QCD equation of state at nonzero densities: Lattice result,''
hep-lat/0208078.


\bibitem{rummu1}
C.~W.~Bernard {\it et al.},
Nucl.\ Phys.\ Proc.\ Suppl.\  {\bf 60A} (1998) 195.

\bibitem{gup4}
R.~V.~Gavai and S.~Gupta,
``Pressure and non-linear susceptibilities in QCD at finite chemical  potentials,''
hep-lat/0303013.



\bibitem{birn1}
J.~P.~Blaizot, E.~Iancu and A.~Rebhan,
Phys.\ Rev.\ Lett.\  {\bf 83} (1999) 2906 [hep-ph/9906340]; Phys.\ Lett.\ B {\bf 470}
(1999) 181 [hep-ph/9910309]; Phys.\ Rev.\ D {\bf 63} (2001) 065003 [hep-ph/0005003]; hep-ph/0303185.


\bibitem{htl2loop}
J.~O.~Andersen, E.~Braaten, E.~Petitgirard and M.~Strickland,
Phys.\ Rev.\ D {\bf 66} (2002) 085016
[hep-ph/0205085].

\bibitem{htl2loopq}
J.~O.~Andersen, E.~Petitgirard and M.~Strickland,
``Two-loop HTL thermodynamics with quarks,''
hep-ph/0302069.

\bibitem{andstr}
J.~O.~Andersen and M.~Strickland,
Phys.\ Rev.\ D {\bf 66} (2002) 105001
[hep-ph/0206196].



\bibitem{moore}
G.~D.~Moore,
JHEP {\bf 0210} (2002) 055
[hep-ph/0209190];
A.~Ipp, G.~D.~Moore and A.~Rebhan,
JHEP {\bf 0301} (2003) 037
[hep-ph/0301057].

\bibitem{ippreb}
A.~Ipp and A.~Rebhan,
``Thermodynamics of large-N(f) QCD at finite chemical potential,''
hep-ph/0305030.


\bibitem{avippreb}
A.~Ipp, A.~Rebhan and A.~Vuorinen,
``Perturbative QCD at non-zero chemical potential: Comparison with the large-Nf limit and apparent convergence,''
hep-ph/0311200.

\bibitem{igr}
A.~Ipp, A.~Gerhold and A.~Rebhan,
Phys.\ Rev.\ D {\bf 69} (2004) 011901
[hep-ph/0309019].



\bibitem{parma}
G.~Burgio, F.~Di Renzo, G.~Marchesini, E.~Onofri, M.~Pepe and L.~Scorzato,
Nucl.\ Phys.\ Proc.\ Suppl.\  {\bf 63} (1998) 808
[hep-lat/9709106];
F.~Di Renzo and L.~Scorzato,
Nucl.\ Phys.\ Proc.\ Suppl.\  {\bf 94} (2001) 567
[hep-lat/0010064];
F.~Di Renzo, A.~Mantovi, V.~Miccio and Y.~Schroder,
``Four loop stochastic perturbation theory in 3d SU(3),''
hep-lat/0309111.



\bibitem{hlp}
A.~Hart, M.~Laine and O.~Philipsen,
Nucl.\ Phys.\ B {\bf 586} (2000) 443
[hep-ph/0004060].



\bibitem{bir}
J.~P.~Blaizot, E.~Iancu and A.~Rebhan,
Phys.\ Lett.\ B {\bf 523} (2001) 143
[hep-ph/0110369].


\bibitem{ac}
T.~Appelquist and J.~Carazzone,
Phys.\ Rev.\ D {\bf 11} (1975) 2856.

\bibitem{huli}
S.~z.~Huang and M.~Lissia,
Nucl.\ Phys.\ B {\bf 438} (1995) 54
[hep-ph/9411293].

\bibitem{lainesewm}
M.~Laine,
``What is the simplest effective approach to hot QCD thermodynamics?,''
hep-ph/0301011.

\bibitem{bir3}
J.~P.~Blaizot, E.~Iancu and A.~Rebhan,
Phys.\ Rev.\ D {\bf 68} (2003) 025011
[hep-ph/0303045].



\bibitem{klrs}
K.~Kajantie, M.~Laine, K.~Rummukainen and M.~E.~Shaposhnikov,
Nucl.\ Phys.\ B {\bf 503} (1997) 357 [hep-ph/9704416].


\bibitem{pslq}
H.~R.~P.~Ferguson, D.~H.~Bailey,
RNR Technical Report RNR-91-032, NASA Ames Research Center, Moffett Field, CA. December 1991.


\bibitem{boyd}
G.~Boyd {\it et al.}, 
Nucl.\ Phys.\  {\bf B469}, 419  (1996)
[hep-lat/9602007];
%
A.~Papa,
Nucl.\ Phys.\  {\bf B478}, 335  (1996)
[hep-lat/9605004];
%
B.~Beinlich, F.~Karsch, E.~Laermann and A.~Peikert,
Eur.\ Phys.\ J.\  {\bf C6}, 133  (1999)
[hep-lat/9707023];
%
M.~Okamoto {\it et al.}  [CP-PACS Collaboration],
Phys.\ Rev.\  {\bf D60}, 094510  (1999)
[hep-lat/9905005].


\bibitem{gup3}
R.~V.~Gavai, S.~Gupta and P.~Majumdar,
Phys.\ Rev.\ D {\bf 65} (2002) 054506
[hep-lat/0110032].

\bibitem{blm}
S.~J.~Brodsky, G.~P.~Lepage and P.~B.~Mackenzie,
Phys.\ Rev.\ D {\bf 28} (1983) 228.

\bibitem{gup1}
S.~Gupta,
Phys.\ Rev.\ D {\bf 64} (2001) 034507
[hep-lat/0010011].




\bibitem{gup2}
R.~V.~Gavai and S.~Gupta,
``Valence quarks in the QCD plasma: Quark number susceptibilities and  screening,''
hep-lat/0211015.



\bibitem{ramond}
P.~Ramond,
``Field theory - a modern primer,''
Benjamin/Cummings, Reading (1981).

\bibitem{wolfram}
http://functions.wolfram.com/10.08.17.0009.01


\bibitem{gr}
I.S.~Gradshteyn and I.M.~Ryzhik,
"Table of Integrals, Series and Products", 6th edition, Academic Press (2000).


\bibitem{avgradu}
A.~Vuorinen,
Master's Thesis, University of Helsinki (2001), \\ http://ethesis.helsinki.fi/julkaisut/mat/fysii/pg/vuorinen/fourloop.pdf

\bibitem{gkp}
S.~Groote, J.~G.~Korner and A.~A.~Pivovarov,
Eur.\ Phys.\ J.\ C {\bf 11} (1999) 279
[hep-ph/9903412].



\bibitem{arttu}
A.~K.~Rajantie,
Nucl.\ Phys.\ B {\bf 480} (1996) 729
[Erratum-ibid.\ B {\bf 513} (1998) 761]
[hep-ph/9606216].



\bibitem{bn2}
E.~Braaten and A.~Nieto,
Phys.\ Rev.\ D {\bf 51} (1995) 6990
[hep-ph/9501375].


\bibitem{kast1}
B.~Kastening,
Phys.\ Rev.\ D {\bf 54} (1996) 3965
[hep-ph/9604311].


\end{thebibliography}
\end{document}